\documentclass[a4paper,aps,showpacs,amsmath,amssymb,twocolumn,pra,superscriptaddress,notitlepage,accepted=2025-06-19]{quantumarticle}
\pdfoutput=1

\usepackage{qcircuit}
\usepackage{amsmath,bm}
\usepackage[dvips]{graphicx}
\usepackage{amsmath,amssymb,amsthm,mathrsfs,amsfonts,dsfont}
\usepackage{subfigure, epsfig}
\usepackage{braket}
\usepackage{bbm}
\usepackage{plain}
\usepackage{bm}
\usepackage{enumerate}
\usepackage{color}
\usepackage{graphicx}
\usepackage{algorithm}
\usepackage{algorithmic}
\usepackage{braket}
\usepackage{comment}
\usepackage{appendix}
\usepackage{here}
\usepackage{tabularx}
\usepackage{todonotes}
\usepackage{enumitem}
\usepackage{tikz,pgfplots}

\usepackage{hyperref}
\usepackage{natbib}

\theoremstyle{plain}

\newtheorem{lemma}{Lemma}

\newtheorem*{thm*}{Theorem}
\newtheorem*{lemma*}{Lemma}
\newtheorem*{remark*}{Remark}

\newtheorem*{corollary*}{Corollary}
\newtheorem{prop}{Proposition}

\newcommand{\tr}{\mathrm{Tr}}

\begin{document}


\title{Density matrix representation of hybrid tensor networks for noisy quantum devices}


\author{Hiroyuki Harada}
\email{hiro.041o\_i7@keio.jp}
\affiliation{Department of Applied Physics and Physico-Informatics, Keio University, Hiyoshi 3-14-1, Kohoku, Yokohama 223-8522, Japan}

\author{Yasunari Suzuki}
\affiliation{NTT Computer and Data Science Laboratories, NTT Corporation, 3-9-11 Midori-cho, Musashino-shi, Tokyo 180-8585, Japan}

\author{Bo Yang}
\affiliation{LIP6, Sorbonne Université, 4 Place Jussieu, Paris 75005, France}
\affiliation{Graduate School of Information Science and Technology, The University of Tokyo, 7-3-1 Hongo, Bunkyo-ku, Tokyo 113-8656, Japan}

\author{Yuuki Tokunaga}
\affiliation{NTT Computer and Data Science Laboratories, NTT Corporation, 3-9-11 Midori-cho, Musashino-shi, Tokyo 180-8585, Japan}

\author{Suguru Endo}
\email{suguru.endou@ntt.com}
\affiliation{NTT Computer and Data Science Laboratories, NTT Corporation, 3-9-11 Midori-cho, Musashino-shi, Tokyo 180-8585, Japan}
\affiliation{JST, PRESTO, 4-1-8 Honcho, Kawaguchi, Saitama, 332-0012, Japan}


\begin{abstract}
 The hybrid tensor network (HTN) method is a general framework allowing for the construction of an effective wavefunction with the combination of classical tensors and quantum tensors, i.e., amplitudes of quantum states. In particular, hybrid tree tensor networks (HTTNs) are very useful for simulating larger systems beyond the available size of the quantum hardware. However, while the realistic quantum states in NISQ hardware are highly likely to be noisy, this framework is formulated for pure states. In this work, as well as discussing the relevant methods, i.e., Deep VQE and entanglement forging under the framework of HTTNs, we investigate the noisy HTN states by introducing the expansion operator for providing the description of the expansion of the size of simulated quantum systems and the noise propagation. This framework enables the general tree HTN states to be explicitly represented and their physicality to be discussed. We also show that the expectation value of a measured observable exponentially vanishes with the number of contracted quantum tensors. Our work will lead to providing the noise-resilient construction of HTN states. 
\end{abstract}

\maketitle

\section{introduction}
 Noisy intermediate-scale quantum (NISQ) computers are expected to have many applications ranging from quantum computational chemistry~\cite{mcardle2020quantum,cao2019quantum} to machine learning~\cite{romero2017quantum,mitarai2018quantum} and even quantum sensing~\cite{kaubruegger2019variational,koczor2020variational}. However, despite the potential significant advantages of NISQ devices~\cite{cerezo2021variational,endo2021hybrid,tilly2022variational}, many obstacles need to be overcome for practical applications.
 One of the significant problems is the scalability of quantum computers: the largest number of integrated qubits is restricted to the order of $10^2$~\cite{arute2019quantum,kim2023evidence}, and we need technical leaps to scale quantum computers beyond. 
 
 Considering the current situation of scalability of NISQ computers, a couple of methods are being developed to effectively enlarge the simulated quantum systems, e.g., entanglement forging~\cite{eddins2022doubling,huembeli2022entanglement,castellanos2023quantum}, Deep variational quantum eigensolver (DeepVQE)~\cite{fujii2022deep,mizuta2021deep,erhart2022constructing}, and hybrid tensor networks \textcolor{black}{(HTN)}~\cite{yuan2021quantum,kanno2021quantum,kanno2023quantum}. Additional operations on quantum computers and classical post-processing of measurement outcomes allow us to virtually emulate the effect of quantum entanglement between subsystems. Note that, while these methods work for the expectation value estimation, a broad range of NISQ algorithms, such as the variational quantum eigensolver (VQE), construct the target cost function from the expectation values of observables~\cite{peruzzo2014variational,mcardle2020quantum,cerezo2021variational}; therefore, they are quite useful for large-scale simulations in the NISQ era.

 In particular, the HTN framework is a quite general framework that comprehensively discusses the hybridization of quantum tensors and classical tensors~\cite{yuan2021quantum}. While classical tensors are the same as conventional tensors, quantum tensors refer to amplitudes of quantum states, which have two types of indices: classical index and quantum index. Suppose an example of a quantum tensor $\psi_{j_1,j_2,...,j_n}^{i}$, corresponding to the amplitude of a quantum state in the set $\{\ket{\psi^{i}}=\sum_{i_1, i_2, ...,i_k} \psi_{j_1,j_2,...,j_n}^{i} \ket{j_1,j_2,...,j_n}\}_{i}$, where $\ket{j_k}$~($j_k \in \{0,1\}~\forall k $) represents a computational basis state; the upper index $i$ denotes a classical index, specifying a quantum state in the set $\{\ket{\psi^{i}}\}_{i}$, while the lower indices $j_{1},j_{2},...,j_{n}$ indicate quantum indices relevant to physical qubits. Accordingly, the classical indices are contracted similarly to a conventional classical tensor, while the quantum indices are contracted through quantum measurements. By connecting both quantum and classical tensors defined above, the HTN framework expresses the quantum-classical hybrid tensor networks.
 Especially, it has been pointed out that hybrid tree tensor networks (HTTNs) are significantly useful for simulating larger quantum systems~\cite{yuan2021quantum,schuhmacher2024hybridtreetensornetworks}.
 
 While the above frameworks for simulating larger quantum systems are appealing for small-scale NISQ devices, another massive problem in NISQ devices is inevitable noise effects due to unwanted system interactions with the environment~\cite{endo2021hybrid,cerezo2021variational}. We remark that those methods for expanding the simulated quantum systems have been analyzed only for pure states under the assumption that noise effects are negligible~\cite{yuan2021quantum,fujii2022deep,eddins2022doubling}.
 However, such a strong assumption does not reflect the reality of the current situation in which NISQ devices suffer from a significant amount of computation errors. 
 
 In the present paper, we try to bridge the gap between the theoretical framework and the actual experiments. To do so, we first extend the previously known contraction method for $2\times2$ local tensors, as studied in prior research~\cite{yuan2021quantum,kanno2021quantum}, to a more general contraction method for $2^n\times2^n$ matrices ($n\geq1$). Then, we show that Deep VQE and entanglement forging can be discussed with the language of HTN, especially HTTNs. Next, we introduce the expansion operator to consistently describe the propagation of noise and expansion of the simulated quantum system in HTN states. 
 Note that we detail the form of the expansion operator depending on the type of tensor contractions. This argument allows us to represent the explicit form of general multiple-layer tree HTN states and discuss their physicality, i.e., whether the noisy HTN states $\tilde{\rho}_{\rm HT}$ satisfy the positivity $\tilde{\rho}_{\rm HT} \geq 0$. In addition, the effect of noise on the resultant expectation value is another important issue. We consider the global depolarizing error for each quantum tensor, and we reveal that the expectation value decreases exponentially with the number of quantum tensors involved in the contraction procedure of HTTNs. This result sheds light on the importance of balancing the number of classical and quantum tensors to extract useful information from the HTN calculation.

 The remainder of this paper is organized as follows. In Sec. \ref{Sec: hybridtensor}, we review the framework of HTN, highlighting the importance of tree tensor networks and classifying the contraction rules of quantum tensors. In Sec. \ref{sec:preliminaries}, we reformulate the Deep VQE and entanglement-forging methods under the framework of HTTNs. In Secs.~\ref{sec:operator_based_representation}--\ref{sec:application}, we study the explicit form of the expansion operator in accordance with the contraction rules and discuss the physicality of the HTN states. In Sec. \ref{Sec: exponentialdecay}, we discuss the exponential decay of the observable with the number of quantum tensors. Finally, we conclude this paper with discussions and conclusions.

\label{section: introduction}
\section{Preliminaries}
\label{Sec: hybridtensor}

\subsection{Notation}
\label{Sec: notation}
In this paper, each notation is defined as it is introduced; however, for the sake of clarity, we use some symbols with a consistent meaning throughout the text. Specifically, $N$ is used associated with the number of systems, $n$ and $b$ the number of qubits (or indices), $\kappa$ and $d$ the dimensions, $\alpha, \beta, \gamma, \lambda$ the tensors, $M, S, O$ the observables, $H$ the Hamiltonian, and $P$ the Pauli operators, $U$ and $V$ unitary operators. Additionally, calligraphic symbols are used to represent superoperators. Also, we let $\mathrm{L}(\mathbb{C}^{d})$ be the set of  square linear operators on $\mathbb{C}^{d}$ and $\mathrm{H}(\mathbb{C}^{d})$ be the set of Hermitian operators on $\mathbb{C}^{d}$. We sometimes add subscripts to operators to indicate the system on which the operator acts, e.g., the operator acting on system $\rm A$ will be denoted by $O_{\rm A}$.

\subsection{Hybrid tensor networks}\label{sec: htn}

 Classical tensor network (TN) methods~\cite{white1992density,white1993density,schollwock2011density,shi2006classical,vidal2007entanglement} try to efficiently describe quantum states of interest with a much smaller subset of the entire Hilbert space on the basis of physical observation. For example, the matrix product state (MPS)~\cite{schollwock2011density} ansatz represents the state as
\begin{equation}\label{eq:mps}
\ket{\rm MPS}=\sum_{j_1,j_2,...,j_n} \tr[\alpha_1^{j_1}\alpha_2^{j_2}...\alpha_{n}^{j_n}] \ket{j_1,j_2,...,j_n},
\end{equation}
 where $j_k $ takes $0$ or $1$ in the case of qubit systems, and $\alpha_1$ and $\alpha_n$ are rank-2 tensors with $\alpha_k$ being rank-3 tensors. This ansatz construction allows us to compress the number of parameters representing the state to $O(n \kappa^2)$ from $O(2^n)$ for the bond dimension $\kappa$. While MPS representations may capture weakly and locally entangled quantum systems, they probably fail to simulate strongly entangled quantum dynamics.  

 While an efficient ansatz representation is necessary for classical simulations of quantum systems, quantum computers allow for an efficient generation of rank-$n$ tensors, which motivated the introduction of quantum tensors~\cite{yuan2021quantum}. Quantum tensors are generally represented as $\psi^{i_1, i_2, ...,i_b}_{j_1,j_2,...,j_n}$, where the upper indices $(i_1, i_2, ...,i_b)$ are classical indices and the lower indices $(j_1,j_2,...,j_n)$ are quantum indices. Quantum tensors are generated from quantum devices; then, the corresponding quantum state is:
\begin{equation}
\ket{\psi^{i_1,i_2,...,i_b}}=\sum_{j_1,j_2,...,j_n} \psi^{i_1, i_2, ...,i_b}_{j_1,j_2,...,j_n} \ket{j_1,j_2,...,j_n}. 
\end{equation}
 The quantum indices are necessarily related to the ``physical" quantum devices. For $j_k \in \{0,1\} ~\forall k$, $\ket{j_1,j_2,...,j_n}$ is a computational basis state in qubit-based quantum computers. Tensor contractions of quantum indices are performed through the measurement of quantum states. 

 As a simple example of the HTN formalism, let us consider a tensor $\tilde{\psi}_{j_1,j_2,...,j_n}=\sum_i \alpha^{i} \psi^{i}_{j_1,j_2,...,j_n}$ for a rank-1 classical tensor $\alpha^{i}$. This hybrid tensor corresponds to a quantum state $\ket{\tilde{\psi}}=\sum_i \alpha^i \ket{\psi^i}$ with $\ket{\psi^i}=\sum_{j_1,j_2,...,j_n}  \psi^{i}_{j_1,j_2,...,j_n}\ket{j_1,j_2,...,j_n}$. A linear combination of quantum states with a classical coefficient is a generalization of quantum subspace expansion (QSE), which can strengthen the representability of quantum states~\cite{mcclean2017hybrid,takeshita2020increasing,yoshioka2022variational} and accordingly works as a quantum error mitigation method~\cite{mcclean2020decoding,yoshioka2022generalized}. To compute an expectation value of an observable $O$ for the state $\ket{\tilde{\psi}}$, we have $\braket{O}_{\tilde{\psi}}=\sum_{i,i'} \alpha^{i*} \alpha^{i'}  M^{i,i'}$, where $M^{i,i'}=\bra{\psi^i} O \ket{\psi^{i'}}$ is evaluated on a quantum computer. For example, when quantum tensors can be represented as $\ket{\psi^i}=P^i \ket{\psi_0}$ with $P^i$ being a Pauli operator, we obtain $M^{i,i'}= \bra{\psi_0}P^i O P^{i'} \ket{\psi_0}$. When we linearly decompose the observable as $O=\sum_k f_k P^k$, $M^{i,i'}$ can be evaluated only from measurements of Pauli operators. As another example, when quantum tensors are described as $\ket{\psi^i}=U^{i} \ket{\psi_0}$ for a unitary operator $U^{i}$, we obtain $M^{i,i'}=\bra{\psi_0}U^{i \dag} O U^{i'}\ket{\psi_0}=\sum_k f_k \bra{\psi_0}U^{i \dag} P^k U^{i'}\ket{\psi_0}$. Note that each term $\bra{\psi_0}U^{i \dag} P^k U^{i'}\ket{\psi_0}$ can be computed with a Hadamard test circuit.

\begin{figure}[tp]
\centering
\begin{center}
\includegraphics[width=80mm]{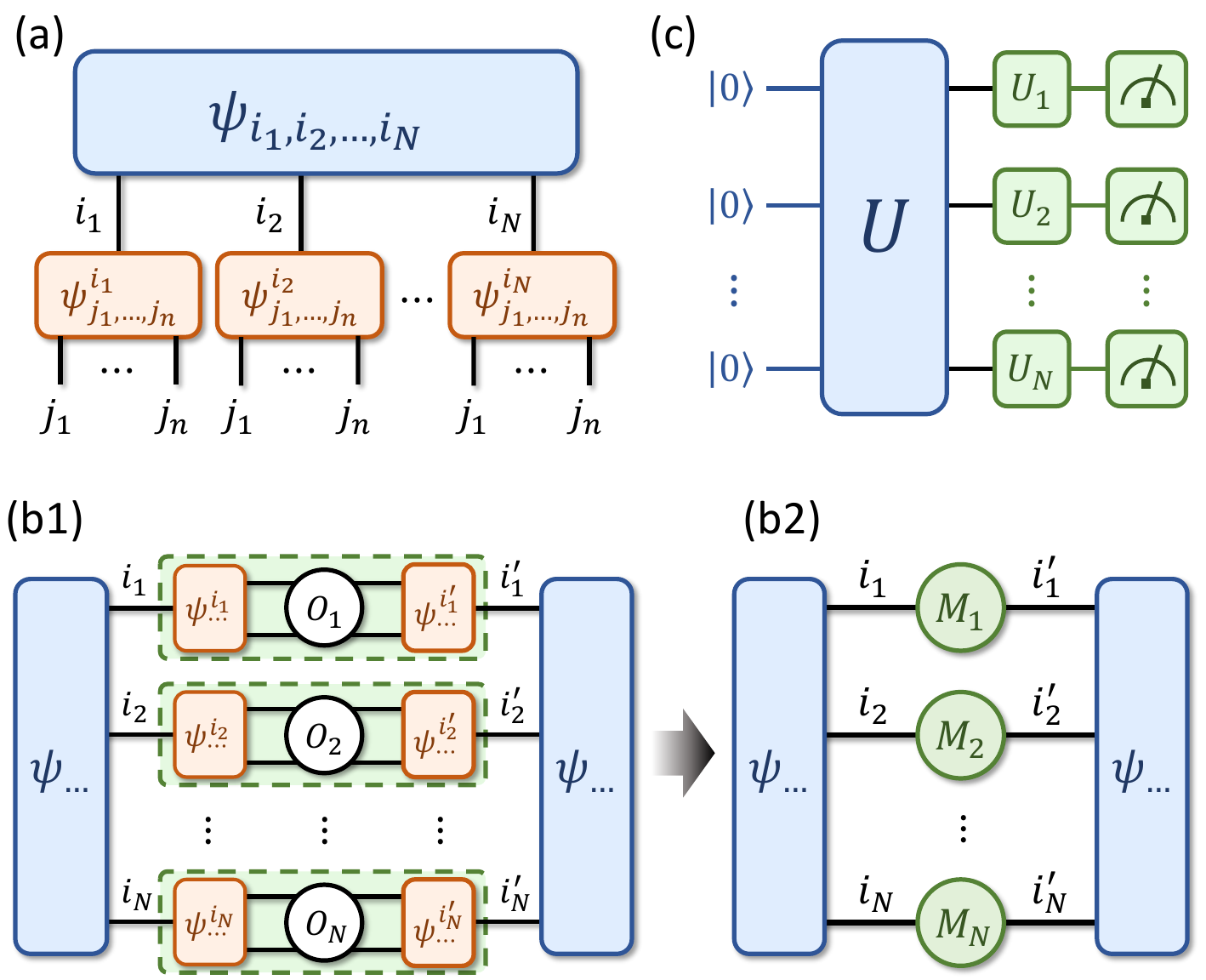}
\end{center}
\caption{\label{fig: HTTNs}(a) Graphical representations of the 2-layer HTTN state~(\ref{eq: HTTN state}). The upper quantum tensor $\psi_{i_1, i_2, ...,i_N}=\braket{i_1, i_2, ..., i_N|U|0^{N}}$ represents the non-local correlation between the subsystems, and the lower quantum tensor $\psi_{j_1,...,j_n}^{i_k}=\braket{j_1, ..., j_n|\psi_{k}^{i_k}}$($k=1,...,N$) represents a quantum state of each subsystem. When calculating the expectation value of the observable $O=\bigotimes_{k=1}^N O_k$ for $\ket{\psi_{\rm HT}}$, (b1) we first contract the local quantum systems (green area) and then (b2) obtain Hermitian operators $\{M_{k}\}_{k=1}^{N}$. Using the quantum circuit in (c) corresponding to the diagram of (b2), we complete the contraction of the whole system.}
\end{figure}

\subsection{Hybrid tree tensor networks}\label{sec:hybrid_tree_tensor_networks}

 Here, we review hybrid tree tensor networks (HTTNs) as an important class of HTNs. The $2$-layer tree TN state consisting of quantum tensors can be expressed by
\begin{equation}\label{eq: HTTN state}
    \ket{\psi_{\rm HT}}=\frac{1}{C}\sum_{i_1, ...,i_N} \psi_{i_1,...,i_N}  \ket{\psi^{i_1}_1} \otimes ... \otimes \ket{\psi^{i_N}_N}.
\end{equation}
Let $\psi_{i_1, ...,i_N}=\braket{i_1,...,i_N|\psi}$ be the probability amplitude of an $N$-qubit state $\ket{\psi}$ (i.e., a rank-$N$ quantum tensor) where $\ket{i_k}$ ($i_{k}\in\{0,1\}~\forall k$) denotes a computational basis state, $\{\ket{\psi^{i_k}_k}\}_{i_k} (k=1,...,N)$ be a set of $n$-qubit states (i.e., a rank-$(n+1)$ quantum tensor), and $C$ be a normalization constant, $\ket{\psi_{\rm HT}}$ is a $Nn$-qubit state. 
The graphical representation of Eq.~(\ref{eq: HTTN state}) is shown in Fig.~\ref{fig: HTTNs}(a).
 Given the HTN representation~(\ref{eq: HTTN state}), we can compute the expectation value of an $Nn$-qubit observable for the state $\ket{\psi_{\rm HT}}$ only with a $\mathrm{max}\{N,n\}$-qubit system. To clarify this, we consider the calculation of the expectation value of an observable $O=\bigotimes_{k=1}^N O_k$, where $O_{k}$ ($k=1,2,...,N$) is a $n$-qubit observable, for the state $\ket{\psi_{\rm HT}}$. Now, the expectation value $\braket{O}_{\psi_{\rm HT}}$ including the normalization constant can be represented as follows:
 \begin{equation}\label{eq: contraction of local tensors}
\begin{aligned}
\braket{O}_{\psi_{\rm HT}} = \frac{1}{C^2}\bra{\psi} \bigotimes_{k=1}^N M_k \ket{\psi}
\end{aligned}
\end{equation}
and 
\begin{equation}\label{eq: normalization constant}
\begin{aligned}
C^2 = \bra{\psi} \bigotimes_{k=1}^N S_{k} \ket{\psi},
\end{aligned}
\end{equation}
 where $\vec{i}=(i_1,i_2,...,i_N)$, $\vec{i}'=(i_1',i_2',...,i_N')$, $M_k^{i_k,i_k'}= \bra{\psi_{k}^{i_k}} O_k \ket{\psi_{k}^{i_k'}}$ and $S_{k}^{i_k,i_k'}= \braket{\psi_{k}^{i_k} | \psi_{k}^{i_k'}}$. Therefore, $\braket{O}_{\psi_{\rm HT}}$ can be evaluated as follows. First, we calculate $\{M_k \}_{k=1}^{N}$ ($\{S_{k} \}_{k=1}^{N}$) via contractions of local quantum states $\{\ket{\psi_{k}^{i_k}}\}_{i_k}$ ($k=1,2,...,N$), as shown in Fig.~\ref{fig: HTTNs}(b). Then, because $\{M_k\}_k$ ($\{S_{k} \}_k$) are Hermitian operators, we have the spectral decomposition $M_k=\hat{U}_k^\dag \Lambda_k \hat{U}_k$ {($S_{k}=\hat{V}_{k}^\dag \Lambda'_{k} \hat{V}_{k}$)} for a unitary operator $\hat{U}_k$ ($\hat{V}_{k}$) and a diagonal operator $\Lambda_k$ ($\Lambda'_{k}$). Thus, by applying $\otimes_{k=1}^N \hat{U}_k$ ($\otimes_{k=1}^N \hat{V}_k$) immediately before measurement and measuring on the $Z$ basis and assigning the elements of $\Lambda_k$ ($\Lambda'_{k}$) to the measurement outcomes, we can measure the expectation value of $\bigotimes_{k=1}^N M_k$ ($\bigotimes_{k=1}^N S_{k}$) for the state $\ket{\psi}$, as in Fig.~\ref{fig: HTTNs}(c). In this way, we can compute $\braket{O}_{\psi_{\rm HT}}$ using a $\mathrm{max}\{N,n\}$-qubit system, because the calculation procedures in Fig.~\ref{fig: HTTNs}(b) and (c) each require quantum computers of only $n$ and $N$ qubits, respectively.
 For the contraction rule for the transition amplitudes $\bra{\psi_{\rm HT}^{(1)}} O \ket{\psi_{\rm HT}^{(2)}}$ for different hybrid tensors $\ket{\psi_{\rm HT}^{(1)}}$ and  $\ket{\psi_{\rm HT}^{(2)}}$, refer to \textcite{kanno2021quantum}. While we considered the HTN state comprising only quantum tensors above, note that we can also consider the HTN model where quantum tensors and classical tensors are used interchangeably~\cite{yuan2021quantum}.

In Eq.~(\ref{eq: HTTN state}), we present a simple model of the 2-layer HTTN state. In the subsequent discussion, for the sake of generality, we redefine and refer to the 2-layer HTTN state $\ket{\psi_{\rm HT}}$ as follows:
\begin{equation}\label{eq: HTTN state_2}
    \ket{\psi_{\rm HT}} = \frac{1}{C} \sum_{\vec{i}_{1},...,\vec{i}_{N}} \psi_{\vec{i}_{1},...,\vec{i}_{N}} \ket{\psi_{1}^{\vec{i}_1}} \otimes ... \otimes \ket{\psi_{N}^{\vec{i}_N}},
\end{equation}
where $\ket{\psi_{\vec{i}_1,...\vec{i}_N}}=\braket{\vec{i}_{1},...,\vec{i}_{N}|\psi}$ is the probability amplitude of an $(\sum_{k=1}^{N} b_{k})$-qubit state $\ket{\psi}$ where $\ket{\vec{i}_k}$ ($\vec{i}_k:=(i_{k1},...,i_{kb_{k}})\in\{0,1\}^{b_k}$) is a $b_k$-qubit computational-basis state, $\{\ket{\psi_{k}^{\vec{i}_k}}\}_{\vec{i}_k}$ ($k=1,...,N$) is a set of $n_k$-qubit states, $C$ is a normalization factor. It is straightforward to check that the expression~(\ref{eq: HTTN state_2}) is a generalization of Eq.~(\ref{eq: HTTN state}) by setting $b_k=1$ and $n_k=n$ for all $k$.

Regarding the calculation of each $M_{k}$ ($S_{k}$), it is important to note that the specific calculation procedure and its efficiency highly depend on how we define the set of quantum states $\{\ket{\psi_{k}^{\vec{i}_{k}}}\}$.
In this subsection, we explain the procedures for computing $M_{k}$ ($S_{k}$) by contracting local observable $O_k$ and quantum states $\{\ket{\psi_k^{\vec{i}_k}}\}$. Letting $\psi_{k,\vec{j}_{k}}^{\vec{i}_{k}}$ be the probability amplitude of $\ket{\psi_k^{\vec{i}_k}}$ where $\vec{j}_{k}=(j_{k1},..., j_{kn_k}) \in\{0,1\}^{n_k}$, the contraction of local tensors can be graphically represented as: 
\begin{center}
    \includegraphics[height=2cm]{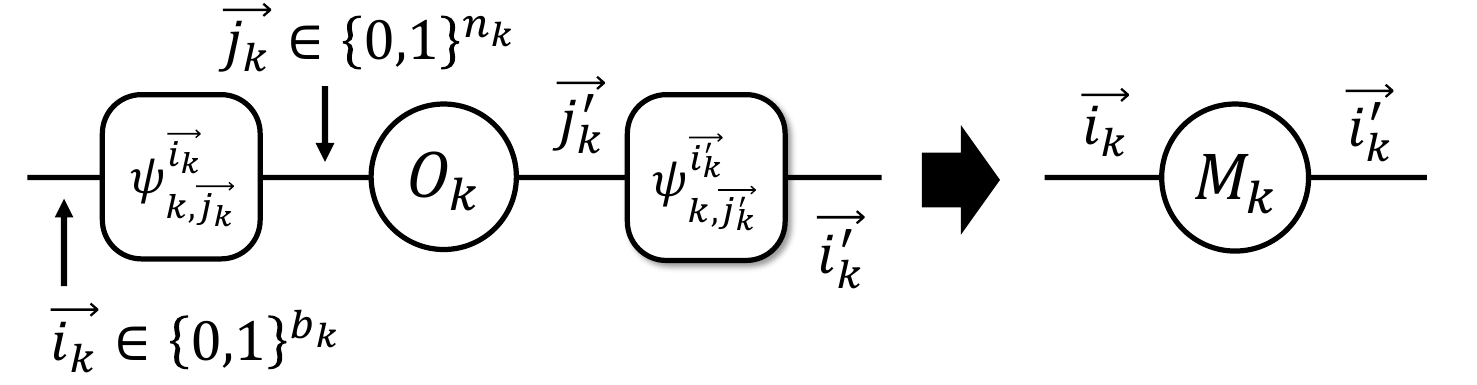}.
\end{center}
When $\psi_{k,\vec{j}_{k}}^{\vec{i}_{k}}$ is a classical tensor, i.e., the set $\{ \ket{\psi_{k}^{\vec{i}_{k}}} \}$ prepared by classical computers, each element of $M_{k}$ ($S_{k}$) is computed on classical computers.
 For example, if the local tensor is expressed in the form of MPS as Eq.~(\ref{eq:mps}),
 the matrix elements $M_{k}^{\vec{i}_{k},\vec{i}'_{k}}=\braket{\psi_{k}^{\vec{i}_{k}}|O_{k}|\psi_{k}^{\vec{i}'_{k}}}$ are obtained by the MPS calculation.  Meanwhile, in the case of $\psi_{k,\vec{j}_k}^{\vec{i}_{k}}$ being a quantum tensor (i.e., the set $\{ \ket{\psi_{k}^{\vec{i}_{k}}} \}$ is prepared by quantum computers), $M_{k}$ ($S_{k}$) would be calculated by the measurements on quantum computers. As already stated, the detailed computational procedure lies on the way of defining the set $\{ \ket{\psi_{k}^{\vec{i}_{k}}} \}$, and
 we consider the case where the set $\{ \ket{\psi_{k}^{\vec{i}_{k}}} \}$ is given as the following four ways:
\begin{description}
    \item[(i) initial states]~~~~~~~ $\ket{\psi_{k}^{\vec{i}_k}}=U_{k}^{(1)}\ket{\vec{i}_k}\ket{0}^{\otimes (n_k-b_k)}$,
    \item[(ii) projections]~~~~~~~~ $\ket{\psi_{k}^{\vec{i}_k}}=\braket{\vec{i}_k|\psi_{k}^{(2)}}$, 
    \item[(iii) Pauli operators]\, $\ket{\psi_{k}^{\vec{i}_k}}=P_{k}^{\vec{i}_k}\ket{\psi_{k}^{(3)}}$,
    \item [(iv) unitary gates]~~~~ $\ket{\psi_{k}^{\vec{i}_k}}=U_{k}^{\vec{i}_k}\ket{0}^{\otimes n_k}$.
\end{description}
$U^{(1)}_{k}$ denotes a $n_k$-qubit unitary operator, $\ket{\psi_{k}^{(2)}}$ and $\ket{\psi_{k}^{(3)}}$ are respectively $(n_k+b_k)$-qubit and $n_k$-qubit quantum states, $P_{k}^{\vec{i}_k} \in \{I,X,Y,Z\}^{\otimes n_k}$ in type (iii) is a tensor product of the Pauli operators, and $U_{k}^{\vec{i}_{k}}$ in type (iv) represent $n_k$-qubit unitary operators.
Note that although certain types can be mathematically represented within the framework of another type (e.g., type (i) and (iii) can be regarded as special examples of type (iv)), we distinguish each type for the practical reason that different characterizations of the state $\ket{\psi^{\vec{i}_k}_{k}}$ by $\vec{i}_{k}$ require the different quantum circuits and the classical post-processing to calculate $M_{k}$ ($S_{k}$). In Appendix~\ref{apdx:calculation of M_k}, as the simplest example of the calculation of $M_k$($S_k$), we consider the case where $b_k=1$, i.e., $M_k$($S_k$) is a $2\times2$ matrix, and provide a detailed explanation of how to calculate the matrices for each type.

\section{Contraction of local tensors}
\label{contraction_of_local_observables}
This section provides detailed procedures for calculating $M_k$ for different types. Although the calculation methods for $M_k$ in the case of $b_k=1$ have already been provided in Refs.~\cite{yuan2021quantum,kanno2021quantum}, here we present the estimation procedures for the more general case where $b_{k}\geq 2$ in order to use for the analysis in the main result. Note that we omit the explanation of calculating $S_k$, as it can be computed by a similar procedure to that of $M_k$ by setting $O_k=I^{\otimes n_k}$. In the remainder of this paper, we use the superscript $\xi_k$ to distinguish the types associated with the $M_k$ and $S_k$, denoting $M_k^{(\xi_k)}$ and $S_k^{(\xi_k)}$, respectively. Here, $\xi_k=1,2,3,4,\rm{c}$ corresponds to $M_k$ ($S_k$) obtained via type-(i) to type-(iv) procedures and classical computations, respectively.

\subsection{Type (i)}\label{sec:2a}

 Let $\mathcal{U}^{(1)}_{k}(\bullet)=U_{k}^{(1)}\bullet U_{k}^{(1)\dagger}$ be a $n_k$-qubit unitary channel, $M_{k}^{(1)}$ is represented as
\begin{equation}\label{eq: Case1, M_k}
    M_{k}^{(1)} = \sum_{\vec{i}_k,\vec{i}'_k} \mathrm{Tr}\left[ O_{k}\,\mathcal{U}_{k}^{(1)} \left(\ket{\vec{i}'_{k}}\bra{\vec{i}_{k}} \otimes \ket{\bar{0}}\bra{\bar{0}} \right) \right]\ket{\vec{i}_k}\bra{\vec{i}'_k},
\end{equation}
where we have defined $\ket{\bar{0}}\bra{\bar{0}} := \ket{0}\bra{0}^{\otimes (n_k-b_k)}$. 
To compute $M_k$ in this case, we use a decomposition of the $\rm SWAP$ operator as a linear combination of Pauli strings.
That is, let $\mathrm{SWAP}=\sum_{\vec{i}'_k,\vec{i}_k}\ket{\vec{i}_k}\bra{\vec{i}'_k}\otimes \ket{\vec{i}'_k}\bra{\vec{i}_k}$ be the $2b_k$-qubit swap operator, it can be decomposed into a linear combination of Pauli operators as $\mathrm{SWAP}=\frac{1}{2^{b_k}} \sum_{x=1}^{4^{b_k}} P_x \otimes P_x$ where $\{P_x\}_{x=1}^{4^{b_k}}=\{I,X,Y,Z\}^{\otimes b_k}$. By plugging the relation in Eq.~(\ref{eq: Case1, M_k}) and using the spectral decomposition $P_x=\sum_{y} \lambda_{x,y}\ket{v_{x,y}}\bra{v_{x,y}}$ ($\lambda_{x,y}\in\{\pm1\}$), Eq.~(\ref{eq: Case1, M_k}) can be rewritten as follows:
\begin{equation}\label{eq: Case1, M_k, 1}
    M_{k}^{(1)} = \sum_{x,y} \mathrm{Tr}\left[ O_{k}\,\mathcal{U}_{k}^{(1)} \left( \ket{v_{x,y}}\bra{v_{x,y}} \otimes \ket{\bar{0}}\bra{\bar{0}} \right) \right] \frac{\lambda_{x,y}}{2^{b_k}} P_x.
\end{equation}
Hence, $M_k^{(1)}$ can be computed as the following procedure: for each $x=1,...,4^{b_{k}}$ and $y=1,...,2^{b_k}$, we prepare the state $\mathcal{U}_{k}^{(1)} \left(\ket{v_{x,y}}\bra{v_{x,y}} \otimes \ket{\bar{0}}\bra{\bar{0}} \right)$ and calculate the expected value of $O_k$ for the state using the circuit in Fig.~\ref{fig: Quantum circuits for M_k}(a). Repeating the above process for every $x,y$, we then combine the outputs with the weights $\lambda_{x,y}P_{x}/2^{b_k}$.
Here, we emphasize that $S_k^{(1)}$ does not need to be calculated because $S_{k}^{(1)} = I^{\otimes b_k}$ holds for any $k$, and consequently $C^2=1$ holds. 

\subsection{Type (ii)}\label{sec: 2b}
\begin{figure}[ht]
\centering
\begin{center}
 \includegraphics[width=80mm]{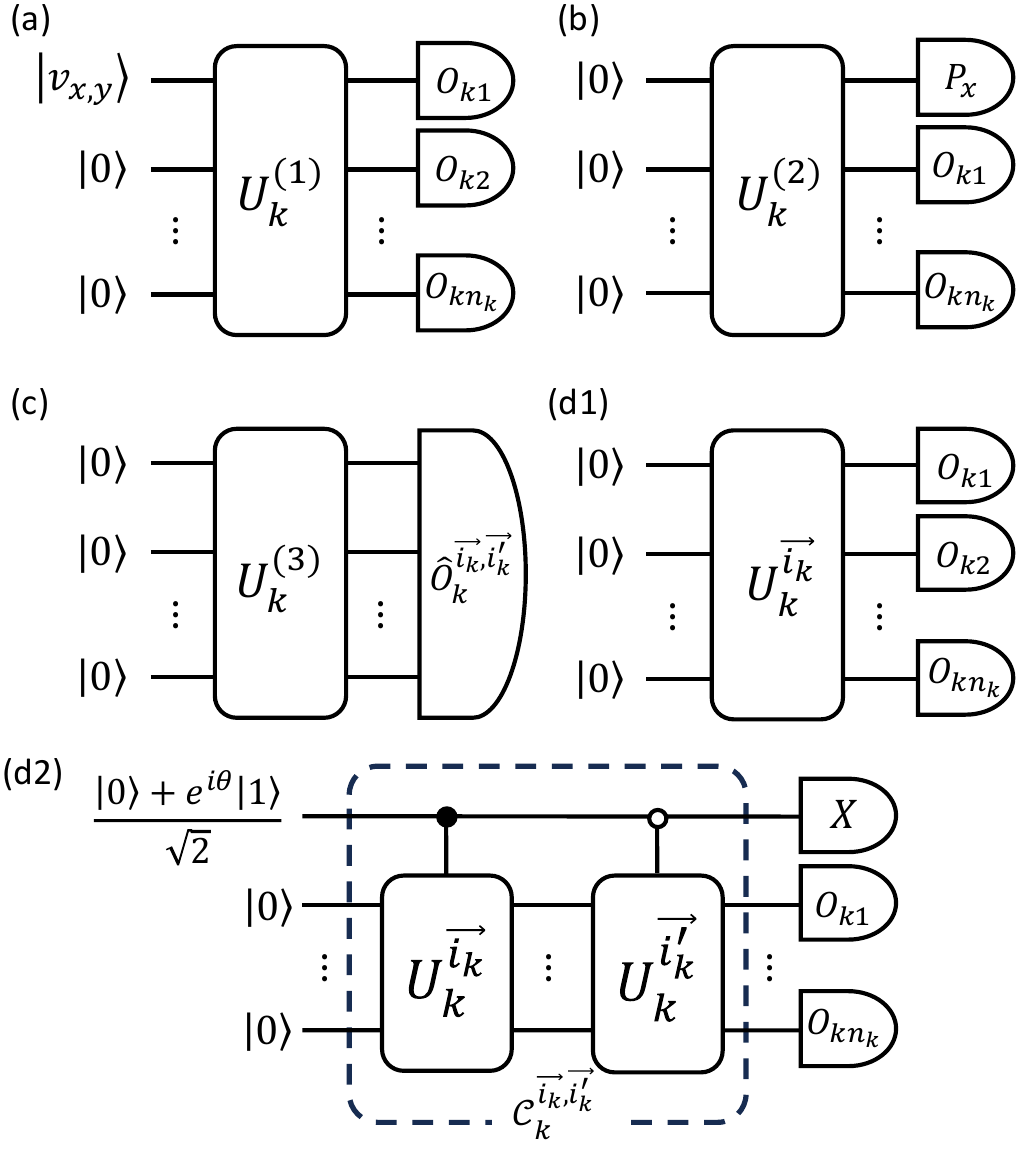}    
\end{center}
\caption{\label{fig: Quantum circuits for M_k} Quantum circuits for computing $M_{k}^{(\xi)}$ ($\xi=1,2,3,4$)  are shown in (a), (b), (c) and (d), respectively. $S_{k}^{(\xi)}$ can be obtained by replacing $O_{k}=\otimes_{m=1}^{n_k} O_{km}$ with $I^{\otimes n_k}$ from types (ii) to (iv). In type (iv), two types of quantum circuits are used to compute $M_k$: (d1) A quantum circuit for calculating diagonal elements $M_{k}^{(4),\vec{i}_{k},\vec{i}'_{k}}$, and (d2) a circuit for calculating off-diagonal elements $M_{k}^{(4),\vec{i}_{k},\vec{i}_{k}}$. A controlled-$U_{k}^{\vec{i}_{k}}$ (controlled-$U_{k}^{\vec{i}'_{k}}$) with a black (white) circle works on the target qubits when the control qubit is $\ket{1}$ ($\ket{0}$).}
\end{figure}

Suppose $\mathcal{U}^{(2)}_{k}(\bullet)=U_{k}^{(2)}\bullet U_{k}^{(2)\dagger}$ is a $(n_k+b_k)$-qubit unitary channel acting as $\mathcal{U}_{k}^{(2)}(\ket{\bar 0}\bra{\bar 0})=\ket{\psi_{k}^{(2)}}\bra{\psi_{k}^{(2)}}$ where $\ket{\bar{0}}\bra{\bar{0}} := \ket{0}\bra{0}^{\otimes (n_k+b_k)}$. 
Then, $M_{k}^{(2)}$ is expressed as
\begin{equation}\label{eq: Case2, M_k}
    M_{k}^{(2)} = \sum_{\vec{i}_k,\vec{i}'_k} \mathrm{Tr}\left[ \bigl( \ket{\vec{i}_{k}}\bra{\vec{i}'_{k}} \otimes O_{k} \bigr) \mathcal{U}_{k}^{(2)}(\ket{\bar{0}}\bra{\bar{0}}) \right] \ket{\vec{i}_{k}}\bra{\vec{i}'_{k}}.
\end{equation}
Here, let $\ket{\rm Bell}\bra{\rm Bell}:=\sum_{\vec{i}'_k,\vec{i}_k}\ket{\vec{i}_k}\bra{\vec{i}'_k}\otimes \ket{\vec{i}_k}\bra{\vec{i}'_k}$ be the $2b_k$-qubit unnormalized bell state, it can be decomposed as $\ket{\rm Bell}\bra{\rm Bell}=\frac{1}{2^{b_k}} \sum_{x=1}^{4^{b_k}} P_x \otimes P_x^{\rm T}$. Applying the decomposition to Eq.~(\ref{eq: Case2, M_k}), Eq.~(\ref{eq: Case2, M_k}) can be equivalently written as follows.
\begin{equation}\label{eq: Case2, M_k, 2}
    M_{k}^{(2)} = \sum_{x} \mathrm{Tr}\left[ \left( P_x \otimes O_{k} \right) \mathcal{U}_{k}^{(2)}(\ket{\bar{0}}\bra{\bar{0}}) \right] \frac{1}{2^{b_k}} P_x^{\rm T}.
\end{equation}
Thus, $M_k^{(2)}$ can be computed as follows:
for each $x=1,...,4^{b_k}$, we prepare $\mathcal{U}_{k}^{(2)}(\ket{\bar 0}\bra{\bar 0})$ and calculate the expected value of $P_{x} \otimes O_{k}$ on the system using the circuit in Fig.~\ref{fig: Quantum circuits for M_k}(b). Collecting the calculation results for $x=1,...,4^{b_k}$, we add the outcomes together with the weights $P_{x}^{\rm T}$.

\subsection{Type (iii)}\label{sec: 2c}

Suppose $\mathcal{U}^{(3)}_{k}(\bullet)=U_{k}^{(3)}\bullet U_{k}^{(3)\dagger}$ is a $n_k$-qubit unitary channel acting as $\mathcal{U}_{k}^{(3)}(\ket{\bar 0}\bra{\bar 0})=\ket{\psi_{k}^{(3)}}\bra{\psi_{k}^{(3)}}$ where $\ket{\bar{0}}\bra{\bar{0}} := \ket{0}\bra{0}^{\otimes n_k}$.
Then, $M_{k}^{(3)}$ is represented as 
\begin{equation}\label{eq: Case3, M_k, 2}
    M_{k}^{(3)} = \sum_{\vec{i}_k,\vec{i}'_k} \mathrm{Tr}\left[ \hat{O}_{k}^{\vec{i}_{k},\vec{i}'_{k}}\, \mathcal{U}_{k}^{(3)}(\ket{\bar 0}\bra{\bar 0})\right] \ket{\vec{i_{k}}}\bra{\vec{i}'_{k}},
\end{equation}
where we have defined $\hat{O}_{k}^{\vec{i}_{k},\vec{i}'_{k}}:=P_{k}^{\vec{i}_{k}} O_{k}P_{k}^{\vec{i}'_{k}}$. Thus, we can evaluate $M_{k}^{(3),\vec{i}_{k},\vec{i}'_{k}}$ by performing direct measurements of Pauli operators $\hat{O}_{k}^{\vec{i}_{k},\vec{i}'_{k}}$ on the system $\mathcal{U}_{k}^{(3)} (\ket{\bar 0}\bra{\bar 0})$, and combing the calculation results with $\ket{\vec{i_{k}}}\bra{\vec{i}'_{k}}$; refer to Appendix~\ref{apdx:calculation of M_k^3} for the detailed explanation where $b_k=1$.

\begin{figure*}[t]
\centering
\begin{center}
 \includegraphics[width=170mm]{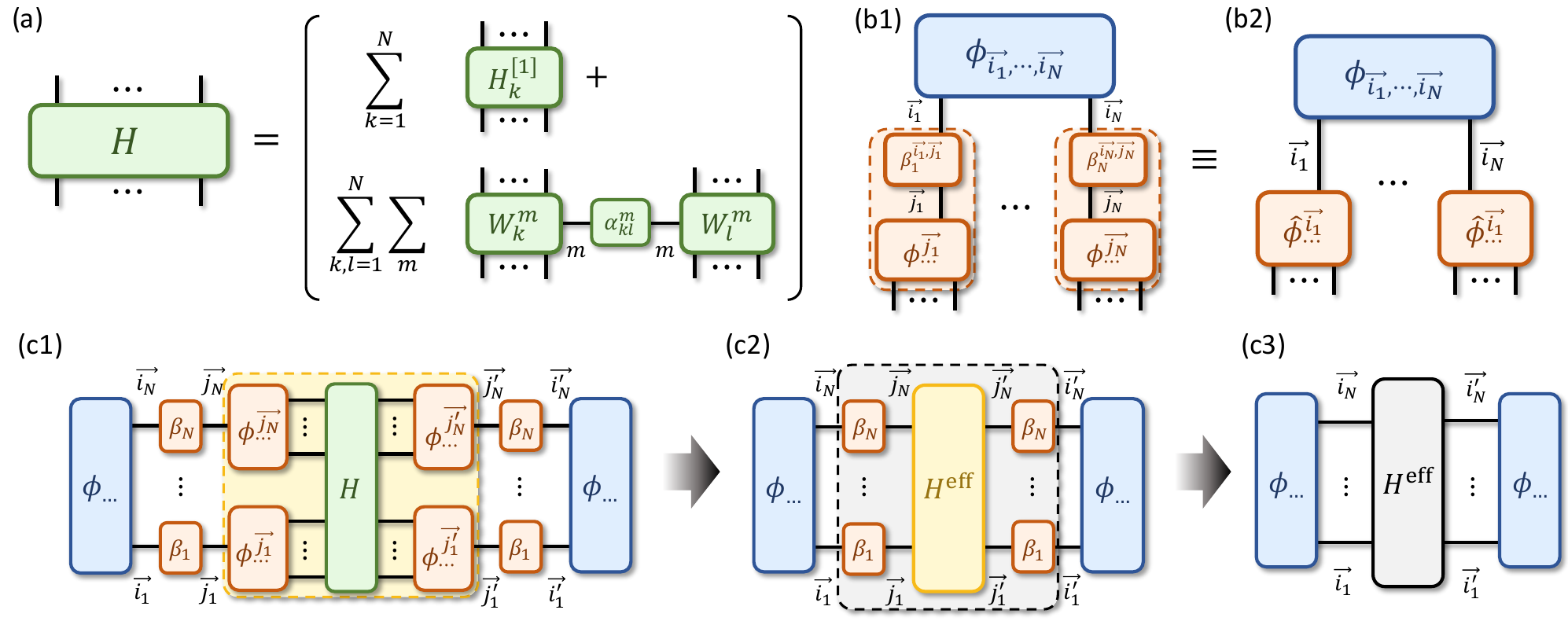}    
\end{center}
\caption{\label{fig:deep_vqe} HTN representations of Deep VQE algorithm. (a) The tensor representation of the Hamiltonian~(\ref{eq: deepVQE hamiltonian}). (b) The graphical representation of the HTTN states~(\ref{eq: HTNs representation of Deep VQE,1}) and (\ref{eq: HTNs representation of Deep VQE,2}) are shown in (b1) and (b2), respectively. The $k$-th classical tensor $\beta_{k}^{\vec{i}_{k},\vec{j}_{k}}$ is calculated so that the $k$-th set of states $\{\ket{\widehat{\phi}^{\vec{i}_{k}}_{k}}\}_{\vec{i}_{k}}$ forms the orthonormal basis of the $k$-th subsystem, where $\ket{\widehat{\phi}^{\vec{i}_{k}}_{k}}$ is represented as a linear combination of the non-orthogonal basis $\{\ket{\phi_{k}^{\vec{j}_{k}}}\}_{\vec{j}_{k}}$, i.e., $\ket{\widehat{\phi}^{\vec{i}_k}_{k}}=\sum_{\vec{j}_{k}}\beta_{k}^{\vec{i}_{k},\vec{j}_{k}}\ket{\phi_{k}^{\vec{j}_{k}}}$. (c) The schematic diagram of the calculation of the expectation value of the Hamiltonian $H$ for the HTTN state $\ket{\phi_{\rm DV}}$. By contracting the local quantum tensors $\phi_{k}^{\vec{j}_{k}}$, $\phi_{k}^{\vec{j}'_{k}}$, and the classical tensor $H$ (yellow area in (c1)), we obtain an contracted tensor $H^{\rm eff}$, corresponding to the unnormalized effective Hamiltonian $H^{\rm eff}$. Then, by contracting the effective Hamiltonian $H^{\rm eff}$ and the classical tensors $\beta_{k}^{\vec{i}_{k},\vec{j}_{k}}$, and $\beta_{k}^{\vec{i}'_{k},\vec{j}'_{k}}$ (gray area in (c2)), we obtain a new tensor $\widehat{H}^\mathrm{eff}$. Finally, contracting the whole tensor in (c3) using a quantum computer, we obtain $\braket{H}_{\phi_{\rm DV}}$}
\end{figure*}

\subsection{Type (iv)}\label{sec: 2d}
In type (iv), $M_{k}^{(4)}$ is represented as  
\begin{equation}\label{eq: Case4, M_k}
    M_{k}^{(4)} = \sum_{\vec{i}_k,\vec{i}'_k} \mathrm{Tr}\left[ O_{k}U_{k}^{\vec{i}'_{k}} \ket{0}\bra{0}^{\otimes n_k} U_{k}^{\vec{i}_{k}\dagger} \right] \ket{\vec{i}_k}\bra{\vec{i}'_k}.
\end{equation}
Now, let $\mathcal{U}_k^{(4),\vec{i}_k}(\bullet)=U_{k}^{\vec{i}_k}\bullet U_{k}^{\vec{i}_k,\dagger}$ be a $n_k$-qubit unitary channel acting as $\mathcal{U}_k^{(4),\vec{i}_k}(\ket{0}\bra{0}^{\otimes n_k})=\ket{\psi_{k}^{\vec{i}_{k}}}\bra{\psi_{k}^{\vec{i}_{k}}}$, $\mathcal{C}_k^{\vec{i}_k,\vec{i}'_k}$ be the controlled unitary chennel as shown in Fig.~\ref{fig: Quantum circuits for M_k}(d2), Eq.~(\ref{eq: Case4, M_k}) can be expressed as
\begin{equation}\label{eq: Case4, M_k, 2}
\begin{split}
    &M_k^{(4)}\\
    &= \sum_{\vec{i}_k = \vec{i}'_k} \mathrm{Tr}[O_k \,\mathcal{U}_{k}^{(4),\vec{i}_{k}}(\ket{0}\bra{0}^{\otimes n_k})]\ket{\vec{i}_k}\bra{\vec{i}'_k}\\
    &~+ \sum_{\vec{i}_k \neq \vec{i}'_k} \sum_{\theta} \mathrm{Tr}[(X\otimes O_k) \mathcal{C}_k^{\vec{i}_k,\vec{i}'_k} \,(\ket{\varphi(\theta)}\bra{\varphi(\theta)})]\ket{\vec{i}_k}\bra{\vec{i}'_k},
\end{split}
\end{equation}
where we have defined $\ket{\varphi(\theta)}:=\frac{\ket{0}+e^{i\theta}\ket{1}}{\sqrt{2}}\otimes\ket{0}^{\otimes n_k}$. Therefore, the ($\vec{i}_k,\vec{i}_k$) elements of $M_{k}^{(4)}$ can be obtained by measuring the expectation value of the observable $O_{k}$ for the state $\mathcal{U}_{k}^{(4),\vec{i}_{k}}(\ket{0}\bra{0}^{\otimes n_k})$ using the quantum circuit shown in Fig.~\ref{fig: Quantum circuits for M_k}(d1). On the other hand, the ($\vec{i}_k,\vec{i}'_k$) elements of $M_{k}^{(4)}$ can be computed by using the quantum circuit shown in Fig.~\ref{fig: Quantum circuits for M_k}(d2): one measure the expectation value $\braket{X \otimes O_k}$ for the initial states $\ket{\varphi(\theta)}$ ($\theta=0,\frac{\pi}{2}$) and combine the results. Refer to Appendix~\ref{apdx:calculation of M_k^4} for the detailed procedure in the case of $b_k=1$. 

\section{Techniques relevant to hybrid tensor networks} \label{sec:preliminaries}
Many novel algorithms have recently been proposed to simulate large quantum systems using devices with limited quality and quantity of qubits~\cite{bravyi2016trading,eddins2022doubling,huembeli2022entanglement, castellanos2023quantum,fujii2022deep,mizuta2021deep,erhart2022constructing,yuan2021quantum,kanno2021quantum,kanno2023quantum,peng2020simulating,mitarai2021constructing,piveteau2022circuit,brenner2023optimal,PRXQuantum.5.040308}.
While these algorithms differ in the specific implementation of quantum circuits and the way of classical processing, they share several key ideals. For example, these algorithms aim to simulate a larger quantum system by dividing the system into some clusters with strong internal interaction but weak interaction among them. From another perspective, their fundamental concepts are closely related to the conventional TN approaches~\cite{white1992density,white1993density,schollwock2011density,shi2006classical,vidal2007entanglement}. 
The strength of the HTN framework, which integrates classical and quantum tensors, lies in its flexibility to provide a unified basis for discussing wide-ranging quantum-classical hybrid approaches, including not only QSE (shown in Sec.~\ref{sec: htn}) but also these new approaches. In this section, focusing on the Deep VQE~\cite{fujii2022deep} and the entanglement forging method~\cite{eddins2022doubling} as representative examples of such algorithms, we show that they can be discussed within the HTN framework~\cite{yuan2021quantum}, along with their brief overview. A more detailed discussion can be found in Appendix~\ref{apdx:detail_in_relevant_techniques}.

\subsection{Deep VQE}\label{sec:deep_vqe}
 Let us consider $\bar{n}_{\rm tot}$-qubit quantum state composed of $N$ subsystems and the $k$-th subsystem consists of $\bar{n}_{k}$ qubits, i.e., $\bar{n}_{\rm tot}=\sum_{k=1}^{N} \bar{n}_{k}$. Note that we are focusing on the following situation: the whole system can be divided into the $N$ clusters with strong internal interactions but weak interactions between them. Then, the Hamiltonian can be written as 
\begin{equation}\label{eq: deepVQE hamiltonian}
    H = \sum_{k=1}^{N} H^{[1]}_{k} + \sum_{k,l=1}^{N} H^{[2]}_{kl},\quad
    H^{[2]}_{kl} = \sum_{m} \alpha_{kl}^{m}\, W^{m}_{k}\otimes W^{m}_{l}
\end{equation}
 where $H^{[1]}_{k}$ corresponds to the $k$-th subsystem's Hamiltonian and $H^{[2]}_{kl}$ to the interaction Hamiltonian acting on the $k$-th and $l$-th subsystems. Here, the interaction term $H^{[2]}_{kl}$ is decomposed into the linear combination of the tensor products of $W^{m}_{k}$ and $W^{m}_{l}$ acting on the $k$-th and the $l$-th subsystems with the coefficients $\alpha_{kl}^{m}$. The tensor representation of the Hamiltonian $H$ is depicted in Fig.~\ref{fig:deep_vqe}(a).

\begin{figure*}[t]
\centering
\begin{center}
 \includegraphics[width=160mm]{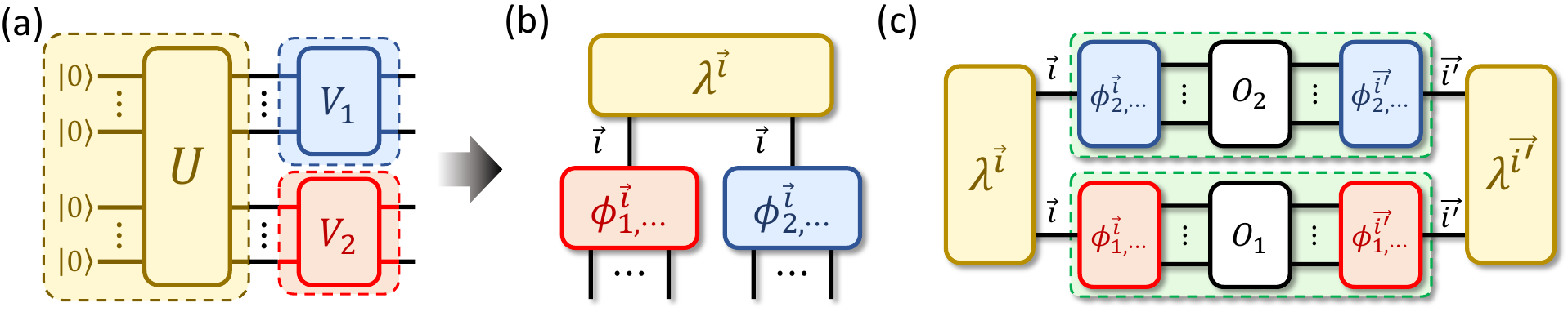}    
\end{center}
\caption{\label{fig: entanglement forging} HTN representations of entanglement forging method. (a) The graphical representation of a $2n$-qubit bipartite quantum system $\ket{\phi_{\rm EF}}$, characterized by the unitary operators $U$, $V_{1}$, and $V_{2}$. (b) Applying Schmidt decomposition to the $2n$-qubit state $U\ket{0}^{\otimes 2n}$ and letting each local operator $V_{1}$ ($V_{2}$) act on the local basis $\{\ket{l_{x}}_{1}\}_{x}$ ($\{\ket{l_{x}}_{2}\}_{x}$), we obtain the HTN representation of (a) as shown in Eq.~(\ref{HTNs rep. of EF}). (c) The graphical diagram of the calculation of the expectation value of the observable $O=O_{1} \otimes O_{2}$ for the state $\ket{\phi_{\rm EF}}$. Sampling $M_{1}^{\vec{i},\vec{i'}} M_{2}^{\vec{i},\vec{i'}}$ (green area) in proportion to the Schmidt coefficients $\lambda^{\vec{i}*}\lambda^{\vec{i'}}$, we can efficiently contract the whole system. }
\end{figure*}

In the Deep VQE algorithm~\cite{fujii2022deep}, the ground state for the Hamiltonian~(\ref{eq: deepVQE hamiltonian}) is calculated by step-by-step optimizations based on the information of its subsystems. This procedure can be understood as the optimization of the following $\bar{n}_{\rm tot}$-qubit HTTN state $\ket{\phi_{\rm DV}}$ (Fig.~\ref{fig:deep_vqe}(b1)):
\begin{equation}\label{eq: HTNs representation of Deep VQE,1}
    \ket{\phi_{\rm DV}} = \sum_{\substack{\vec{i}_1,...,\vec{i}_N\\\vec{j}_1,...,\vec{j}_N}} \phi_{\vec{i}_1,...,\vec{i}_N} \beta_{1}^{\vec{i}_{1},\vec{j}_{1}}\ket{\phi_{1}^{\vec{j}_{1}}} \otimes ... \otimes \beta_{N}^{\vec{i}_{N},\vec{j}_{N}} \ket{\phi_{N}^{\vec{j}_{N}}},
\end{equation}
where $\vec{i}_k:=(i_{k,1},...,i_{k,\bar{b}_{k}})\in\{0,1\}^{\bar{b}_{k}}$ ($k=1,...,N$) and $\vec{j}_k:=(j_{k,1},...,j_{k,\bar{b}_{k}})\in\{0,1\}^{\bar{b}_{k}}$ ($k=1,...,N$) denote $\bar{b}_k$ two-dimensional indices, and $\phi_{\vec{i}_1,...,\vec{i}_N}=\braket{\vec{i}_1,...,\vec{i}_N|\phi}$ is the probability amplitude of a $\bar{b}_{\rm tot}$-qubit quantum state $\ket{\phi}$ where $\bar{b}_{\rm tot} :=\sum_{k} \bar{b}_k$ (i.e., a rank-$\bar{b}_{\rm tot}$ quantum tensor with the quantum indices $\vec{i}_k$). 
Here, each set of states $\{\ket{\phi_{k}^{\vec{j}_{k}}}\}$ is prepared by quantum computers with type (iii) state preparations as
\begin{equation}
    \ket{\phi_{k}^{\vec{j}_{k}}} = \begin{cases} W^{\vec{j}_{k}}_{k} \ket{\phi_{k}^{\mathrm{G}}}, & (\vec{j}_{k})_{(10)} \leq \bar{d}_{k} - 1, \\
    \vec{0}, & (\vec{j}_{k})_{(10)} \geq \bar{d}_{k},
    \end{cases}
\end{equation}
where $\ket{\phi_k^{\rm G}}$ is the approximated ground state for $H_{k}^{[1]}$, $(\vec{j}_{k})_{(10)}(k=1,...,N)$ the decimal representations of the binary indices $\vec{j}_{k}$, $\bar{d}_{k}$ the dimension of local basis, and $\vec{0}$ a vector in which all elements are zero. Since the $k$-th set $\{\ket{\phi_{k}^{\vec{j}_{k}}}\}$ is not necessarily an orthogonal basis, the $k$-th classical coefficients $\beta_{k}^{\vec{i}_{k},\vec{j}_{k}}$ are chosen so that a set
$\{\ket{\widehat{\phi}^{\vec{i}_k}_{k}}:=\sum_{\vec{j}_{k}}\beta_{k}^{\vec{i}_{k},\vec{j}_{k}}\ket{\phi_{k}^{\vec{j}_{k}}}\}_{\vec{i}_k}$ form the orthonormal basis of the $k$-th subsystem. This can be accomplished by the Gram-Schmidt process using a list of inner products of the unnormalized basis $\{\braket{\phi_k^{\vec{j}_k}|\phi_k^{\vec{j}'_k}}\}_{\vec{j}_k, \vec{j}'_{k}}$ calculated on quantum computers; see Appendix~\ref{apdx:deep_vae} for the detailed calculations. We also note that by contracting the indices $\vec{j}_k$ in Eq.~(\ref{eq: HTNs representation of Deep VQE,1}), it can be rewritten as follows:
\begin{equation}\label{eq: HTNs representation of Deep VQE,2}
    \ket{\phi_{\rm DV}} = \sum_{\vec{i}_1,...,\vec{i}_N} \phi_{\vec{i}_1,...,\vec{i}_N} \ket{\widehat{\phi}_{1}^{\vec{i}_{1}}} \otimes ... \otimes \ket{\widehat{\phi}_{N}^{\vec{i}_{N}}}.
\end{equation}
The graphical representation of the HTTN state~(\ref{eq: HTNs representation of Deep VQE,1}) is presented in Fig.~\ref{fig:deep_vqe}(b2).

Next, we elaborate on the way of obtaining the expectation value of the Hamiltonian $H$ for the state $\ket{\phi_{\rm DV}}$, i.e., $\braket{\phi_{\rm DV}|H|\phi_{\rm DV}}$, through contraction of the indices of HTTN state in Eq.~(\ref{eq: HTNs representation of Deep VQE,1}). The contraction process consists of three steps. In the first step, we contract the quantum tensors $\phi_{k}^{\vec{j}_{k}}$, $\phi_{k}^{\vec{j}'_{k}}$ and the classical tensor $H$ (Fig.~\ref{fig:deep_vqe}(c1)), and we obtain a new Hamiltonian $H^{\rm eff}$. This can be achieved by calculation on quantum computers. 
 In the next step, by contracting the indices $\vec{j}_{k}$ and $\vec{j}_{k}'$ ($k=1,...,N$) in Fig.~\ref{fig:deep_vqe}(c2), we obtain the normalized one $\widehat{H}^{\rm eff}$. Lastly, the contraction of the indices $\vec{i}_{k}$ and $\vec{i}_{k}'$ with the similar approach in Eq.~(\ref{eq: contraction of local tensors}) gives us the desired expectation value (Fig.~\ref{fig:deep_vqe}(c3)).

\subsection{Entanglement forging}\label{sec:entanglement_forging}

 Let us consider the state of a bipartite $2n$-qubit system $\ket{\phi_{\rm EF}}$, and decompose it into $n$-qubit subsystems using Schmidt decomposition, i.e.,
\begin{equation}\label{entanglement forging ansatz}
    \ket{\phi_{\rm EF}} = (V_{1} \otimes V_{2}) \sum_{x=0}^{2^n-1} \lambda_{x} \ket{l_{x}} \otimes \ket{l_{x}},
\end{equation}
 where $V_{1}$ and $V_{2}$ are the $n$-qubit unitary operators acting on the 1st and 2nd subsystems respectively, $\lambda_{x}$ are the Schmidt coefficients, and $\ket{l_{x}} \in \{\ket{0},\ket{1}\}^{\otimes n}$ are the computational basis corresponding to indices $x$. Note that this hybrid ansatz structure could vary depending on how one truncates the Schmidt decomposition $\lambda_{x}$, similar to the MPS ansatz construction. Importantly, by choosing only the leading $m (<2^n)$ bitstrings $\ket{l_{x}}$ that is sufficient for capturing the physical property of a target system, rather than considering all possible bitstrings, we can mitigate sampling overheads and the technical difficulties to execute a number of the different structure of quantum circuits in the real quantum devices.

 From the perspective of the HTN, the wavefunction in Eq.~(\ref{entanglement forging ansatz}) can be rewritten as
\begin{equation}\label{HTNs rep. of EF}
    \ket{\phi_{\rm EF}}= \sum_{\vec{i}} \lambda^{\vec{i}} \ket{\phi^{\vec{i}}_{1}} \otimes \ket{\phi^{\vec{i}}_2},
\end{equation}
 where $\lambda^{\vec{i}}$ is the classical tensor representing the non-local correlation between the 1st and 2nd subsystems with the classical index  $\vec{i}=(i_1,i_2,...,i_n)\in\{0,1\}^n$, and $\{\ket{\phi_{k}^{\vec{i}}}\}_{\vec{i}}$ is the $k$-th set of $n$-qubit quantum states with classical tensor $\vec{i}$. Each state $\ket{\phi_{k}^{\vec{i}}}$ in the set has the classical index $\vec{i}$ associated with the state input on the computational basis $\ket{\vec{i}}$; the set is prepared by the type (i) state preparation as $\ket{\phi_{k}^{\vec{i}}}=V_{k}\ket{\vec{i}}$.
 The HTN representation of Eq.~(\ref{HTNs rep. of EF}) is shown in Fig.~\ref{fig: entanglement forging}. 

 Here, our purpose is to calculate the expectation value of the $2n$-qubit observable $O=O_1 \otimes O_2$ for the state $\ket{\phi_{\rm EF}}$ as
\begin{equation}
    \braket{O}_{\phi_{\rm EF}}= \sum_{\vec{i},\vec{i'}} \lambda^{\vec{i}\ast}\lambda^{\vec{i'}}  M_{1}^{\vec{i},\vec{i'}}
    M_{2}^{\vec{i},\vec{i'}},
\end{equation}
 where $M_{1}^{\vec{i},\vec{i'}}=\braket{\phi^{\vec{i}}_{1}|O_{1}|\phi^{\vec{i'}}_{1}}$ and $M_{2}^{\vec{i},\vec{i'}}=\braket{\phi^{\vec{i}}_{2}|O_{2}|\phi^{\vec{i'}}_{2}}$.
 We can estimate $\braket{O}_{\phi_{\rm EF}}$ using the weighted sampling strategy, or by computing $M_{1}$ and $M_{2}$ in advance followed by summing them with $\lambda^{\vec{i}\ast}\lambda^{\vec{i'}}$.
 When the number of coefficients $\lambda^{\vec{i}}$ is small through the truncation, the coefficients $\lambda^{\vec{i}}$ can be updated by computing the eigenvector with the minimum eigenvalue of the contracted operator $M_{1}^{\vec{i},\vec{i'}}M_{2}^{\vec{i},\vec{i'}}$. 

\color{black}

\section{Main result}\label{noise analysis}

 In the previous section, we reviewed recently proposed algorithms that effectively enlarge the size of simulatable quantum systems, such as Deep VQE~\cite{fujii2022deep} and entanglement forging~\cite{eddins2022doubling}, and they can be discussed within the framework of tree HTNs~\cite{yuan2021quantum}. One notable characteristic of HTNs lies in its step-by-step construction of new observables, which requires not only the classical post-processing but also quantum computations, e.g., the contraction of the effective Hamiltonian in Deep VQE~\cite{fujii2022deep} and the relevant algorithms~\cite{mizuta2021deep,erhart2022constructing} need the measurements on quantum computers. In applying these techniques to practical applications, one of the most critical issues is that the presence of physical noise in the quantum computations affects the overall outputs.

To address this issue, we investigate the effect of physical noise on HTTN states. 
Our main results are divided into two parts: in Secs.~\ref{sec:operator_based_representation}--\ref{sec:application}, we focus on the physical properties of an effectively simulated HTTN state under noisy conditions, and in Sec.~\ref{Sec: exponentialdecay}, we build on those findings by considering a specific noise model to analyze how noise impacts the estimation expectation values in an HTTN. Specifically,
\begin{itemize}
    \item In Sec.~\ref{sec:operator_based_representation}, we first introduce the map called {\it expansion map}, which provides a more general representation of existing HTNs in noise-free situations and will be a crucial foundation of the discussion even in noisy situations.
    \item In Sec.~\ref{sec:2-layer} and Sec.~\ref{sec:phys_prop_A}, we analyze the propagation of physical noise in a 2-layer HTTN state. In these two sections, we examine the physical properties of the effectively simulated HTTN state $\tilde{\rho}_{\rm HT}$ by attributing the physical noise in the quantum computational process to the expansion map that comprises $\tilde{\rho}_{\rm HT}$. In Sec.~\ref{sec:generalization}, we extend our discussion to multiple-layer quantum-classical HTTN states, where the types of state preparation (i)-(iii) are interchangeably used. 
    \item In Sec.~\ref{sec:application}, we apply our discussion to analyzing noise propagation in the Deep VQE and entanglement forging method. 
\end{itemize}
As a natural consequence of this argument, we can show that as long as type (i)-(iii) state preparation is used, the physicality of the HTN state $\tilde{\rho}_{\rm HT}$, i.e., $\tr[\tilde{\rho}_{\rm HT} H] \geq E_{\rm min}$ holds for the minimum energy $E_{\rm min}$.

\subsection{Density matrix representation of HTNs}\label{sec:operator_based_representation}

\begin{table*}[t]
\renewcommand{\arraystretch}{2.1}
\centering
\begin{tabular}{ccc}
\hline
~Type,~~$\xi_k$~~ & ~~Quantum states,~~$\ket{\psi_{k}^{\vec{i}_k}}$~~ &~~~Expansion map,~~$\mathcal{A}_{k}^{(\xi_k)}(\bullet)$~~~\\
\hline\hline
1 & $U_{k}^{(1)}\ket{\vec{i}_k}\ket{0}^{\otimes (n_k-b_k)}$ & $\mathcal{U}_k^{(1)}(\,\bullet\, \otimes \,\ket{0}\bra{0}^{\otimes (n_k-b_k)}\,)$ \\
2 & $\braket{\vec{i}_k|\psi_{k}^{(2)}}$ & $\mathrm{Tr}_{12} \left[ \ket{\rm Bell}\bra{\rm Bell}_{12} \left( \bullet_1 \otimes \ket{\psi^{(2)}_k}\bra{\psi_{k}^{(2)}}_{23}\right) \right]$ \\
3 & $P_{k}^{\vec{i}_k}\ket{\psi_{k}^{(3)}}$ & ~~$2^{b_k}\,\mathrm{Tr}_{12} \left[ \ket{\rm Bell}\bra{\rm Bell}_{12} \, \left(\bullet_{1} \, \otimes \mathcal{CP}_{23}\Bigl(\, \ket{+}\bra{+}^{\otimes b_k}_{2} \otimes \ket{\psi^{(2)}_k}\bra{\psi_{k}^{(3)}}_{3} \Bigr) \right)  \right]$~~ \\
4 & $U_{k}^{\vec{i}_k}\ket{0}^{\otimes n_k}$ & ~~$2^{b_k}\,\mathrm{Tr}_{12} \left[ \ket{\rm Bell}\bra{\rm Bell}_{12} \, \left(\bullet_{1} \, \otimes \mathcal{CU}_{23}\Bigl(\, \ket{+}\bra{+}^{\otimes b_k}_{2} \otimes \ket{0}\bra{0}_3^{\otimes n_k} \Bigr) \right)  \right]$~~ \\
\hline
\end{tabular}
\caption{\label{tab:expansion_maps} Relations between the set of states $\{\ket{\psi^{\vec{i}_k}_k}\}$ defined in Sec.~\ref{sec:hybrid_tree_tensor_networks} and the corresponding expansion map $\mathcal{A}_k^{(\xi_k)}$ for each type $\xi$. The subscripts $1,2,3$ represent a system each operator acts on, and we have defined the unnormalized Bell state $\ket{\mathrm{Bell}}_{12}:=\sum_{\vec{i}_{k}} \ket{\vec{i}_{k}}_{1} \ket{\vec{i}_{k}}_{2}$, a controlled Pauli channel $\mathcal{CP}_{23}(\bullet):= (\sum_{\vec{i}_k} \ket{\vec{i}_k}\bra{\vec{i}_k}_{2} \otimes P_{k,3}^{\vec{i}_k}) \bullet_{23} (\sum_{\vec{j}_k} \ket{\vec{j}_k}\bra{\vec{j}_k}_{2} \otimes P_{k,3}^{\vec{j}_k})$, and a controlled unitary channel $\mathcal{CU}_{23}(\bullet):= (\sum_{\vec{i}_k} \ket{\vec{i}_k}\bra{\vec{i}_k}_{2} \otimes U_{k,3}^{\vec{i}_k}) \bullet_{23} (\sum_{\vec{j}_k} \ket{\vec{j}_k}\bra{\vec{j}_k}_{2} \otimes U_{k,3}^{\vec{j}_k,\dagger})$. 
Note that the expansion maps of type 3 and 4 are accompanied by the coefficient $2^{b_k}$, but this can effectively be ignored because it is canceled out due to the appearance of $\mathcal{A}_k^{(\xi_k)}$ in both the numerator and the denominator in Eq.~(\ref{eq: Op-based rep}).
} 
\label{intro_table}
\end{table*}

\subsubsection{The expansion map}\label{sec:expansion_operator}

In the original proposal of hybrid tensor networks~\cite{yuan2021quantum}, the HTN states are represented as a tensor network constructed by contracting indices of classical or quantum tensors, as shown in Fig.~\ref{fig: HTTNs}.
However, this representation requires the quantum tensor to be pure states, making it insufficient for describing the systems in a noisy environment. To capture the influence of physical noise, we introduce a new expression formalism to represent the HTN state based on the {\it expansion operators}.
Our representation describes the HTTN state as an expanded system due to the action of some operators, called expansion operators. 

To be specific, we first introduce the notion of the expansion operator: let $\mathcal{H_{\rm in}}$, $\mathcal{H_{\rm out}}$ be finite-dimensional Hilbert spaces of input and output systems, the expansion operator $A^{(\xi)}: \mathcal{H}_{\rm in} \rightarrow \mathcal{H}_{\rm out}$ 
is defined as follows:
\begin{equation}\label{eq:expansion_operator}
    A^{(\xi)} := \sum_{j} \ket{\psi^{j}}_{\rm out} \bra{j}_{\rm in}
\end{equation}
where $\{\ket{j}_{\rm in}\}$ is the set of computational-basis states on $\mathcal{H}_{\rm in}$ and $\{\ket{\psi^{j}}_{\rm out}\}$ are the set of states on $\mathcal{H}_{\rm out}$, which is not necessarily normalized. The upper subscript $\xi$ takes five possible values 1,2,3,4 and $\mathrm{c}$, and each value indicates the type (i)-(iv) and classical way of state preparation of the set $\{\ket{\psi^{j}}_{\rm out}\}$, respectively. 
As can be seen from Eq.~(\ref{eq:expansion_operator}), $A^{(\xi)}$ transforms the input computational-basis state $\ket{j}_{\rm in}$ on $\mathcal{H}_{\rm in}$ into the state $\ket{\psi^{j}}_{\rm out}$ on $\mathcal{H}_{\rm out}$ as $A^{(\xi)}\ket{j}_{\rm in}=\ket{\psi^j}_{\rm out}$.

\begin{figure}[t]
\begin{center}
\centering
 \includegraphics[width=80mm]{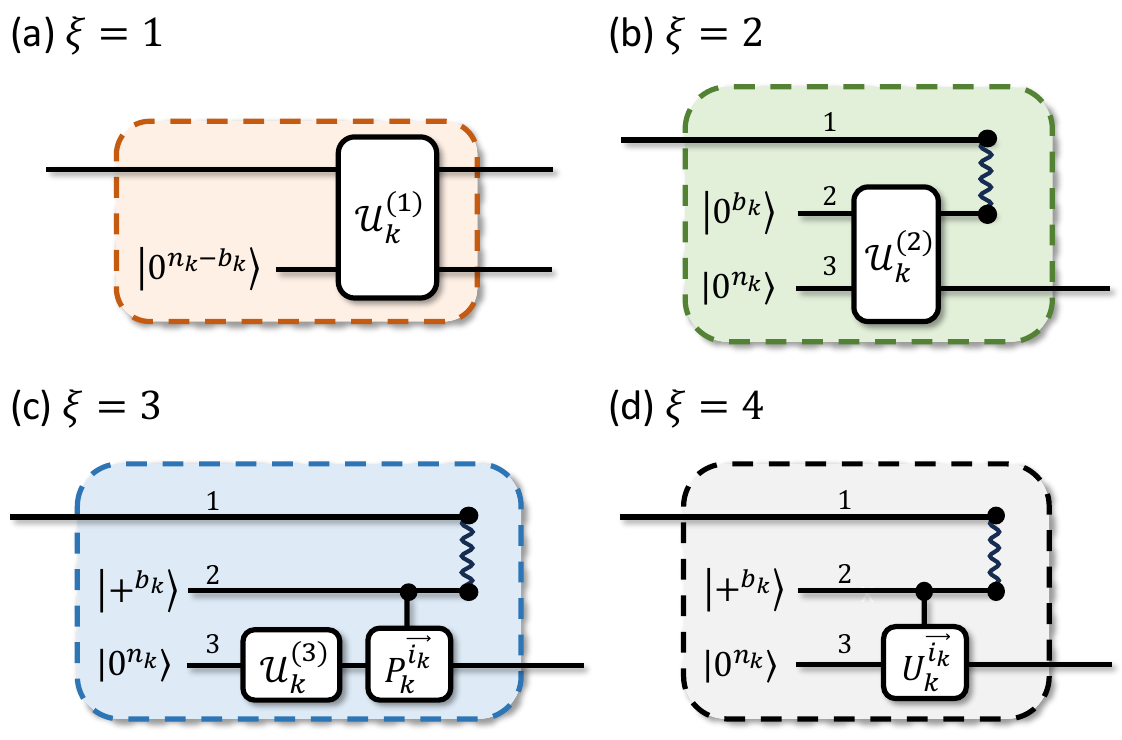}
\end{center}
\caption{\label{Fig:exp_maps} A schematic of the expansion maps $\mathcal{A}_k^{(\xi)}$ ($\xi=1,2,3,4$). The wavy line represents a projective measurement onto the unnormalized Bell state $\ket{\rm Bell}$.}
\end{figure}

By using the concept of the expansion operator introduced above, the HTN state $\ket{\psi_{\rm HT}}$ in Eq.~(\ref{eq: HTTN state_2}) can be regarded as a state where the state $\ket{\psi}$ is expanded by the expansion operators $A_{k}^{(\xi_k)} = \sum_{\vec{i}_k} \ket{\psi_{k}^{\vec{i}_k}} \bra{\vec{i}_{k}}$ ($k=1,...,N$) as
\begin{equation}\label{eq: HTTN state_2_density}
    \ket{\psi_{\rm HT}} = \frac{1}{C} A_{1}^{(\xi_1)} \otimes \cdots \otimes A_{N}^{(\xi_N)} \ket{\psi}.
\end{equation}
The proof of the equivalence between Eqs.~(\ref{eq: HTTN state_2}) and (\ref{eq: HTTN state_2_density}) are given in Appendix~\ref{sec: appendix A}.
One important feature of this representation is the absence of the summations that appeared in Eq.~(\ref{eq: HTTN state_2}), which allows for a simpler description of the HTN state $\ket{\psi_{\rm HT}}$. Another important feature of this representation is that it can be easily extended to the description of the density operators $\rho_{\rm HT} := \ket{\psi_{\rm HT}}\bra{\psi_{\rm HT}}$ while maintaining its simplicity. That is, defining a map 
\begin{equation}\label{eq:def_of_exp_map}
    \mathcal{A}_{k}^{(\xi_k)}(\bullet) := A_{k}^{(\xi_k)} \bullet A_{k}^{(\xi_k)\dagger}
\end{equation}
for $k=1,...,N$, the density operator $\rho_{\rm HT}$ can be represented as a state expanded by the maps $\mathcal{A}_{k}^{(\xi_k)}$ as
\begin{equation}\label{eq: Op-based rep}
    \rho_{\rm HT} =\frac{\mathcal{A}_{1}^{(\xi_{1})} \otimes ... \otimes \mathcal{A}_{N}^{(\xi_{N})} (\ket{\psi}\bra{\psi})}{\mathrm{Tr}[\mathcal{A}_{1}^{(\xi_{1})} \otimes ... \otimes \mathcal{A}_{N}^{(\xi_{N})} (\ket{\psi}\bra{\psi})]}.
\end{equation}
We note that the detailed descriptions of the expansion map $\mathcal{A}_{k}^{(\xi_k)}$ are different from the label $\xi_k$. Table~\ref{tab:expansion_maps} summarize the expansion maps $\mathcal{A}^{(\xi_k)}_{k}$ for $\xi_k=1,2,3,4$ derived from the definition of $\mathcal{A}_k^{(\xi_k)}$ in Eq.~(\ref{eq:def_of_exp_map}); see Appendx~\ref{sec:detail_derive} for the detailed deviations. Fig.~\ref{Fig:exp_maps} provides the quantum circuit diagrams for the expansion maps in Table~\ref{tab:expansion_maps}.

\subsubsection{Application to the expectation value calculation}\label{sec:application_expansion}

Similar to the discussion in Sec.~\ref{sec:hybrid_tree_tensor_networks}, we now consider applying the state $\rho_{\rm HT}$ to estimate the expectation value of an observable. 
Suppose that 
$O=\bigotimes_{k=1}^{N}O_k$ is an $(\sum_k n_k)$-qubit observable where each observable $O_k$ acts on the $k$-th subsystem~\footnote{For simplicity, we consider the case where the observable $O$ can be described as a tensor product of subsystem observables. However, by considering linear combinations of these components, we can extend our discussion to encompass the general observable.}, the expectation value $\braket{O}_{\rho_{\rm HT}}$ of the observable $O$ on the state $\rho_{\rm HT}$ can be written as follows:
\begin{equation}\label{eq: Effective hamiltonian in Op-based rep_1}
    \braket{O}_{\rho_{\rm HT}}
    = \frac{\mathrm{Tr}[O \bigl(\mathcal{A}_{1}^{(\xi_{1})} \otimes ... \otimes \mathcal{A}_{N}^{(\xi_{N})} \bigr)(\ket{\psi}\bra{\psi})]}{\mathrm{Tr}[\mathcal{A}_{1}^{(\xi_{1})} \otimes ... \otimes \mathcal{A}_{N}^{(\xi_{N})} (\ket{\psi}\bra{\psi})]}.
\end{equation}
Here, we define the new local observable $M_k^{(\xi_k)}$ ($S_k^{(\xi_k)}$) as the Heisenberg evolution of the local observable $O_k$ ($I_k$) via the adjoint action of the expansion map $\mathcal{A}_k^{(\xi_k)}$, i.e., $M^{(\xi_{k})}_{k} = \mathcal{A}_{k}^{(\xi_k)\dagger}(O_k)$ where $\mathcal{A}_{k}^{(\xi_k)\dagger}$ denotes the adjoint action. Introducing the observables into Eq.~(\ref{eq: Effective hamiltonian in Op-based rep_1}), we have
\begin{equation}\label{eq: Effective hamiltonian in Op-based rep_2}
   \braket{O}_{\rho_{\rm HT}}
    = \frac{\mathrm{Tr}[(M^{(\xi_{1})}_{1} \otimes ... \otimes M^{(\xi_{N})}_{N}) \ket{\psi}\bra{\psi} ]}{\mathrm{Tr}[(S^{(\xi_{1})}_{1} \otimes ... \otimes S^{(\xi_{N})}_{N}) \ket{\psi}\bra{\psi}]}.
\end{equation}
We remark that the observables $M_k^{(\xi_k)}$ and $S_k^{(\xi_k)}$ ($k=1,...,N$) in Eq.~(\ref{eq: Effective hamiltonian in Op-based rep_2}) are exactly the same as those introduced in Sec.~\ref{sec: htn}.
This can be easily confirmed from the simple calculation of $M^{(\xi_{k})}_{k} = \mathcal{A}_{k}^{(\xi_k)\dagger}(O_k)$ and $S^{(\xi_{k})}_{k} = \mathcal{A}_{k}^{(\xi_k)\dagger}(I_k)$, using the definition $A_{k}^{(\xi_k)} = \sum_{\vec{i}_k} \ket{\psi_{k}^{\vec{i}_k}} \bra{\vec{i}_{k}}$.

Based on the above equations, we can evaluate $\braket{O}_{\rho_{\rm HT}}$ as follows: we first calculate $M_{k}^{(\xi_k)}$ and $S_{k}^{(\xi_k)}$ for all $k$ using the procedure in Sec.~\ref{contraction_of_local_observables}, estimate the expectation values of $\bigotimes_{k=1}^{N} M_k^{(\xi_k)}$ and $\bigotimes_{k=1}^{N} S_k^{(\xi_k)}$ on the system $\ket{\psi}\bra{\psi}$, and then combine the expectation values as Eq.~(\ref{eq: Effective hamiltonian in Op-based rep_2}).

\subsection{2-layer HTTN states under physical noise}\label{sec:2-layer}

 As described in Sec.~\ref{sec:operator_based_representation}, the calculation of $\braket{O}_{\rho_{\rm HT}}$
 is carried out by repeatedly transforming the local observables $O_k$ ($I_k$) into the new local observables $M_k^{(\xi_k)}$ ($S_k^{(\xi_k)}$) via the adjoint action of the classical/quantum tensors $\mathcal{A}_{k}^{(\xi_{k})}$, i.e., $M^{(\xi_{k})}_{k} = \mathcal{A}_{k}^{(\xi_k)\dagger}(O_k)$. 
 This calculation is guaranteed to work well in a noise-free environment, but when physical noise is present in the local computational processes, its effect accumulates in the resulting observables $M_k^{(\xi_k)}$ and propagates to the subsequent calculations.
 
 In Secs.~\ref{sec:2-layer}--\ref{sec:generalization}, we analyze the effect of the physical noise on the HTTN state with the density matrix representation of HTNs. We begin by focusing on the case where the HTTN state is the 2-layer model represented by Eq.~(\ref{eq: Op-based rep}) with the restriction of $\xi_1=\cdots=\xi_N=\xi\in\{1,2,3,4\}$ in Sec.~\ref{sec:2-layer} and Sec.~\ref{sec:phys_prop_A}. These two sections are structured as follows. In Sec.~\ref{sec:noise_assumption}, we introduce the noise assumptions for our analysis, and in Sec.~\ref{sec:2-layer_httn_state}, we define the noisy version $\tilde{\mathcal{A}}_k^{(\xi)}$ of the expansion map to obtain the explicit expression for the HTTN state $\tilde{\rho}_{\rm HT}$ that is effectively simulated under physical noise. In Sec.~\ref{sec:phys_prop_A}, we investigate the physical properties of $\tilde{\mathcal{A}}_k^{(\xi)}$ for examining whether $\tilde{\rho}_{\rm HT}$ is a valid quantum state.

\subsubsection{Noise assumption}\label{sec:noise_assumption}
As for the assumption of physical noise, we mainly focus on the effects of gate noise, assuming that the measurement and statistical noise are sufficiently small (For an extended discussion that includes the effect of these errors during the measurement process, refer to Appendix~\ref{sec:effect_of_meas_and_stat_noise}). Specifically, we consider a situation where the unitary channels $\mathcal{U}_{k}^{(\xi)}$ ($\xi=1,2,3$), $\mathcal{U}_{k}^{(4),\vec{i}_k}$, and $\mathcal{C}^{\vec{i}_k,\vec{i}'_k}_k$ in Fig.~\ref{fig: Quantum circuits for M_k} are transformed into noisy channels $\mathcal{W}_{k}^{(\xi)}$ ($\xi=1,2,3$), $\mathcal{W}_{k}^{(4),\vec{i}_k}$, and $\tilde{\mathcal{C}}^{\vec{i}_k,\vec{i}'_k}_k$, respectively.
In addition, suppose that the Hermitian observables $\{M_{k}^{(\xi)}\}_{k}$($\{S_{k}^{(\xi)}\}_{k}$) and the pure state $\ket{\psi}\bra{\psi}$ are transformed into the noisy ones $\{\tilde{M}_{k}^{(\xi)}\}_{k}$($\{\tilde{S}_{k}^{(\xi)}\}_{k}$) and the mixed state $\rho$ due to the effect of physical noise (see Fig.~\ref{Fig:noise_assumption}(a))

\subsubsection{2-layer HTTN state under physical noise}\label{sec:2-layer_httn_state}

Given the assumptions introduced above, the noisy expectation value, denoted as $\braket{\tilde{O}}_{\rho_{\rm HT}}$, can be written as
\begin{equation}\label{eq: Noisy effective hamiltonian in Op-based rep}
    \braket{\tilde{O}}_{\rho_{\rm HT}} = \frac{\mathrm{Tr}[\,(\tilde{M}^{(\xi)}_{1} \otimes ... \otimes \tilde{M}^{(\xi)}_{N}) \,\rho\, ]}{\mathrm{Tr}[\,(\tilde{S}^{(\xi)}_{1} \otimes ... \otimes \tilde{S}^{(\xi)}_{N}) \,\rho\,]}.
\end{equation}
Here, our interest lies in the noisy state $\tilde{\rho}_{\rm HT}$, which we effectively simulate under the physical noise, for the observable $O$ in Eq.~(\ref{eq: Noisy effective hamiltonian in Op-based rep}), i.e., 
\begin{equation}\label{eq: def_of_noisy_rho}
    \braket{\tilde{O}}_{\rho_{\rm HT}} = \braket{O}_{\tilde{\rho}_{\rm HT}} = \mathrm{Tr}[\,O\, \tilde{\rho}_{\rm HT} \,],
\end{equation}
where $\tilde{\rho}_{\rm HT}$ denotes the noisy effective HTTN state induced by the physical noise. 
Note that how physical noise accumulates in quantum computation depends on the choice of a state preparation method determined by $\xi$. In the following, for each $\xi$, we derive the explicit representation of the noisy effective HTTN state $\tilde{\rho}_{\rm HT}$ and examine whether $\tilde{\rho}_{\rm HT}$ satisfies the condition to be a density operator (i.e., $\mathrm{Tr}(\tilde{\rho}_{\rm HT})=1$ and $\tilde{\rho}_{\rm HT} \ge 0$).  

\begin{figure}[t]
\begin{center}
\centering
 \includegraphics[width=80mm]{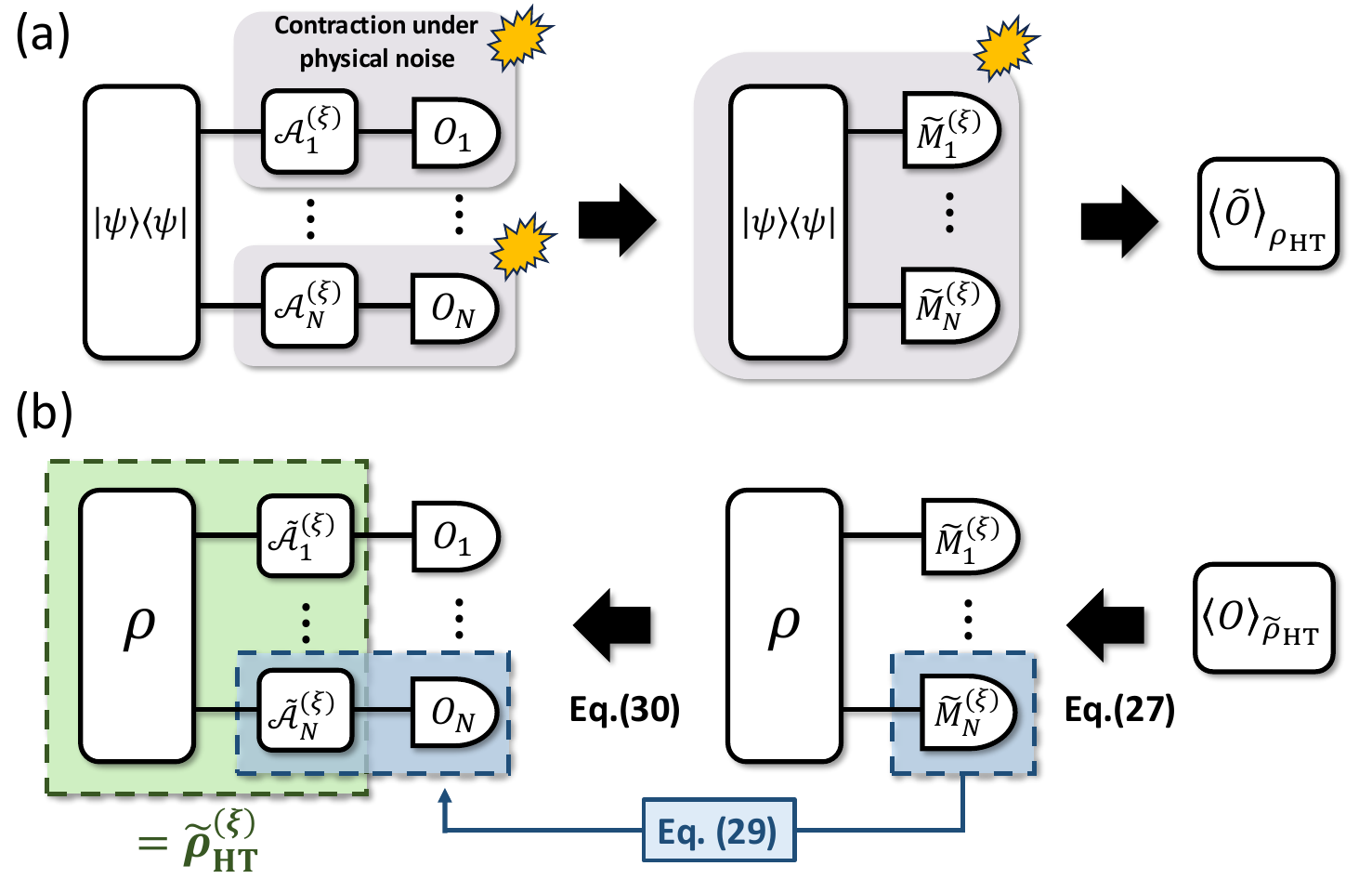}
\end{center}
\caption{\label{Fig:noise_assumption}(a) Contraction processes for calculating the expectation value of the observable $O$ for $\rho^{(\xi)}_{\rm HT}$ under a noisy environment.
(b) A summary of our approach for obtaining the explicit expression of the noisy effective HTTN state 
$\tilde{\rho}^{(\xi)}_{\rm HT}$.
}
\end{figure}

\begin{table*}[tp]
\centering
\renewcommand{\arraystretch}{2.4}
\begin{tabular}{ccc}
\hline
~Type,~~$\xi$~~ &~~~Noisy expansion map,~~$\tilde{\mathcal{A}}_{k}^{(\xi)}(\bullet)$~~~&Physicality\\
\hline\hline
1 & $\mathcal{W}_k^{(1)}(\,\bullet\, \otimes \,\ket{0}\bra{0}^{\otimes (n_k-b_k)}\,)$ & CP \\
2 & $\mathrm{Tr}_{12} \left[ \ket{\rm Bell}\bra{\rm Bell}_{12} \left( \bullet_1 \otimes \sigma_{k,23}^{(2)}\right) \right]$ & CP \\
3 & ~~$2^{b_k}\,\mathrm{Tr}_{12} \left[ \ket{\rm Bell}\bra{\rm Bell}_{12} \, \left(\bullet_{1} \, \otimes \mathcal{CP}_{23}\Bigl(\, \ket{+}\bra{+}^{\otimes b_k}_{2} \otimes \sigma_{k,3}^{(3)} \Bigr) \right)  \right]$ & CP \\
\hline
\end{tabular}
\caption{\label{tab:noisy_expansion_maps} A summary of noisy expansion maps $\tilde{\mathcal{A}}_{k}^{(\xi)}$ ($\xi=1,2,3$). We have defined $\sigma_{k}^{(2)}:=
\mathcal{W}^{(2)}_{k}(\ket{0}\bra{0}^{\otimes (n_k+b_k)})$ and $\sigma_k^{(3)}:=\mathcal{W}^{(3)}_{k}(\ket{0}\bra{0}^{\otimes n_k})$.
The `CP' means that $\tilde{\mathcal{A}}_k^{(\xi)}$ is completely positive (CP) map. The proof of $\tilde{\mathcal{A}}_{k}^{(\xi)}$ being a CP map can be found in Appendix~\ref{sec:physicality_app}.
}
\end{table*}

The main idea for obtaining a concrete description of $\tilde{\rho}_{\rm HT}$ from Eq.~(\ref{eq: Noisy effective hamiltonian in Op-based rep}) is to track back the inverse of the HTTN contraction process (Fig.~\ref{Fig:noise_assumption}(b)). Specifically, once we derive the noisy local observables $\tilde{M}_{k}^{(\xi)}$ ($\tilde{S}_k^{(\xi)}$) by performing a certain estimation procedure for the observable $O_k$ ($I_k$), we consider the following equalities:
\begin{equation}\label{eq: transformation}
\begin{split}
    \mathrm{Tr}[\, \tilde{M}_{k}^{(\xi)} \bullet \,] &= \mathrm{Tr} [\,O_{k}\, \tilde{\mathcal{A}}_{k}^{(\xi)} (\bullet)\,],\\
    \mathrm{Tr}[\, \tilde{S}_{k}^{(\xi)} \bullet\, ] & = \mathrm{Tr} [\, \tilde{\mathcal{A}}_{k}^{(\xi)} (\bullet)\,],
\end{split}
\end{equation}
where $\tilde{\mathcal{A}}_k^{(\xi)}$ is some superoperator satisfying the above equalities for $\tilde{M}_{k}^{(\xi)}$, $\tilde{S}_k^{(\xi)}$, $O_k$ and $I_k$.
The transformations from the left-hand side to the right-hand side correspond to the inverse of the estimation procedure described in Sec.~\ref{contraction_of_local_observables}, where
$M_k^{(\xi)}$ is obtained from $\mathcal{A}^{(\xi)}_k$ and $O_k$. By applying Eq.~(\ref{eq: transformation}) to each $\tilde{M}_{k}^{(\xi)}$ and $\tilde{S}_{k}^{(\xi)}$ in Eq.~(\ref{eq: Noisy effective hamiltonian in Op-based rep}), we have
\begin{equation}\label{eq: transformation2}
    \braket{\tilde{O}}_{\rho_{\rm HT}} = \frac{\mathrm{Tr}[O \bigl(\tilde{\mathcal{A}}_{1}^{(\xi)} \otimes ... \otimes \tilde{\mathcal{A}}_{N}^{(\xi)} \bigr)(\rho)]}{\mathrm{Tr}[\tilde{\mathcal{A}}_{1}^{(\xi)} \otimes ... \otimes \tilde{\mathcal{A}}_{N}^{(\xi)} (\rho)]}.
\end{equation}
Comparing Eq.~(\ref{eq: transformation2}) with Eq.~(\ref{eq: def_of_noisy_rho}), we obtain
\begin{equation}\label{eq:noisy_httn_state}
    \tilde{\rho}_{\rm HT}^{(\xi)} = \frac{ \tilde{\mathcal{A}}_{1}^{(\xi)} \otimes ... \otimes \tilde{\mathcal{A}}_{N}^{(\xi)} (\rho)}{\mathrm{Tr}[\tilde{\mathcal{A}}_{1}^{(\xi)} \otimes ... \otimes \tilde{\mathcal{A}}_{N}^{(\xi)} (\rho)]},
\end{equation}
where the superscript $\xi$ in $\tilde{\rho}_{\rm HT}^{(\xi)}$ indicates the type of state preparation used in each local tensor. 
By comparing Eq.~(\ref{eq: Op-based rep}) with Eq.~(\ref{eq:noisy_httn_state}), we find that $\tilde{\mathcal{A}}_k^{(\xi)}$ corresponds to the expansion operator $\mathcal{A}_k^{(\xi)}$ under the noise-free situation. Hence, we define the {\it noisy expansion map} as the superoperator $\tilde{\mathcal{A}}_k^{(\xi)}$, which satisfies Eq.~(\ref{eq: transformation}) for the observables $\tilde{M}_k^{(\xi)}$ and $\tilde{S}_k^{(\xi)}$ derived under the noisy condition, and for the observabes $O_k$ and $I_k$.

\begin{figure*}[tp]
\begin{center}
\centering
 \includegraphics[width=170mm]{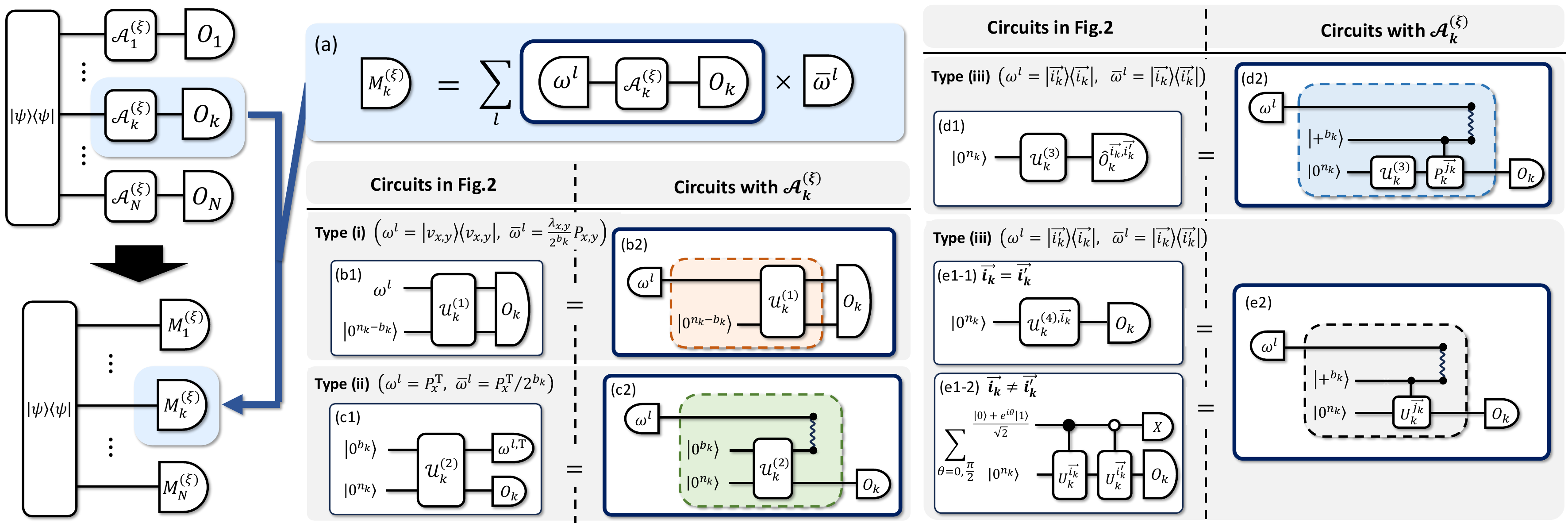}
\end{center}
\caption{\label{Fig:exp_map_relation} Relation between the expansion map $\mathcal{A}_k^{(\xi)}$ and the estimation procedures for $M_k^{(\xi)}$ and $S_k^{(\xi)}$ introduced in Sec.~\ref{contraction_of_local_observables}. (a) A schematic representation of the estimation procedure for $M_k^{(\xi)}$ using $\mathcal{A}_k^{(\xi)}$. The procedure can be identified with the model where for each label $l$, one prepares a matrix $\omega^{l}$, evolves it under the map $\mathcal{A}_k^{(\xi)}$, measures $\mathcal{A}_k^{(\xi)}(\omega^{l})$ with the observable $O_k$, and then weight the outcomes by matrices $\bar{\omega}^{l}$. (b,c,d,e) The relation between $\mathcal{A}_k^{(\xi)}$ and the quantum circuits for estimating $M_k$. (b1), (c1), (d1), and (e1-1)/(e1-2) are the quantum circuits used in types (i) - (iv), and (b2), (c2), (d2), and (e2) are the corresponding rewritten circuits as the maps $\mathcal{A}_k^{(\xi)}$ explicitly appear.
}
\end{figure*}

\subsection{Physical properties of noisy expansion maps}\label{sec:phys_prop_A}

To check the physicality of $\tilde{\rho}_{\rm HT}^{(\xi)}$, we turn our attention to deriving the noisy maps $\tilde{\mathcal{A}}_k^{(\xi)}$ that satisfies Eq.~(\ref{eq: transformation}),
as Eq.~(\ref{eq:noisy_httn_state}) indicates that $\tilde{\rho}^{(\xi)}_{\rm HT}$ satisfies the requirement of a density operator when each $\tilde{\mathcal{A}}_{k}^{(\xi)}$ is a completely positive (CP) map.
To find the map $\tilde{\mathcal{A}}^{(\xi)}_k$ satisfying Eq.~(\ref{eq: transformation}), we start by considering the relation between the expansion maps $\mathcal{A}_k^{(\xi)}$ and the estimation models in Sec.~\ref{contraction_of_local_observables} under noise-free conditions. As a common feature of the estimation procedures for $M_k$ and $S_k$ in types (i) - (iv) (given in Eqs.(\ref{eq: Case1, M_k, 1}), (\ref{eq: Case2, M_k, 2}), (\ref{eq: Case3, M_k, 2}), and (\ref{eq: Case4, M_k, 2}), respectively), each of them can be expressed by using the expansion maps $\mathcal{A}_{k}^{(\xi)}$ as
\begin{equation}\label{eq:estimation_form}
\begin{split}
    M_{k}^{(\xi)} &= \mathcal{A}_{k}^{(\xi)\dagger}(O_k) = \sum_{l} \mathrm{Tr}[ O_k \mathcal{A}_{k}^{(\xi)} (\omega^{l})]~ \bar{\omega}^l,\\
    S_{k}^{(\xi)} &= \mathcal{A}_{k}^{(\xi)\dagger}(I_k) = \sum_{l} \mathrm{Tr}[ \mathcal{A}_{k}^{(\xi)} (\omega^{l})]~ \bar{\omega}^l,
\end{split}
\end{equation}
where $\{\omega^{l},\bar{\omega}^{l}\}_{l}$ is a set of matrices satisfying $\mathrm{SWAP}=\sum_{l} \omega^{l}\otimes \bar{\omega}^{l}$. The graphical representation can be found in Fig.~\ref{Fig:exp_map_relation}(a).
Here, based on the relation between $\mathcal{A}_{k}^{(\xi_k)}$ and the estimation procedure in the noise-free situation, we can use the following proposition to find the noisy expansion map $\tilde{\mathcal{A}}^{(\xi)}_k$ that fulfills Eq.~(\ref{eq: transformation}).
\begin{prop}\label{prop:fact_1}
When a single map $\mathcal{M}$ can express a given estimation procedure in the form of 
\begin{equation}\label{eq:local observable estimation_2}
\begin{split}
    \tilde{M}_{k}^{(\xi)} &= \sum_l \mathrm{Tr}[ O_{k} \, \mathcal{M}(\omega^{l})] \,\bar{\omega}^{l}\\
    \tilde{S}_{k}^{(\xi)} &= \sum_l \mathrm{Tr}[ \mathcal{M}(\omega^{l})] \,\bar{\omega}^{l}\\
\end{split}
\end{equation}
using a set of operators $\{\omega_{l},\bar{\omega}^{l}\}_{l}$ 
that satisfies $\mathrm{SWAP}=\sum_{l} \omega^{l}\otimes \bar{\omega}^{l}$, then the map $\mathcal{M}$ can be identified with the noisy expansion map $\tilde{\mathcal{A}}_k^{(\xi)}$ defined in Eq.~(\ref{eq: transformation}), i.e., 
the version of Eq.~(\ref{eq: transformation}) with $\tilde{\mathcal{A}}^{(\xi)}_k$ replaced by $\mathcal{M}$ holds. 
\end{prop}
The detailed proof can be found in Appendix~\ref{sec:proof_of_prop1}.
Thus, for obtaining the map $\tilde{\mathcal{A}}_k^{(\xi)}$, we only need to seek a single map $\mathcal{M}$ that represents the estimation procedure of each type under noisy conditions.
Before presenting the detailed derivation based on Proposition~\ref{prop:fact_1}, we first summarize the noisy expansion maps for $\xi=1,2,3$ in Table~\ref{tab:noisy_expansion_maps}, which are obtained by applying the noise assumptions (Sec.~\ref{sec:noise_assumption}) to the procedures described in Sec.~\ref{contraction_of_local_observables}.
As shown in Table~\ref{tab:noisy_expansion_maps}, the noisy expansion maps $\tilde{\mathcal{A}}^{(\xi)}_k$ for each of three types are CP maps, indicating that the physical noise occurring in the quantum circuit does not affect the physical properties of the simulated HTTN state $\tilde{\rho}_{\rm HT}^{(\xi)}$. In contrast, as will be detailed later, there is no guarantee that the noisy HTTN state $\tilde{\rho}_{\rm HT}^{(\xi)}$ remains its physicality. 

Here, we outline the discussion in the following.
In Sec.~\ref{sec:relation_between_A_and_sec3}, as a preparation for the discussion in the presence of physical noise, we establish the connection between the expansion maps $\mathcal{A}_k^{(\xi)}$ and the quantum circuits introduced in Sec.~\ref{contraction_of_local_observables}, which are used for estimating $M_k^{(\xi)}$ and $S_k^{(\xi)}$ under noise-free conditions.
In the subsequent sections, based on this connection, we examine how noise in each quantum circuit for each type can be attributed to noise within $\mathcal{A}_k^{(\xi)}$. Specifically, we discuss the cases of types (i) - (iii) in Sec.~\ref{sec:physicality} and the case of type (iv) in Sec.~\ref{sec: noise analysis, case4}.
In Sec.~\ref{sec: noise analysis, classical}, we discuss the physicality of $\tilde{\rho}_{\rm HT}^{(\rm c)}$ for the general discussion in Sec.~\ref{sec:generalization}.

\subsubsection{Relation between the expansion maps and the procedures in Sec.~\ref{contraction_of_local_observables}}\label{sec:relation_between_A_and_sec3}

As already mentioned, the estimation procedures for $M_k^{(\xi)}$ and $S_k^{(\xi)}$ in types (i) - (iv) can be represented with the expansion maps $\mathcal{A}_{k}^{(\xi)}$ as Eq.~(\ref{eq:estimation_form}). Hence, in the absence of noise, the type (i) - (iv) estimation procedures in Sec.~\ref{contraction_of_local_observables} are equivalent to a model in which one computes $M_k^{(\xi)}$ by preparing the matrix $\omega^l$, evolving it under the expansion map $\mathcal{A}_{k}^{(\xi)}$, measuring the resultant object $\mathcal{A}_{k}^{(\xi)}(\omega^{l})$ with the observable $O_k$, and combine the measurement outcomes with the matrices $\bar{\omega}^{l}$. 

For example, recall the estimation procedure for $M_k^{(1)}$ in Sec.~\ref{sec:2a}. In this estimation, we estimate $M_k^{(1)}$ by summing over the measurement outcomes of the observable $O_k$ from the quantum circuit in Fig.~\ref{fig: Quantum circuits for M_k} (or Fig.~\ref{Fig:exp_map_relation}(b1)) with the weight $\bar{\omega}^{l}=\frac{\lambda_{x,y}P_{x}}{2^{b_k}}$ according to Eq.~(\ref{eq: Case1, M_k, 1}). Here, noticing that the quantum circuit in Fig.~\ref{Fig:exp_map_relation}(b1) explicitly includes the expansion map 
\begin{equation}
    \mathcal{A}_{k}^{(1)}(\bullet)=\mathcal{U}_{k}^{(1)}(\bullet\otimes\ket{0}\bra{0}^{\otimes(n_k-b_k)}),
\end{equation}
as shown in Fig.~\ref{Fig:exp_map_relation}(b2), the above estimation procedure can be interpreted as follows: for each $l=(x,y)$, we first prepare the state $\omega^l=\ket{v_{x,y}}\bra{v_{x,y}}$, evolve it under the map $\mathcal{A}_k^{(1)}$, measure the resulting state with the observable $O_k$ (Fig.~\ref{Fig:exp_map_relation}(b2)), and then weight the outcome using the matrix $\bar{\omega}^{l}$ (Fig.~\ref{Fig:exp_map_relation}(a)).

With regard to types (ii) and (iii), although the expansion maps $\mathcal{A}_{k}^{(2)}$ and $\mathcal{A}_{k}^{(3)}$ do not explicitly appear in the quantum circuits (Fig.~\ref{Fig:exp_map_relation}(c1) and (d1)) for estimating $M_k^{(\xi)}$, we can modify each original quantum circuit with the one which includes $\mathcal{A}_{k}^{(2)}$ and $\mathcal{A}_{k}^{(3)}$ as shown in Fig.~\ref{Fig:exp_map_relation}(c2) and (d2), thanks to the following relations
\begin{equation}\label{eq:trans_a2}
    \mathrm{Tr}[O_k \mathcal{A}_k^{(2)}(P_x^{\rm T})]=\mathrm{Tr}[(P_x \otimes O_k) \mathcal{U}_{k}^{(2)}(\ket{0}\bra{0}^{\otimes (n_k+b_k)})]
\end{equation}
and 
\begin{equation}\label{eq:trans_a3}
    \mathrm{Tr}[O_k \mathcal{A}_k^{(3)}(\ket{\vec{i}'_k}\bra{\vec{i}_k})] =\mathrm{Tr}[\hat{O}_k^{\vec{i}_k,\vec{i}'_k} \mathcal{U}_k^{(3)}(\ket{0}\bra{0}^{\otimes n_k})].
\end{equation}
The proof of the Eqs.~(\ref{eq:trans_a2}) and (\ref{eq:trans_a3}) can be found in Appendix~\ref{sec:relation_A_and_circuit}.
By choosing
\begin{equation}
    \{\omega^{l},\bar{\omega}^{l}\}_{l}=\left\{P_x^{\rm T},\frac{P_x^{\rm T}}{2^{b_k}}\right\}_{x}, ~~\text{and}~~ \{\ket{\vec{i}'_k}\bra{\vec{i}_k},\ket{\vec{i}_k}\bra{\vec{i}'_k}\}_{\vec{i}_k,\vec{i}'_k},
\end{equation}
for the type (ii) and (iii) estimations, respectively, we can also regard the original procedures as the model which indirectly simulates $\mathcal{A}_k^{(\xi)}$, computes $\mathrm{Tr}[ O_k \mathcal{A}_{k}^{(\xi)} (\omega^{l})]$, and then weights the result with the matrix $\bar{\omega}^{l}$. Note that the above discussion also holds when estimating $S_k^{(\xi)}$ by setting $O_k=I_k$; see Appendix~\ref{sec:relation_A_and_circuit}.

Lastly, we mention the relationship between $\mathcal{A}_k^{(4)}$ and the estimation procedure for $M_k^{(4)}$ and $S_k^{(4)}$ in Sec.~\ref{sec: 2d}. Similar to the types (ii) - (iii), we can rewrite the two types of original quantum circuits in Fig.~\ref{Fig:exp_map_relation}(e1-1) and (e1-2) with the circuit including $\mathcal{A}_{k}^{(4)}$ as in Fig.~\ref{Fig:exp_map_relation}(e2); see Appendix~\ref{apsec:a4_2} for the detailed calculations for the rewriting. Thus, by setting
\begin{equation}
    \{\omega^l,\bar{\omega}^l\}_l = \{\ket{\vec{i}'_k}\bra{\vec{i}_k},\ket{\vec{i}_k}\bra{\vec{i}'_k}\}_{\vec{i}_k,\vec{i}'_k},
\end{equation}
the estimation procedure of type (iv) can also be represented in the form of Eq.~(\ref{eq:estimation_form}).
However, there exists a key distinction between the procedures of types (i) - (iii) and type (iv) in a way of simulating the map $\mathcal{A}_k^{(\xi)}$. While the former simulates
the map $\mathcal{A}_k^{(\xi)}$ ($\xi=1,2,3$) which includes the same unitary channel $\mathcal{U}_k^{(\xi)}$ for the label $l$, 
the latter realize $\mathcal{A}_k^{(4)}$ by using different structures of quantum circuits depending on $l$. In other words, while the former simulates the outcome of $\mathrm{Tr}[ O_k \mathcal{A}_{k}^{(\xi)} (\omega^{l})]$ in a single type of quantum circuit as shown in Fig.~\ref{Fig:exp_map_relation}(b1),(c1), and (d1), the latter simulates the outcome of $\mathrm{Tr}[ O_k \mathcal{A}_{k}^{(\xi)} (\omega^{l})]$ in multiple types of quantum circuit as shown in Fig.~\ref{Fig:exp_map_relation}(e1-1) and (e1-2). As will be explained in the next section, this difference is related to how the map $\mathcal{A}_k^{(\xi)}$ incurs the effect of noise from the quantum circuits.

\subsubsection{Physicality in types (i) - (iii)}\label{sec:physicality}

We here consider the case where the physical noise occurs on each estimation circuit for types (i) - (iii), i.e., $\xi=1,2,3$. As discussed in the previous section, under noise-free conditions, the estimation procedures for $M_k^{(\xi)}$ and $S_k^{(\xi)}$ in type (i) - (iii) can be viewed as models that, for each label $l$, directly (type (i)) or indirectly (types (ii) and (iii)) simulate the expansion map $\mathcal{A}_k^{(\xi)}$ on a quantum circuit (Fig.~\ref{Fig:exp_map_relation}(b1),(c1),(d1)), compute the expectation value of the observable $O_k$ for $\mathcal{A}_k^{(\xi)}(\omega^{l})$, and summing over the results with the weight $\bar{\omega}^{l}$ as Eq.~(\ref{eq:estimation_form}). Now, consider the situation where physical noise arises in the quantum circuit causing $\mathcal{U}_k^{(\xi)}\rightarrow\mathcal{W}_k^{(\xi)}$. By using the correspondence between the actually used quantum circuits (Fig.~\ref{Fig:exp_map_relation}(b1),(c1),(d1)) and those modified to explicitly include $\mathcal{A}_k^{(\xi)}$ (Fig.~\ref{Fig:exp_map_relation}(b2),(c2),(d2)), this circuit-level noise can be viewed as noise acting on $\mathcal{A}_k^{(\xi)}$.

\begin{figure}[t]
\centering
\begin{center}
 \includegraphics[width=80mm]{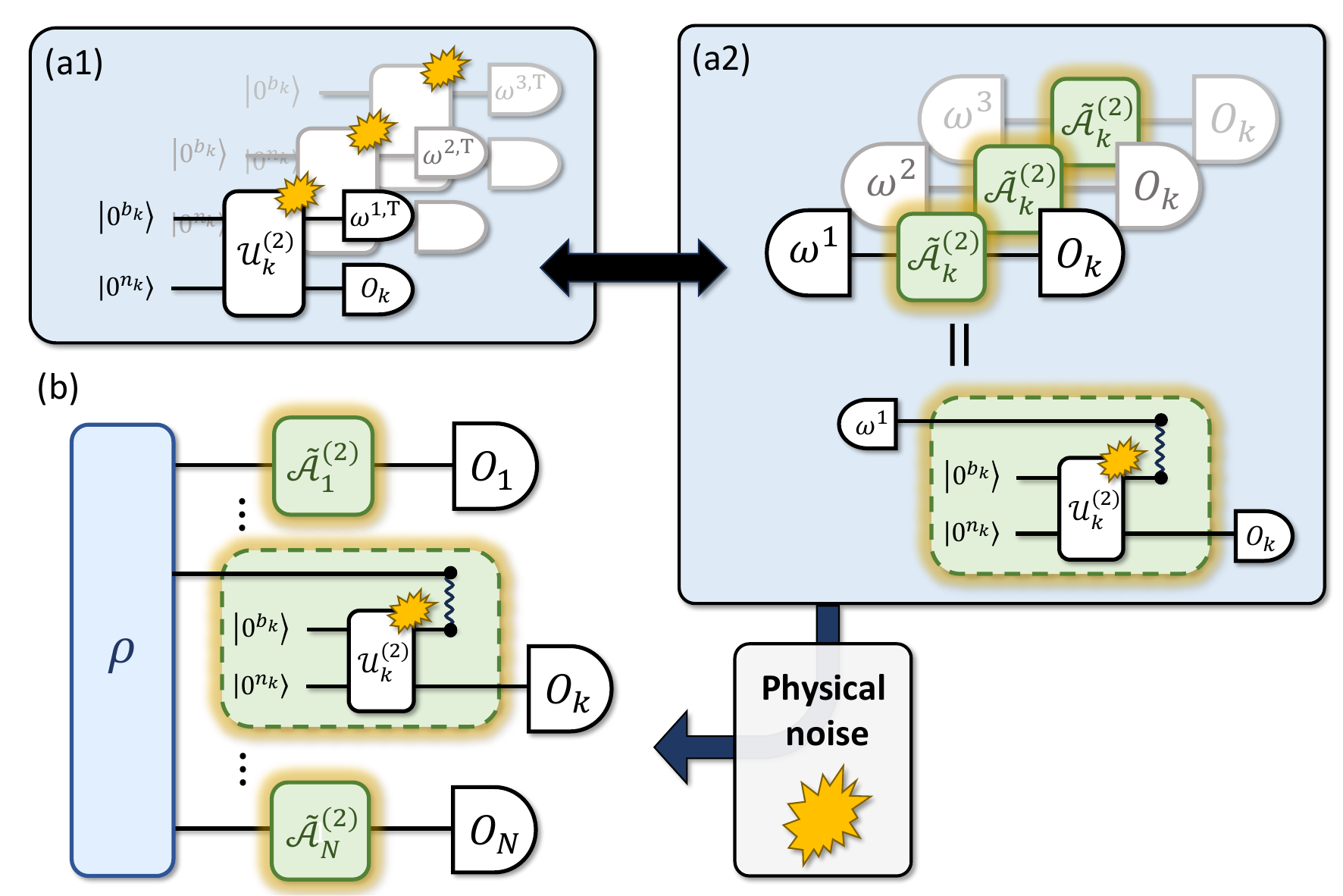}
\end{center}
\caption{Illustrative explanations of how noise in the quantum circuits for estimating $M_k^{(2)}$ and $S_k^{(2)}$ is attributed to noise in the expansion map in type (ii). (a) Each circuit (a1) for estimating $M_k^{(2)}$ and $S_k^{(2)}$ can be rewritten as a circuit (a2) that explicitly includes $\mathcal{A}_k^{(2)}$. Due to the correspondence, the estimation procedure can be interpreted as effectively simulating a modified expansion map where the internal $\mathcal{U}_k^{(2)}$ is replaced by $\mathcal{W}_k^{(2)}$. (b) Since the noise in the modified map is physical, it does not alter the physical properties of the HTTN state.
\label{fig: exp_formalism_1}}
\end{figure}

Specifically, since the noise $\mathcal{U}_k^{(\xi)}\rightarrow\mathcal{W}_k^{(\xi)}$ affects $\mathcal{A}_k^{(\xi)}$ regardless of $l$, each circuit labeled with $l$ can be regarded as simulating the map $\mathcal{A}^{(\xi)}_{k,\,\mathcal{U}\rightarrow \mathcal{W}}$, in which the unitary $\mathcal{U}_k^{(\xi)}$ inside $\mathcal{A}_k^{(\xi)}$ is replaced with $\mathcal{W}_k^{(\xi)}$. Consequently, under noisy conditions, the estimation procedure can be expressed as
\begin{equation}\label{eq:estimation_form_noisy}
\begin{split}
    \tilde{M}_{k}^{(\xi)} &= \sum_{l} \mathrm{Tr}[ O_k \,\mathcal{A}^{(\xi)}_{k,\,\mathcal{U}\rightarrow \mathcal{W}} (\omega^{l})]~ \bar{\omega}^l,\\
    \tilde{S}_{k}^{(\xi)} &= \sum_{l} \mathrm{Tr}[ \,\mathcal{A}^{(\xi)}_{k,\,\mathcal{U}\rightarrow \mathcal{W}} (\omega^{l})]~ \bar{\omega}^l.
\end{split}
\end{equation}
Thus, by Proposition~\ref{prop:fact_1} we can derive an explicit description of the noisy expansion map $\tilde{\mathcal{A}}_k^{(\xi)}$ defined in Eq.~(\ref{eq: transformation}) as $\mathcal{A}^{(\xi)}_{k,\,\mathcal{U}\rightarrow \mathcal{W}}$.
Moreover, because the change from $\mathcal{A}_k^{(\xi)}$ to $\tilde{\mathcal{A}}_k^{(\xi)}$ comes from physical noise $\mathcal{U}_k^{(\xi)}\rightarrow\mathcal{W}_k^{(\xi)}$, the map $\tilde{\mathcal{A}}_k^{(\xi)}$ is a necessary CP map, resulting in $\tilde{\rho}_{\rm HT}^{(\xi)}$ being a valid quantum state. Fig.~\ref{fig: exp_formalism_1} illustrates the above discussion using the type (ii) estimation as an example.

\begin{figure}[tp]
\centering
\begin{center}
 \includegraphics[width=80mm]{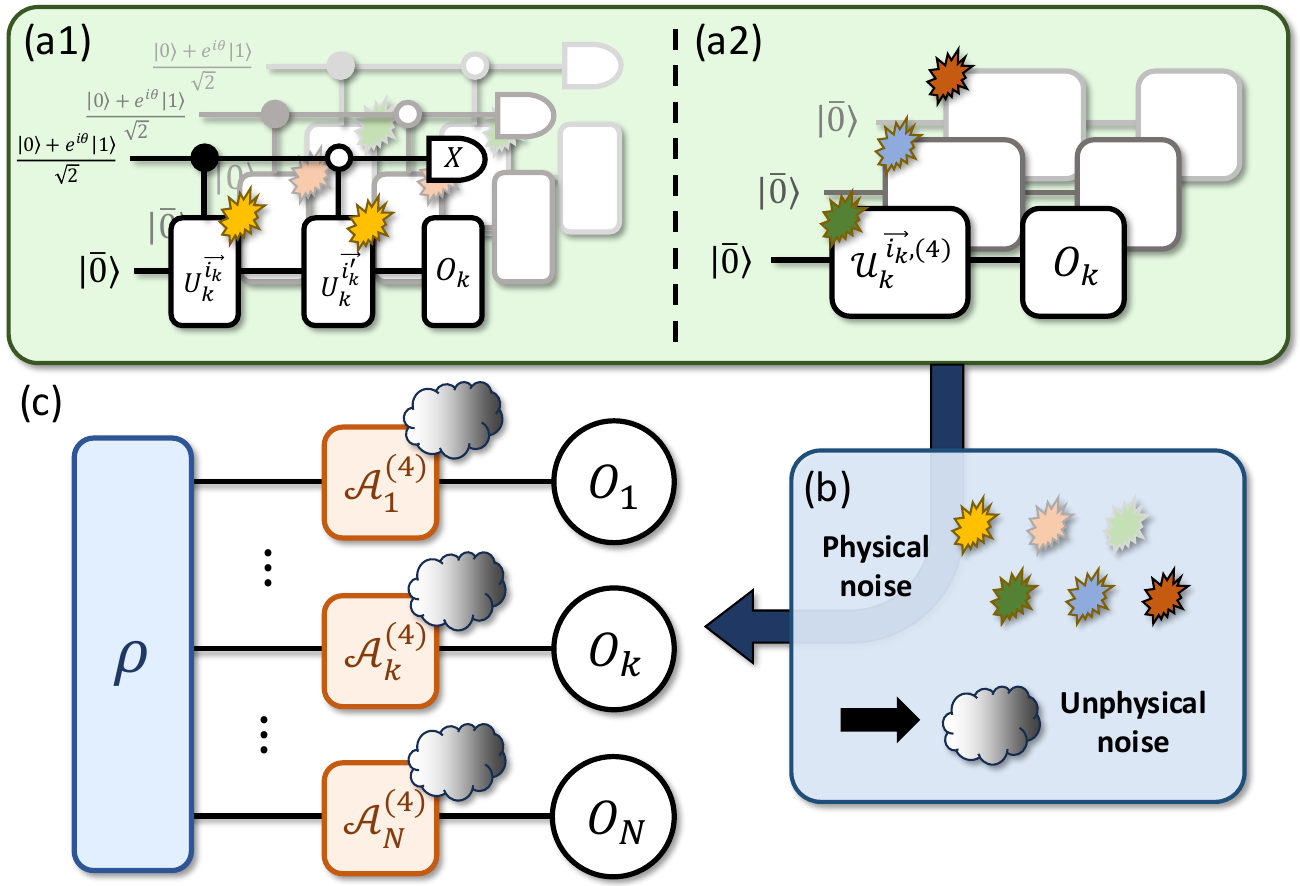}    
\end{center}
\caption{Graphical representations of estimating $M_k^{(4)}$ using the quantum circuits (a) introduced in Sec.~\ref{contraction_of_local_observables}: (a1) the circuits for estimating the diagonal element and (a2) the circuits for estimating the off-diagonal elements of $M_k^{(4)}$. Each term of $M_k^{(4)}$ is estimated using circuits equipped with different unitary channels and affected by different types of physical noise. (b) The physical errors from different error sources, i.e., different types of quantum circuits, are virtually combined into a single unphysical noise. (c) The unphysical noise is mapped onto the ideal HTTN state $\rho_{\rm HT}^{(4)}$.
\label{fig: exp_formalism_4}}
\end{figure}

\subsubsection{Unphysicality in Type (iv)}\label{sec: noise analysis, case4}

Now, we discuss the case where the physical noise affects the type (iv) estimation circuits in Sec.~\ref{sec: 2d}. 
As noted in Sec.~\ref{sec:phys_prop_A}, the type-(iv) estimation procedure can be understood as a model that indirectly simulates the expansion map $\mathcal{A}_k^{(4)}$. 
However, unlike the estimation models for type (i) - (iii), type (iv) uses quantum circuits with different structures for each label $l=(\vec{i}_k,\vec{i}'_k)$ to simulate the expansion map. 
In other words, it simulates $\mathcal{A}_k^{(4)}$ using quantum circuits that vary with $l$, thereby inducing noise in $\mathcal{A}_k^{(4)}$ in a way that also depends on $l$. Concretely, let $\mathcal{A}_{k,\mathcal{U}^{l}\rightarrow\mathcal{W}^{l}}^{(4)}$ denote the map obtained by attributing circuit-level noise $\mathcal{U}_k^{(4),\vec{i}_k}\rightarrow\mathcal{W}_k^{(4),\vec{i}_k}$ and $\mathcal{C}^{\vec{i}_k,\vec{i}'_k}\rightarrow\tilde{\mathcal{C}}^{\vec{i}_k,\vec{i}'_k}$ to $\mathcal{A}_k^{(4)}$. Under a noisy environment, the type-(iv) estimation can then be written as
\begin{equation}\label{eq:estimation_form_noisy_4}
\begin{split}
    \tilde{M}_{k}^{(4)} &= \sum_{l} \mathrm{Tr}[ O_k \,\mathcal{A}_{k,\mathcal{U}^{l}\rightarrow\mathcal{W}^{l}}^{(4)} (\omega^{l})]~ \bar{\omega}^l,\\
    \tilde{S}_{k}^{(4)} &= \sum_{l} \mathrm{Tr}[ \,\mathcal{A}_{k,\mathcal{U}^{l}\rightarrow\mathcal{W}^{l}}^{(4)} (\omega^{l})]~ \bar{\omega}^l.
\end{split}
\end{equation}
Hence, in type (iv), one effectively simulates the family of maps$\{
\mathcal{A}_{k,\mathcal{U}^{l}\rightarrow\mathcal{W}^{l}}^{(4)}\}_l$ depending on inputs $\omega^{l}$. 

Note that because each label $l$ involves a different map $\mathcal{A}_{k,\mathcal{U}^{l}\rightarrow\mathcal{W}^{l}}^{(4)}$, we can not apply Proposition~\ref{prop:fact_1} to recover a single $\tilde{\mathcal{A}}_k^{(4)}$.
In Appendix~\ref{sec:unphysical_in_type_4}, we show that, under the assumption of the simple noise model and observable, the noisy HTTNs state $\tilde{\rho}_{\rm HT}^{(4)}$ resulting from incorporating the noise effects $\{ \mathcal{A}_k^{(4)} \rightarrow
 \mathcal{A}_{k,\mathcal{U}^{l}\rightarrow\mathcal{W}^{l}}^{(4)}\}_l$ can become unphysical, i.e., $\tilde{\rho}_{\rm HT}^{(4)}$ could have negative eigenvalues. We provide an intuitive illustration of the discussion in Fig.~\ref{fig: exp_formalism_4}.

\subsubsection{Classical case}\label{sec: noise analysis, classical}
 
 For the general discussion in the next section, let us consider the case where $\xi=\mathrm{c}$. Given that errors in classical computations are sufficiently small and negligible, the contracted operators $\tilde{M}_{k}^{(\mathrm{c})}$ and $\tilde{S}_{k}^{(\mathrm{c})}$ can be represented as $\tilde{M}^{(\mathrm{c})}_{k}=M^{(\mathrm{c})}_{k}={A_{k}^{(\mathrm{c})}}^{\dagger} O_{k} A^{(\mathrm{c})}_{k}$ and $\tilde{S}^{(\mathrm{c})}_{k}=S^{(\mathrm{c})}_{k}={A_{k}^{(\mathrm{c})}}^{\dagger} A^{(\mathrm{c})}_{k}$. Thus, the following equations clearly hold:
\begin{equation}\label{eq: Classical, transformation}
\begin{split}
    \mathrm{Tr}[\, \tilde{M}_{k}^{(\mathrm{c})} \bullet \,] &= \mathrm{Tr} [\,O_{k}\, \tilde{\mathcal{A}}^{(c)}_{k} (\bullet)\,], \\
    \mathrm{Tr}[\, \tilde{S}_{k}^{(\mathrm{c})} \bullet \,] &= \mathrm{Tr} [\, \tilde{\mathcal{A}}^{(\mathrm{c})}_{k} (\bullet)\,], 
\end{split}
\end{equation}
 where we defined $\tilde{\mathcal{A}}^{(\mathrm{c})}_{k} (\bullet)=A^{(\mathrm{c})}_{k} \bullet{A_{k}^{(\mathrm{c})}}^{\dagger}$. Hence, we can easily obtain
\begin{equation}
    \tilde{\rho}_{\rm HT}^{(\mathrm{c})}
    = \frac{(\bigotimes_{k=1}^{N} A_{k}^{(\mathrm{c})}) \,\rho\, (\bigotimes_{k=1}^{N} A_{k}^{(\mathrm{c})})^\dagger}{\mathrm{Tr}[(\bigotimes_{k=1}^{N} A_{k}^{(\mathrm{c})}) \,\rho\, (\bigotimes_{k=1}^{N} A_{k}^{(\mathrm{c})})^\dagger]}.
\end{equation}
 Since $\rho$ is a density operator, $\tilde{\rho}_{\rm HT}^{(\mathrm{c})}$ is obviously a density operator.

\subsection{Quantum-classical mixed-type HTTN state preparations with multiple layers}\label{sec:generalization}

\begin{figure}[t]
\begin{center}
\centering
 \includegraphics[width=80mm]{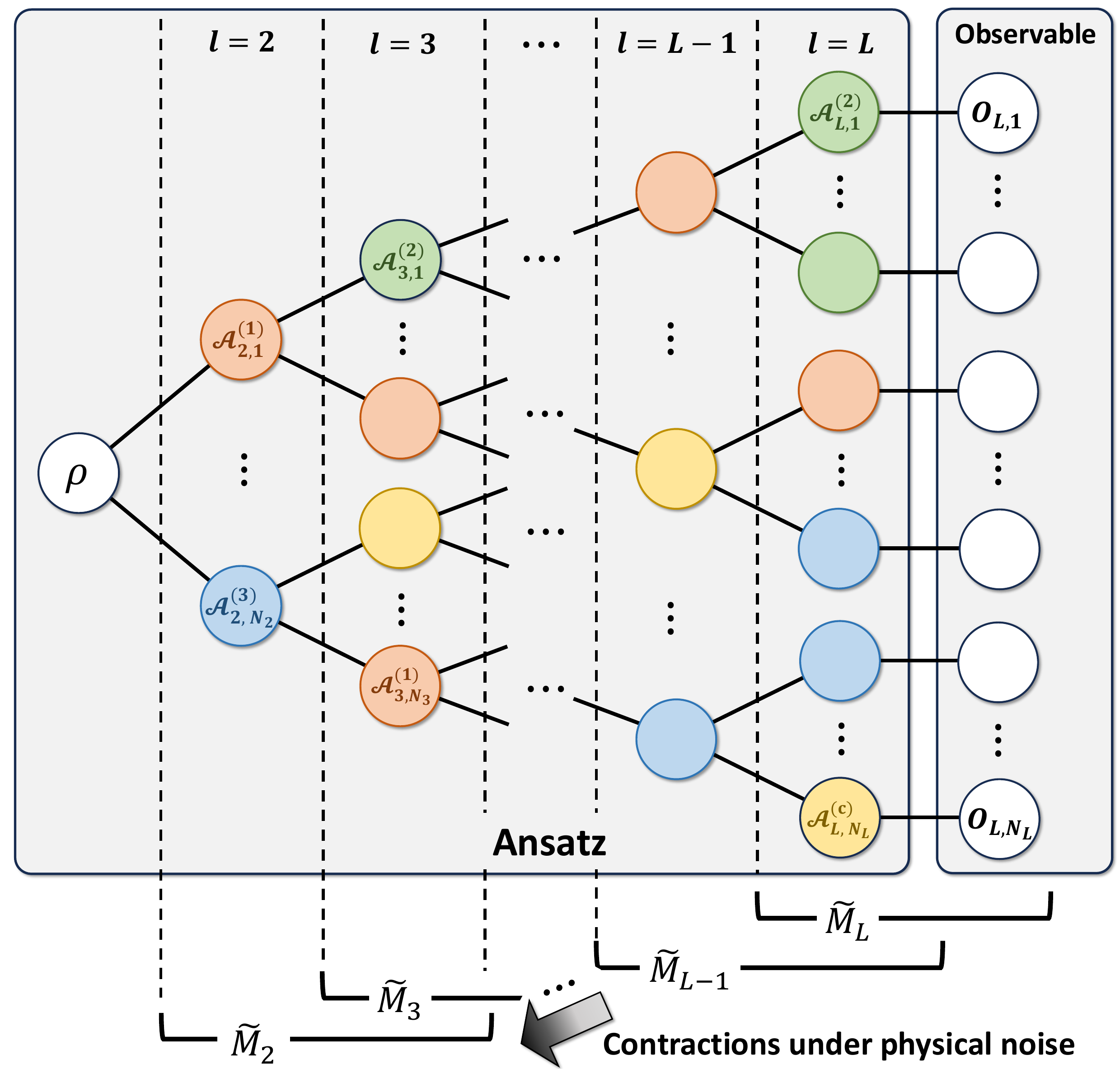}
\end{center}
\caption{\label{Fig: multi_layer_mixed_ansatz}Graphical diagrams of the quantum-classical mixed-type HTTN state $\rho_{\rm HT}$ with $L$ layers and the observable $O$. The ansatz is composed of the parent classical or quantum tensor $\rho$, and the $l$-th layer's $\mu$-th operators $\mathcal{A}_{l,\mu}^{(\xi_{l,\mu})}$ where $\xi_{l,\mu}$ takes four different values: $\xi_{l,\mu}=1$(orange vertices), $2$(green vertices), $3$(blue vertices), and $\mathrm{c}$ (yellow vertices). In actual calculations, the contractions are performed sequentially from $\tilde{M}_{L}$ down to $\tilde{M}_{2}$, and at each contraction step, physical noise propagates to the upper layer.}
\end{figure}

 In the preceding discussions, we have investigated a noisy effective HTTN state $\tilde{\rho}_{\rm HT}$ with two layers. In those analyses, we have considered all subsystems to share the same type of state preparation. Here, extending the previous discussion, let us consider the effect of physical noise on a quantum-classical mixed-type HTTN state with $L$ layers. To begin with, we consider the density matrix representation of a noiseless tree tensor state:
\begin{equation}\label{eq:general ansatz}
    \rho_{\rm HT} = \frac{\mathcal{A}_{L} \circ \mathcal{A}_{L-1} \circ \cdots \mathcal{A}_{2} (\rho)}{\mathrm{Tr}[\mathcal{A}_{L} \circ \mathcal{A}_{L-1} \circ \cdots \mathcal{A}_{2} (\rho)]},
\end{equation}
 where $\rho$ is a parent classical or quantum tensor. In Eq.~(\ref{eq:general ansatz}), we define the $l$-th layer's operator $\mathcal{A}_{l}(\bullet)$ for $l=2,...,L$ as 
\begin{equation}\label{eq:lth operator}
    \mathcal{A}_{l}(\bullet) = \mathcal{A}_{l,1}^{(\xi_{l,1})} \otimes \mathcal{A}_{l,2}^{(\xi_{l,2})} \otimes \cdots \otimes \mathcal{A}_{l,N_{l}}^{(\xi_{l,N_{l}})}(\bullet),
\end{equation}
 where $\mathcal{A}_{l,\mu}^{(\xi_{l,\mu})}(\bullet)$ is the $l$-th layer's $\mu$-th operator defined as $\mathcal{A}_{l,\mu}^{(\xi_{l,\mu})}(\bullet) = A_{l,\mu}^{(\xi_{l,\mu})} \bullet {A_{l,\mu}^{(\xi_{l,\mu})}}^{\dagger}$ with $A_{l,\mu}^{(\xi_{l,\mu})} = \sum_{\vec{i}_{l,\mu}} \ket{\psi_{l,\mu}^{\vec{i}_{l,\mu}}} \bra{\vec{i}_{l,\mu}}$ for $\mu=1,...,N_{l}$. The graphical description of the state $\rho_{\rm HT}$ is shown in Fig.~\ref{Fig: multi_layer_mixed_ansatz}. Hereafter, we only consider types (i)-(iii) and classical tensor preparations because type (iv) may induce unphysicality.

 The assumptions of physical noise in the computation for the main system and each subsystem follow the  statement in Sec.~\ref{sec:noise_assumption}: the physical noise in quantum computation converts an ideal unitary channel to a noisy one. Additionally, for simplicity, we consider the case where the observable $O$ is written as $O=O_{L,1} \otimes \cdots \otimes O_{L,N_{L}}$, and note that the same line of discussion holds for observables without tensor product structure.

 In the calculations of this setting, we sequentially construct the contracted operators $\tilde{M}_{l}$ and $\tilde{S}_{l}$ from the $L$-th layer to the 2nd layer, i.e., $\tilde{M}_{L} \rightarrow \tilde{M}_{L-1} \rightarrow \cdots \rightarrow \tilde{M}_{2}$ and $\tilde{S}_{L} \rightarrow \tilde{S}_{L-1} \rightarrow \cdots \rightarrow \tilde{S}_{2}$. Using the lastly obtained operators $\tilde{M}_{2}$, $\tilde{S}_{2}$, and $\rho$, we obtain the noisy expectation value $\braket{\tilde{O}}_{\rho_{\rm HT}}$ as
\begin{equation}\label{eq:base}
    \braket{\tilde{O}}_{\rho_{\rm HT}} = \frac{\mathrm{Tr}[\, \tilde{M}_{2} \, \rho \,]}{\mathrm{Tr}[\, \tilde{S}_{2} \, \rho \,]}.
\end{equation}
Our objective is to explicitly express the noisy effective state $\tilde{\rho}_{\rm HT}$ defined as $\braket{\tilde{O}}_{\rho_{\rm HT}} = \braket{O}_{\tilde{\rho}_{\rm HT}}$. To this end, from Eq.~(\ref{eq: transformation}) for $\xi=1,2,3$ and Eq.(\ref{eq: Classical, transformation}),  we define the noisy expanion maps $\tilde{\mathcal{A}}_{l}(\bullet)$($l=2,...,L$) via the following equations:
\begin{equation}\label{eq:transform1}
\begin{split}
    \mathrm{Tr}[\, \tilde{M}_{l} \, \bullet \,] &= \mathrm{Tr} [\, \tilde{M}_{l+1}\, \tilde{\mathcal{A}}_{l} (\bullet)\,], \\
    \mathrm{Tr}[\, \tilde{S}_{l} \, \bullet \,] &= \mathrm{Tr} [\, \tilde{S}_{l+1}\, \tilde{\mathcal{A}}_{l} (\bullet)\,],
\end{split}
\end{equation}
for $l= 2,...,L-1$, and 
\begin{equation}\label{eq:transform2}
\begin{split}
    \mathrm{Tr}[\, \tilde{M}_{L} \bullet \,] &= \mathrm{Tr} [\, O\, \tilde{\mathcal{A}}_{L} (\bullet)\,],\\
    \mathrm{Tr}[\, \tilde{S}_{L} \bullet \,] &= \mathrm{Tr} [\, \tilde{\mathcal{A}}_{L} (\bullet)\,].
\end{split}
\end{equation}
 By iteratively employing Eqs.~(\ref{eq:transform1}) and (\ref{eq:transform2}) on Eq.~(\ref{eq:base}), we obtain
\begin{equation}
\begin{split}
    \braket{\tilde{O}}_{\rho_{\rm HT}} 
    &= \frac{\mathrm{Tr}[\, \tilde{M}_{2} \, \rho \,] }{ {\mathrm{Tr}[\, \tilde{S}_{2} \, \rho \,] }} \\
    &= \frac{\mathrm{Tr}[\, \tilde{M}_{3} \,\tilde{\mathcal{A}}_{2} (\rho) \,]}{\mathrm{Tr}[\, \tilde{S}_{3}\, \tilde{\mathcal{A}}_{2} (\rho) \,]} \\
    &= \cdots \\
    &= \frac{\mathrm{Tr}\Bigl[\, O\, \Bigl( \tilde{\mathcal{A}}_{L} \circ \tilde{\mathcal{A}}_{L-1} \circ \cdots \circ \tilde{\mathcal{A}}_{2} (\rho) \Bigr) \,\Bigr]}{ \mathrm{Tr}\Bigl[\, \tilde{\mathcal{A}}_{L} \circ \tilde{\mathcal{A}}_{L-1} \circ \cdots \circ \tilde{\mathcal{A}}_{2} (\rho) \,\Bigr] }.
\end{split}
\end{equation}
Thus, the effective state can be expressed as 
\begin{equation}\label{eq:general noisy state}
    \tilde{\rho}_{\rm HT} = \frac{ \tilde{\mathcal{A}}_{L} \circ \tilde{\mathcal{A}}_{L-1} \circ \cdots \circ \tilde{\mathcal{A}}_{2} (\rho) } { \mathrm{Tr}\Bigl[ \tilde{\mathcal{A}}_{L} \circ \tilde{\mathcal{A}}_{L-1} \circ \cdots \circ \tilde{\mathcal{A}}_{2} (\rho) \Bigr] }.
\end{equation}
 Here, since the operator $\rho$ is quantum state by definition and the mappings $\tilde{\mathcal{A}}_{l}(\bullet)$($l=2,...,L$) defined in Eqs.~(\ref{eq:transform1}) and (\ref{eq:transform2}) are CP maps (see Appendix~\ref{sec:physicality_app}), the numerator of Eq.~(\ref{eq:general noisy state}) is thereby a positive semidefinite operator. Taking into account the normalization term in Eq.~(\ref{eq:general noisy state}), $\tilde{\rho}_{\rm HT}$ is a density operator. 

\subsection{Application to previously proposed methods}\label{sec:application}
 
 Here, we apply the expansion operator method to Deep VQE and the entanglement forging method. Then, we see the explicit density matrix representation and discuss the physicality of the HTN state under the effect of hardware noise.

\subsubsection{Deep VQE}\

 In Sec.~\ref{sec:deep_vqe}, we have captured the Deep VQE algorithm in the framework of HTTNs. Here, using the density matrix representation of HTTNs, we express Deep VQE ansatz $\rho_{\rm DV}$ corresponding to Eq.~(\ref{eq: HTNs representation of Deep VQE,1}) as 
\begin{equation}\label{eq: deep_vqe_density}
    \rho_{\rm DV} = \mathcal{A}_{3} \circ \mathcal{A}_{2} ( \ket{\phi}\bra{\phi} ),
\end{equation}
 where $\ket{\phi}\bra{\phi}$ is a parent quantum tensor, and $\mathcal{A}_{2}$ and $\mathcal{A}_{3}$ are the 2nd and 3rd layer's operators defined as
\begin{align} 
    \mathcal{A}_{2}(\bullet) &= \mathcal{A}_{2,1}^{(\xi_{2,1})} \otimes ... \otimes \mathcal{A}_{2,N}^{(\xi_{2,N})}(\bullet),\\
    \mathcal{A}_{3}(\bullet) &= \mathcal{A}_{3,1}^{(\xi_{3,N})} \otimes ... \otimes \mathcal{A}_{3,N}^{(\xi_{3,N})}(\bullet).\label{eq: def_A3_1}
\end{align}
$\mathcal{A}_{2,k}^{(\xi_{2,k})}(\bullet)=A_{2,k}^{(\xi_{2,k})} \bullet{A_{2,k}^{(\xi_{2,k})}}^{\dagger}$ are the 2nd layer's $k$-th operators where $A_{2,k}^{(\xi_{2,k})}$ is defined as
\begin{equation}
    A_{2,k}^{(\xi_{2,k})}
    = \sum_{\vec{i}_{k} \in\{0,1\}^{b_{k}}} \ket{\phi_{2,k}^{\vec{i}_{k}}}\bra{\vec{i}_{k}},
\end{equation}
and $\ket{\phi_{2,k}^{\vec{i}_{k}}} = \sum_{\vec{j}_{k} \in\{0,1\}^{b_{k}}} \beta^{\vec{i}_{k},\vec{j}_{k}}_{k} \ket{\vec{j}_{k}}$. 
Since each element of $\ket{\phi_{2,k}^{\vec{i}_{k}}}$ is stored in a classical register, the operator $A_{2,k}^{(\xi_{2,k})}$ is a classical tensor, i.e., $\xi_{2,k}=\mathrm{c}$ for all $k$.
$\mathcal{A}_{3,k}^{(\xi_{3,k})}(\bullet)=A_{3,s}^{(\xi_{3,k})} \bullet{A_{3,k}^{(\xi_{3,k})}}^{\dagger}$ is the 3rd layer's $k$-th operator where
\begin{equation}\label{eq: def_A3_2}
    A_{3,k}^{(\xi_{3,k})} = \sum_{\vec{j}_{k}\in\{0,1\}^{b_{k}}} \ket{\phi_{3,k}^{\vec{j}_{k}}} \bra{\vec{j}_{k}},
\end{equation}
and 
\begin{equation}\label{eq: def_A3_3}
    \ket{\phi_{3,k}^{\vec{j}_{k}}} = \begin{cases} D^{\vec{j}_{k}}_{k} \ket{\phi_{k}^{\mathrm{G}}}, & (\vec{j}_{k})_{(10)} \leq \bar{d}_{k} - 1, \\
    \vec{0}, & (\vec{j}_{k})_{(10)} \geq \bar{d}_{k}.
    \end{cases}
\end{equation}
 When setting the operator $D^{\vec{j}_k}_{k}$ as the Pauli operators, it is clear that $\xi_{3,k}=3$ for all $k$. Alternatively, when we adopt the general local excitation operator $D^{\vec{j}_{k}}_{k}$ as $W^{\vec{j}_{k}}_{k}$, the computational process to calculate the local tensor $H^{\rm eff}$(see Appendix~\ref{apdx:deep_vae} and Eq.~(\ref{eq: calculation of Heff})) remains the same as type (iii) in Sec.~\ref{sec: 2c}. Thus, without loss of generality, we can consider $\xi_{3,k}=3$ for all $k$.

 Now, we consider the effect of physical noise. From the discussions in Sec.~\ref{sec:phys_prop_A}, $\tilde{\mathcal{A}}_{2}$ and $\tilde{\mathcal{A}}_{3}$ are CP maps, so the effective noisy state $\tilde{\rho}_{\rm DV}= \tilde{\mathcal{A}}_{3} \circ \tilde{\mathcal{A}}_{2} ( \rho )$ is a positive semidefinite operator. However, the operator $\mathcal{A}_{2}$, playing a role in normalizing $\rho_{\rm DV}$, is composed of $\beta^{\vec{i}_{k},\vec{j}_{k}}_{k}$, which is also susceptible to physical noise, $\mathrm{Tr}[\tilde{\rho}_{\rm DV}]=1$ may not be maintained. 

 To address this problem, one possible solution is to modify the ansatz in Eq.~(\ref{eq: deep_vqe_density}) as follows:
\begin{equation}\label{eq: deep_vqe2_ansatz}
    \rho_{\rm DV2} = \frac{\mathcal{A}_{3}(\ket{\phi}\bra{\phi})}{\mathrm{Tr}[\mathcal{A}_{3}(\ket{\phi}\bra{\phi})]},
\end{equation}
 where the definition of $\mathcal{A}_{3}$ follows Eqs.~(\ref{eq: def_A3_1}),(\ref{eq: def_A3_2}), and (\ref{eq: def_A3_3}). The denominator in Eq.~(\ref{eq: deep_vqe2_ansatz}) is derived from the list of inner products $\{\braket{\phi_{k}^{\vec{j}_{k}}|\phi_{k}^{\vec{j}'_{k}}}\}$, which is used to construct $\mathcal{A}_{2}$ in Eq.~(\ref{eq: deep_vqe_density}). As $\rho_{\rm DV2}$ is composed solely of type (iii) operators, it follows from the discussion in Sec.~\ref{sec:generalization} that the noisy state $\tilde{\rho}_{\rm DV2}$ can maintain the properties of the density operator.

\subsubsection{Entanglement forging}

 Following the notation in Sec.~\ref{sec:entanglement_forging}, the operator representation $\rho_{\rm EF}$ of entanglement forging ansatz is represented as
\begin{equation}
    \rho_{\rm EF} = \mathcal{A}_{2}(\ket{\phi}\bra{\phi}),
\end{equation}
 where $\ket{\phi}$ is a vector whose elements are derived from the Schmidt coefficients and stored in a classical register. The operator $\mathcal{A}_{2}$ is defined as follows:
\begin{equation}
    \mathcal{A}_{2}(\bullet) = \mathcal{A}_{2,1}^{(\xi_{2,1})} \otimes \mathcal{A}_{2,2}^{(\xi_{2,2})} (\bullet).
\end{equation}
$\mathcal{A}_{2,k}^{(\xi_{2,k})}(\bullet)=A_{2,k}^{(\xi_{2,k})} \bullet{A_{2,k}^{(\xi_{2,k})}}^{\dagger}$ is the 2nd layer's $k$-th operator where $A_{2,k}^{(\xi_{2,k})}$ is defined as
\begin{equation}
    A_{2,k}^{(\xi_{2,k})} = \sum_{\vec{i}_{k}\in\{0,1\}^{N}} \ket{\phi^{\vec{i}_{k}}_{2,k}} \bra{\vec{i}_{k}},
\end{equation}
and $\ket{\phi^{\vec{i}_{k}}_{2,k}}=V_{k}\ket{\vec{i}_{k}}$. From the structure of the state 
$\ket{\phi^{\vec{i}_{k}}_{2,k}}$, we see that $\xi_{k}=1$ for $k=1,2$.

 Here, we make the same assumptions as before, where the unitary channels of the first and second subsystems, $\mathcal{V}_{1}(\bullet)=V_{1} \bullet V_{1}^{\dagger}$ and $\mathcal{V}_{2}(\bullet)=V_{2}\bullet V_{2}^{\dagger}$, are transformed into noisy channels $\mathcal{W}_{1}$ and $\mathcal{W}_{2}$, respectively. Then, as $\rho_{\rm EF}$ is constructed from type (i) operators, the analysis in Sec.~\ref{noise analysis} shows that noisy state $\tilde{\rho}_{\rm EF}=\tilde{\mathcal{A}}_{2}(\ket{\phi}\bra{\phi})$ preserves its property as a density operator.

\subsection{Analysis of noise propagation for specific noise models}\label{Sec: exponentialdecay}


We further delve into the expansion operator by specifying the noise model to the global depolarizing model for each quantum tensor. Note that error mitigation techniques based on the assumption of global depolarizing noise successfully reduce computation errors~\cite{tsubouchi2022universal,qin2023error,urbanek2021mitigating,vovrosh2021simple}, which indicates that global depolarizing noise is a good model for noise analysis.

 Specifically, using the noisy expansion maps $\tilde{\mathcal{A}}_k^{(\xi)}$ shown in Table~\ref{tab:noisy_expansion_maps}, which enables the representation of the HTTN state that $\tilde{\rho}_{\rm HT}$ as effectively simulated under physical noise, we show that the expectation value of observables exponentially vanishes with the number of quantum tensors.
 Here, we assume that the global depolarizing channel acts just after the unitary channel in Fig.~\ref{fig: Quantum circuits for M_k}. In this case, the expansion operators $\tilde{\mathcal{A}}^{(\xi_{k})}_k(\xi_{k}=1,2,3)$ for a noisy quantum tensor can be represented as:

\begin{equation}
\tilde{\mathcal{A}}^{(\xi_{k})}_{k}(\bullet)=(1-\varepsilon_{k}) \mathcal{A}^{(\xi_{k})}_k(\bullet) + \varepsilon_{k} \mathcal{K}^{(\xi_{k})}_k(\bullet),
\end{equation}
where $\varepsilon_{k}$ are noise rates, $\mathcal{A}^{(\xi_{k})}_k$ is the noiseless expansion operator, and
\begin{equation}
    \mathcal{K}^{(\xi_{k})}_k(\bullet) = \begin{cases} \frac{I^{\otimes n_{k}}}{2^{n_{k}}}, & \xi_{k}=1, \\
    \mathrm{Tr}(\bullet)\frac{I^{\otimes n_{k}}}{2^{n_{k}}}, & \xi_{k}=2,\\
    \mathcal{P}_{k,12}^{\dagger} \bigl\{ \bullet_1 \otimes \frac{I_2^{\otimes n_{k}}}{2^{n_{k}}}\bigr\} \mathcal{P}_{k,12}, & \xi_{k}=3,
    \end{cases}
\end{equation}
 where we have defined $\mathcal{P}_{k,12} = \sum_{\vec{i}_{k}}\ket{\vec{i}_{k}}_{1} \otimes P_{k,2}^{\vec{i}_k}$
 By assuming that the HTN state consists only of noisy quantum tensors and that $\varepsilon_{k}=\varepsilon$ for simplicity, we obtain
\begin{equation}
\begin{aligned}
\bigcirc_{k=2}^{N_{\mathrm{tot}}} \tilde{\mathcal{A}}^{(\xi_{k})}_k (\rho) &= (1-\varepsilon)^{N_{\mathrm{tot}}-1} \bigcirc_{k=2}^{N_{\mathrm{tot}}} \mathcal{A}^{(\xi_{k})}_k (\rho) \\
&+\sum_{m=1}^{N_{\mathrm{tot}}-1} p(N_\mathrm{tot}-1,m, \varepsilon) \mathcal{B}_m,
\end{aligned}
\end{equation}
 where $p(N_\mathrm{tot}-1,m, \varepsilon)=\binom{N_{\mathrm{tot}}-1}{m} (1-\varepsilon)^{N_{\mathrm{tot}}-m-1} \varepsilon^{m}$ is the  binomial distribution, and $\mathcal{B}_m$ is the process in which the error $\mathcal{K}_k^{(\xi_{k})}$ occurs $m$ times. Here, $\bigcirc_{k=1}^N \mathcal{E}_k=\mathcal{E}_N \circ \mathcal{E}_{N-1} \circ ...\circ \mathcal{E}_{1}$ for a quantum process $\mathcal{E}_{k}~(k=1,2,...,N)$. The process $\mathcal{B}_m$ renders the HTN states highly mixed, which makes expectation values of non-local Pauli observables negligible. 
 Then we can regard $\bigcirc_{k=2}^{N_{\mathrm{tot}}} \tilde{\mathcal{A}}^{(\xi_{k})}_k (\rho) \sim (1-\varepsilon)^{N_{\mathrm{tot}}-1} \bigcirc_{k=2}^{N_{\mathrm{tot}}} \mathcal{A}^{(\xi_{k})}_k (\rho)$.

 Considering the $L$-layer TTNs consisting of noisy $N$-rank quantum tensors, since the total number of the contracted quantum tensor is $N_{\rm tot}=1+N+N^2+...+N^{L-1}=\frac{1-N^L}{1-N}$, we have
\begin{equation}
\braket{O}_{\rm noisy}\sim (1-\varepsilon)^{\frac{1-N^L}{1-N}} \braket{O}_{\rm ideal},
\label{Eq: multilayer}
\end{equation}
where we write $\braket{O}_{\rm noisy}$ and $\braket{O}_{\rm ideal}$ to denote the noisy and error-free expectation values of the observable $O$ on the system $\bigcirc_{k=2}^{N_{\mathrm{tot}}} \mathcal{A}^{(\xi_{k})}_k (\rho)$.

\begin{figure}[b]
\begin{center}
    \includegraphics[width=0.5\columnwidth]{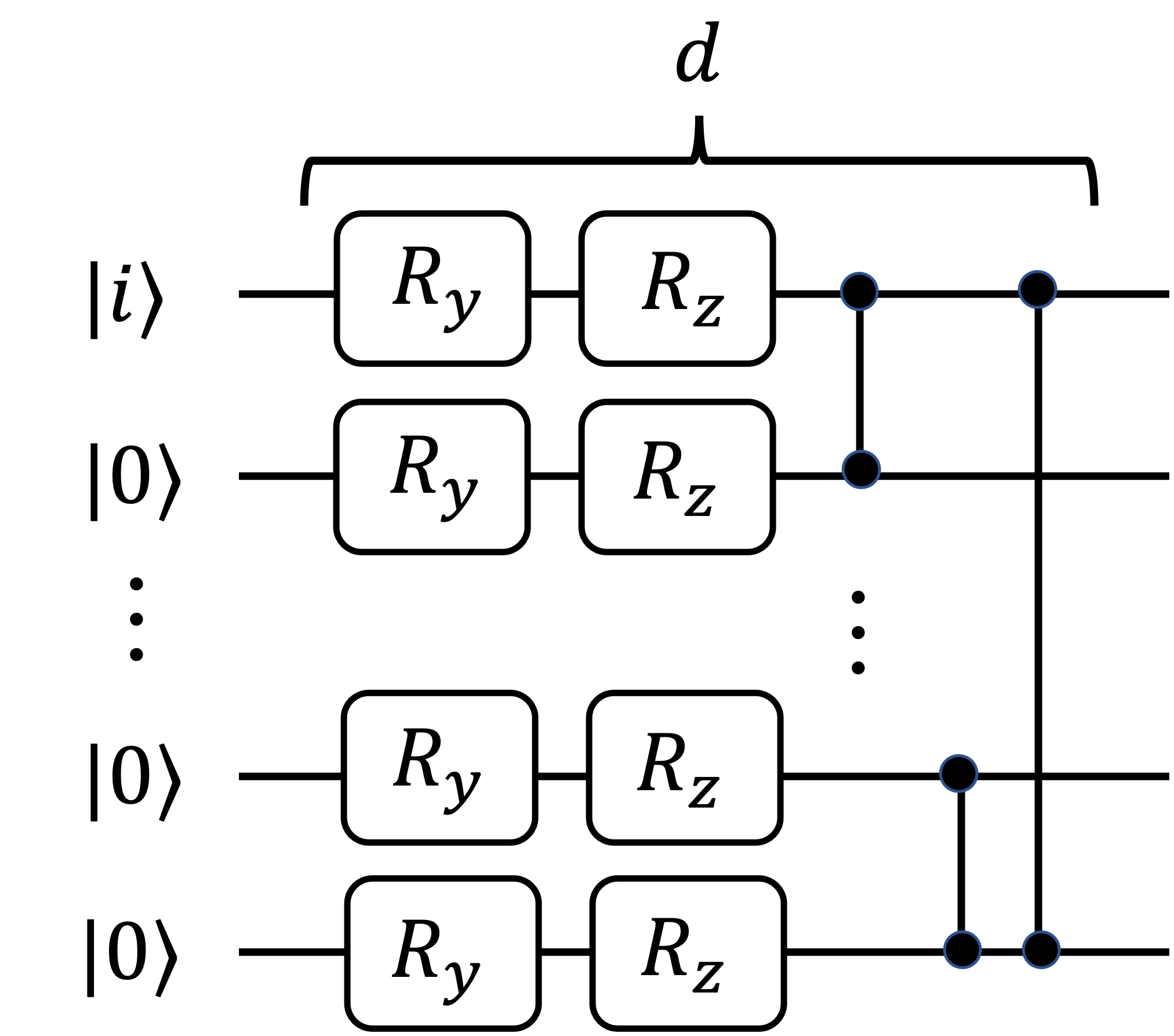}
    \caption{Hardware-efficient ansatz used in numerical simulations. Because we simulate type~(i) contraction, we set the first register to $i=0,1$. To avoid extremely small computational results that induce numerical errors, we restrict variational rotation angles to the range of $[-\pi/10^3, \pi/10^3]$. We set $d=2$. }
    \label{Fig:ansatz}
\end{center}
\end{figure}

 We numerically verify Eq. \eqref{Eq: multilayer} by simulating a $L$-layer TTN comprising of noisy rank-$N$ tensors $\rho_{\vec{i}, \vec{i}'}$ for $\vec{i}=(i_1, i_2, ...,i_N),~i_k \in \{0,1\}$ for $O=Z^{\otimes  N^{L-1}}$ with the type (i) state preparation. Each tensor is generated by a two-layer hardware-efficient ansatz shown in Fig. \ref{Fig:ansatz}, accompanied by the global depolarizing noise. For efficiently simulating noisy TTNs, once we obtain a contracted matrix $\tilde{M}$ in some layer, we use $\tilde{M}^{\otimes N}$ for the contraction in the next layer.  

\begin{figure}[t]
\begin{center}
    \includegraphics[width=\columnwidth]{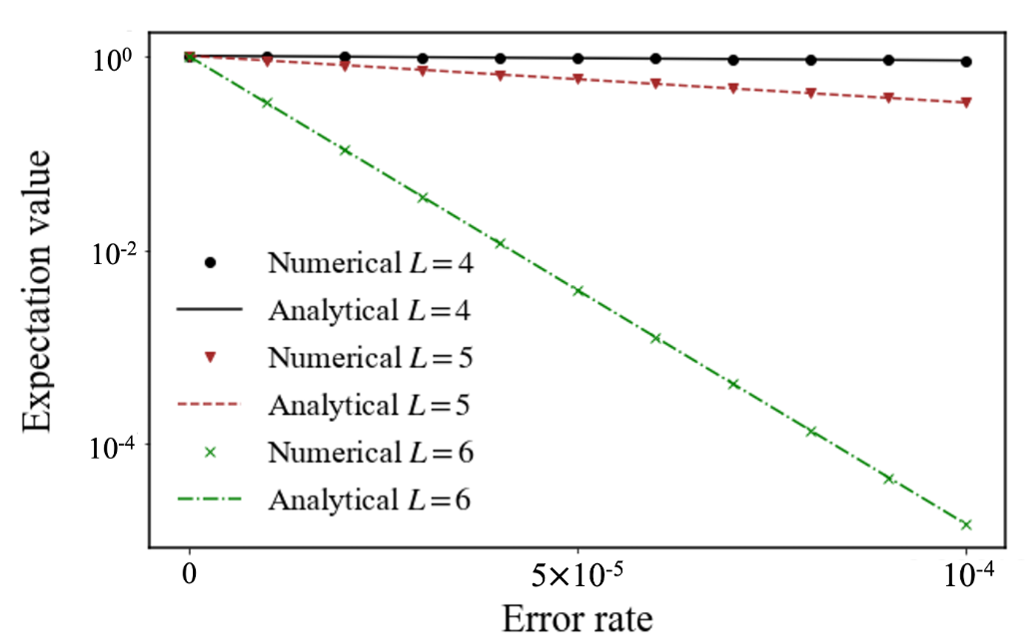}
    \caption{The ratio of the noisy and the noise-free computation results $r(\varepsilon)$ versus the error rate $\varepsilon$. The numerically simulated result $r(\varepsilon)$ and its approximation $\tilde{r}(\varepsilon)$ in Eq.~(\ref{Eq: approximatedratio}) exactly coincide. We simulate $N=10$.  }
    \label{Fig:ratio}
\end{center}
\end{figure}

 We then show the ratio between the noise-free and the noisy value $r_L(\varepsilon)=\braket{O}_{\rm noisy}/\braket{O}_{\rm ideal}$ for $L=4, 5, 6$ with $N=10$ in Fig. \ref{Fig:ratio}. In all the cases, we can see
\begin{equation}
\tilde{r}_L(\varepsilon)= (1-\varepsilon)^{\frac{1-N^L}{1-N}}
\label{Eq: approximatedratio}
\end{equation}
 exactly coincides with the numerical results, which indicates Eq. (\ref{Eq: multilayer}) is a very good approximation. 

 When we can characterize the value of $\varepsilon$, division of $\braket{O}_{\rm noisy}$ by $\tilde{r}_L(\varepsilon)$ yields:
\begin{equation}
\braket{O}_{\rm QEM}= \braket{O}_{\rm noisy}/\tilde{r}_L(\varepsilon),
\label{Eq: QEM}
\end{equation}
 which effectively works as quantum error mitigation~\cite{temme2017error,endo2018practical,endo2021hybrid,cai2022quantum}. However, for the variance of the error mitigated result $\braket{O}_{\rm QEM}$, we obtain
\begin{equation}
\mathrm{Var}[\braket{O}_{\rm QEM}]=\tilde{r}_L(\varepsilon)^2 \mathrm{Var}[\braket{O}_{\rm noisy}], 
\end{equation}
 resulting in the necessity to have $\tilde{r}_L(\varepsilon)^2$ times more samples to achieve the same accuracy as the error-free case. 

Here, we provide remarks on the additional effect of truncation errors in classical tensors on Eq.~(\ref{Eq: multilayer}). Our derivation accounts only for the effect of quantum tensors on the ideal expectation value, assuming that no truncation errors arise from the contraction of classical tensors. However, when considering truncations related to maximal bond dimensions, additional errors may occur. For instance, if one employs the entanglement forging ansatz in Eq.~(\ref{entanglement forging ansatz}) and retains only the leading $m(\leq 2^n)$ components, the truncated errors may happen. In such cases, since classical tensors can be regarded as trace non-increasing maps, the expectation value
the $\braket{O}_{\rm noisy}$ in Eq.~(\ref{Eq: multilayer}) may undergo further decay without appropriate normalization.

 The above discussions indicate that, although the simulatable scale of the quantum systems can be significantly expanded, we need exponential resources with the number of contracted quantum tensors. To avoid this problem, we must take the following strategies: 
\begin{enumerate}
    \item[(1)] Adjusting the number of classical and quantum tensors in accordance with the magnitude of the error;
    \item[(2)] Suppressing the error rate itself with error-suppression techniques such as quantum error correction~\cite{devitt2013quantum,lidar2013quantum} and dynamical decoupling~\cite{viola1999dynamical}. 
\end{enumerate}
 In terms of (1), replacing the quantum tensors with the classical tensors that exactly match their respective matrix representations in the $L$-layer, for example, yields $\braket{O'}_{\rm noisy}\sim (1-\varepsilon)^{\frac{1-N^{L-1}}{1-N}} \braket{O}_{\rm ideal} \sim (1-\varepsilon)^{-N^{L-1}} \braket{O}_{\rm noisy}$, and the problem of exponential decay of the observable is mitigated. 
 However, if the classical tensors used in the replacement are not constrained to correspond to the original quantum tensors and are allowed some freedom, extra normalization factors may be required during the observable‑contraction process, leading to an increased variance in the estimator.  
 This issue can, however, be circumvented by suitably restricting the class of replacement tensors; for example, by limiting them to unitary operators.

\section{Conclusions and Discussions}
\label{conclusion}

 In this paper, we first reviewed the hybrid tensor network (HTN) framework and discussed the related methods, i.e., the Deep VQE and entanglement forging under the language of HTN. Then, whereas these methods are formulated for pure states, the quantum states should be described with the density matrices due to the effect of physical noise in actual experiments. Therefore, we introduced the density matrix representation for HTNs \textcolor{black}{by introducing the expansion operator} and discussed the HTN states' physicality by considering each contraction method case. We found that as long as we use type (i)-(iii) state preparation, the HTN state is physical, while type (iv) may induce unphysical states. We also revealed that the exponential decay of the computation results occurs with the number of contracted quantum tensors in the tree network.

 While we mainly analyzed the expectation values, some algorithms use transition amplitudes in the HTN framework, e.g., quantum-inspired Monte-Carlo simulations~\cite{kanno2023quantum}; therefore, we cannot directly apply our arguments to this case, although the exponential decay of the measured values with the number of quantum tensors is likely to be observed. In addition, we note that the theoretical analysis of exponential decay assumes that classical tensors are unaffected by physical noise. When an HTN state includes a classical tensor constructed via quantum computation (e.g., the classical tensor $\beta_{k}^{\vec{i}_k,\vec{j}_k}$ in Deep VQE), the theoretical analysis based on Eq.~(\ref{Eq: multilayer}) would be affected by this additional noise.
 Also, based on the results of the explicit formulas for HTN states under noise, it may be interesting to apply quantum error mitigation (QEM), e.g., probabilistic error cancellation with finite noise characterization errors~\cite{suzuki2022quantum}, to HTN states and see the resulting effective quantum states and their physicality. Finally, although we derived the exponential decay of the observables for tree tensor networks with noisy quantum states and the exponential growth of the variance of the error-mitigated observable for a specific strategy in Eq. (\ref{Eq: QEM}), the application of information-theoretic analysis of QEM~\cite{takagi2022fundamental,takagi2022universal,tsubouchi2022universal,quek2022exponentially} for HTN states may be useful to verify the efficiency of this QEM protocol.

\section{Acknowledgments}
 This project is supported by Moonshot R\&D, JST, Grant No.\,JPMJMS2061; MEXT Q-LEAP Grant No.\,JPMXS0120319794, and No.\, JPMXS0118068682 and PRESTO, JST, Grant No.\, JPMJPR2114,  No. \,JPMJPR1916, and JST CREST Grant No. JPMJCR23I4, Japan. We would like to thank the summer school on "A novel numerical approach to quantum field theories" at the Yukawa Institute for Theoretical Physics (YITP-W-22-13) for providing us with the opportunity to deepen our ideas. We thank Shu Kanno, Nobuyuki Yoshioka and Hideaki Hakoshima for useful discussions.

\bibliography{main}

\onecolumngrid
\appendix

\section{Procedures for contracting local tensors in the case of \texorpdfstring{$b_k=1$}{bk=1}}\label{apdx:calculation of M_k}

In this section, we review the estimation procedures for $M_k$ for types (i) -- (iv) in the case of $b_k=1$, given in Refs.\cite{yuan2021quantum,kanno2021quantum}. As stated in Sec.~\ref{contraction_of_local_observables}, $S_k$ can be computed by setting $O_k=I_k$ in the estimation of $M_k$.

\subsection{Type (i)}\label{apdx:calculation of M_k^1}

\begin{figure}[tp]
\centering
\begin{center}
 \includegraphics[width=160mm]{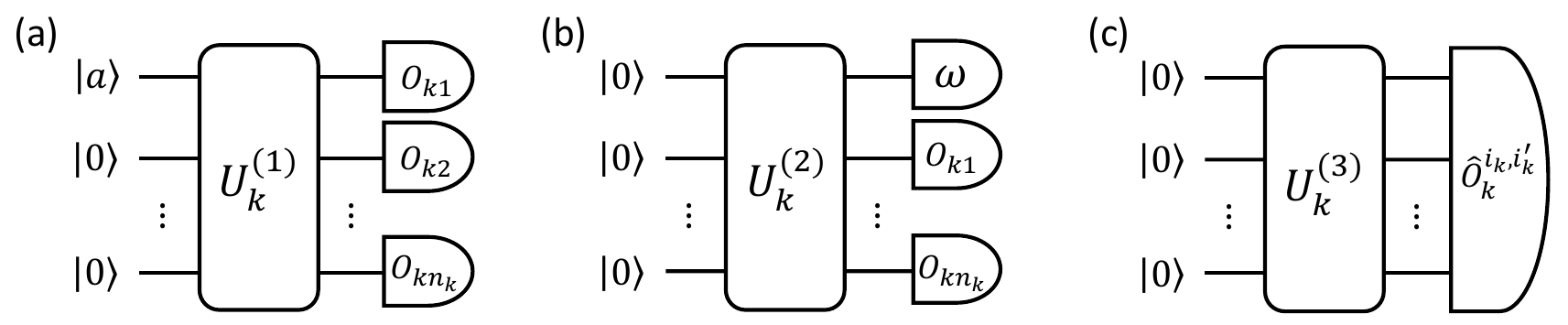}    
\end{center}
\caption{\label{fig: Quantum circuits for M_k_app}Quantum circuits for calculating $M_k$ and $S_k$ in types (i) -- (iii). (a), (b), and (c) show the quantum circuits for types (i), (ii), and (iii), respectively.}
\end{figure}

Each element of $M_{k}$ is represented as 
\begin{equation}\label{apeq: Case1, M_k}
    M_{k}^{i_{k},i'_{k}} = \mathrm{Tr}\left[ O_{k}\,U_{k}^{(1)} \left(\ket{i'_{k}}\bra{i_{k}} \otimes \ket{\bar{0}}\bra{\bar{0}}\right) U_k^{(1)\dagger}\right],
\end{equation}
where we have defined $\ket{\bar 0}\bra{\bar 0}:=\ket{0}\bra{0}^{\otimes (n_k-1)}$, and the indices $i_k$ and $i'_k$ takes $0$ or $1$. Here, let us define $E_{k}(\ket{a})=\mathrm{Tr}[ O_{k}\,U_{k}^{(1)} (\ket{a}\bra{a} \otimes \ket{\bar{0}}\bra{\bar{0}})U_{k}^{(1),\dagger}]$. Since $\ket{i'_{k}}\bra{i_{k}}$ can be expressed by a linear combination of the eigenstates of the Pauli operators, we can calculate each element of $M_{k}$ by appropriately combining the measurement results $E_{k}(\ket{a})$ for various inputs $\ket{a}$. More specifically, the diagonal elements of $M_{k}$ by measuring the expectation values of the observable $O_{k}$ for the states prepared by the quantum circuit in Fig.~\ref{fig: Quantum circuits for M_k_app}(a) with the input state $\ket{a}$ being $\ket{0}$ and $\ket{1}$, i.e.,
\begin{equation}
    M_{k}^{0,0}=E_{k}(\ket{0}),~~~~~~~~~~~~~ M_{k}^{1,1}=E_{k}(\ket{1}).
\end{equation}
On the other hand, the off-diagonal elements of $M_{k}$ can be calculated by using the quantum circuit in Fig.~\ref{fig: Quantum circuits for M_k_app}(a) as
\begin{equation}\label{apeq: Case1, M_k^01}
    M_{k}^{0,1}= \frac{E_{k}(\ket{+}) - E_{k}(\ket{-}) -i\{E_{k}(\ket{+i}) - E_{k}(\ket{-i}) \}}{2}
\end{equation}
 and $M_{k}^{1,0}=M_{k}^{0,1\ast}$. Here, we define $\ket{+}, \ket{-}, \ket{+i},$ and $\ket{-i}$ as the eigenvectors with eigenvalues $+1$ and $-1$ of Pauli operators $X$ and $Y$, respectively. Noticing that $E_{k}(\ket{0})+E_{k}(\ket{1})=E_{k}(\ket{+})+E_{k}(\ket{-})=E_{k}(\ket{+i})+E_{k}(\ket{-i})$, we can reuse the calculation results $E_{k}(\ket{0})$ and $E_{k}(\ket{1})$ used in the estimation of the diagonal elements to calculate $M_{k}^{0,1}$, that is,
\begin{equation}\label{apeq: Case1, M_k^01 v2}
    M_{k}^{0,1}= E_{k}(\ket{+})-iE_{k}(\ket{+i})+\frac{i-1}{2}\{E_{k}(\ket{0})+E_{k}(\ket{1})\}.
\end{equation}
This modification allows a reduction from six to four in the number of distinct input states needed to estimate $M_k$. As mentioned in the main text, we do not need to calculate the matrix $S_{k}$ because it holds that $S_{k} = I$ for any $k$, and consequently, $C^2=1$ holds. 

\subsection{Type (ii)}\label{apdx:calculation of M_k^2}

Each element of $M_{k}$ is represented as  
\begin{equation}\label{apeq: Case2, M_k}
    M_{k}^{i_{k},i'_{k}} = \mathrm{Tr}\left[ \left( \ket{i_{k}}\bra{i'_{k}} \otimes O_{k} \right) U^{(2)}_{k}\ket{\bar{0}}\bra{\bar{0}} U_k^{(2)\dagger} \right],
\end{equation}
where we have defined $\ket{\bar 0}\bra{\bar 0}:=\ket{0}\bra{0}^{\otimes (n_k+1)}$. Since $\ket{i_{k}}\bra{i'_{k}}$ can be decomposed into a linear combination of the Pauli operators, we can estimate each element of $M_{k}$ by combining the measurement results $E_{k}(\omega)=\mathrm{Tr}\left[ \left( \omega \otimes O_{k} \right) U^{(2)}_{k}\ket{\bar{0}}\bra{\bar{0}} U_k^{(2),\dagger} \right]$ for various Pauli observables $\omega\in\{I,X,Y,Z\}$. Specifically, each element of $M_k$ can be estimated by measuring the expectation values of $\omega \otimes O_{k}$ for the state $\ket{\psi^{(2)}_{k}}\bra{\psi^{(2)}_{k}}$ prepared by the quantum circuit in Fig.~\ref{fig: Quantum circuits for M_k_app}(b), i.e.,
\begin{equation}
    M_{k}^{0,0} = \frac{E_{k}(I)+E_{k}(Z)}{2},~~~
    M_{k}^{1,1} = \frac{E_{k}(I)-E_{k}(Z)}{2},~~~
    M_{k}^{0,1} = \frac{E_{k}(X)+iE_{k}(Y)}{2},~~~
    M_{k}^{1,0} = M_{k}^{0,1\ast}.
\end{equation}
In the same way, we can calculate $S_{k}$ by replacing $O_{k}$ with $I$ in the calculation of $M_{k}$.

\subsection{Type (iii)}\label{apdx:calculation of M_k^3}

 Each element of $M_{k}$ is represented as
\begin{equation}\label{apeq:mk3_1}
    M_{k}^{i_k,i'_k} = \mathrm{Tr}\left[ P_{k}^{i_{k}} O_{k}P_{k}^{i'_{k}} \, U_k^{(3)} \ket{0}\bra{0}^{\otimes n_k} U_k^{(3)\dagger} \right].
\end{equation}
Each matrix element $M_{k}^{i_{k},i'_{k}}$ can be evaluated by direct measurements of Pauli operators on the state $U_k^{(3)} \ket{0}\bra{0}^{\otimes n_k} U_k^{(3)\dagger}$, using the circuit Fig.~\ref{fig: Quantum circuits for M_k_app}(c); we first decompose the observable $\hat{O}_{i_{k},i'_{k}}:=P_{k}^{i_{k}} O_{k}P_{k}^{i'_{k}}$ into a linear combination of the Pauli operators as $\hat{O}_{i_{k},i'_{k}}=\sum_{l_{k}} f_{l_{k}}P_{l_{k}}$, calculate the expected values of $P_{l_k}$ on the state $U_k^{(3)} \ket{0}\bra{0}^{\otimes n_k} U_k^{(3)\dagger}$ for all $l_k$, and combine the results with the weights $f_{l_k}$. Note that once we estimate $M_{k}^{i_k,i'_k}$, we do not need to calculate $M_{k}^{i'_{k},i_{k}}$ due to the relation $M_{k}^{i'_{k},i_{k}}=M_{k}^{i_{k},i'_{k}\ast}$. For the calculation of $S_{k}$, we can set each diagonal element $S_{k}^{i_{k},i'_{k}}$ as 1 since $S_{k}^{i_k,i_k} = \mathrm{Tr}\left[ P_{k}^{i_{k}} P_{k}^{i_{k}} \, U_k^{(3)} \ket{0}\bra{0}^{\otimes n_k} U_k^{(3)\dagger} \right]=1$. Each off-diagonal element $S_{k}^{i_{k},i'_{k}}$ can be calculated by directly measuring the Pauli observable $P^{i_{k}}P^{i'_{k}}$ for the state $\mathcal{U}_{k}^{(3)}(\ket{\bar{0}}\bra{\bar{0}})$.

\subsection{Type (iv)}\label{apdx:calculation of M_k^4}
\begin{figure}[tp]
\centering
\begin{center}
\includegraphics[width=160mm]{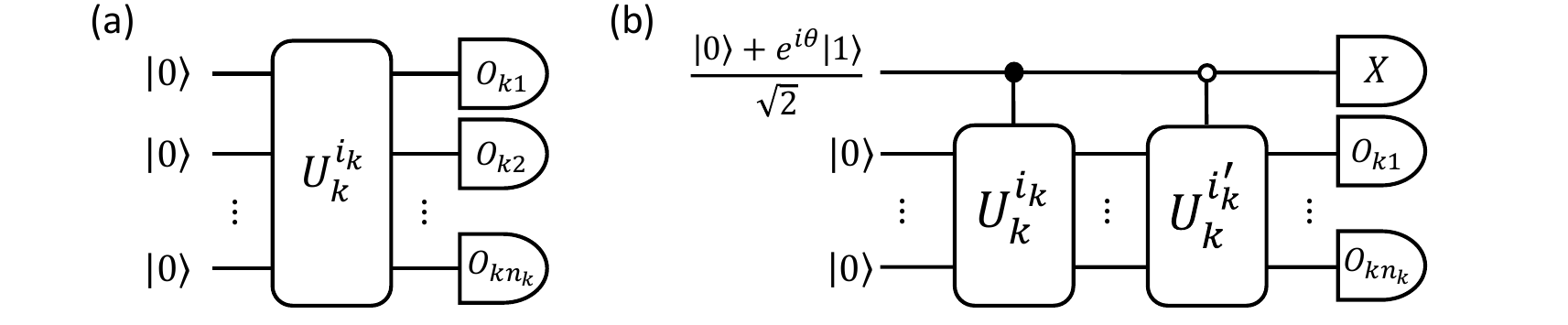}
\end{center}
\caption{\label{fig: Quantum circuits for M_k_app2} Quantum circuits for calculating $M_k$ and $S_k$ in types (iv). (a) and (b) show the quantum circuits for estimating the diagonal elements $M^{i_k,i_k}$ and the off-diagonal elements $M^{i_k,i'_k}$ of $M_k$, respectively. $S_k$ can be estimated by setting $O_k=I^{\otimes n_k}$ in (b) and assigning $S^{0,0}=S^{1,1}=1$.
}
\end{figure}

Each element of the Hermitian matrix $M_{k}$ of type (iv) is represented as  
\begin{equation}\label{apeq:mk4_1}
    M_{k}^{i_{k},i'_{k}} = 
    \begin{cases}
        \mathrm{Tr}\left[ O_k \,U_k^{i_k}\ket{0}\bra{0}^{\otimes n_k}U_k^{i_k\dagger} \right], & i_k = i'_k \\
        \mathrm{Tr}\left[ O_{k}U_{k}^{i'_{k}} \ket{0}\bra{0}^{\otimes n_k} U_{k}^{i_{k}\dagger} \right], & i_k \neq i'_k
    \end{cases}.
\end{equation}
The diagonal and off-diagonal elements of $M_k$ ($S_k$) are computed using different quantum circuits. The estimation procedure of the diagonal elements $M_{k}^{i_{k},i_{k}}$ is quite simple because it can be obtained by measuring the expectation value of $O_{k}$ for the state $U_k^{i_k}\ket{0}\bra{0}^{\otimes n_k}U_k^{i_k\dagger}$ as shown in Fig.~\ref{fig: Quantum circuits for M_k_app2}(a). On the other hand, the estimation of off-diagonal uses the Hadamard test circuit, as shown in Fig.~\ref{fig: Quantum circuits for M_k_app2}(b). We set the initial state as $\frac{\ket{0}+e^{i\theta}\ket{1}}{\sqrt{2}}\otimes\ket{0}^{\otimes n_k}$ where $\theta$ is the phase. When setting $\theta=0$ for the parameterized initial state, we obtain Re$(M_{k}^{i_{k},i'_{k}})$ by measuring the expectation value $\braket{X \otimes O_k}$. Similarly, we obtain Im$(M_{k}^{i_{k},i'_{k}})$ by setting $\theta=\pi/2$ and calculating the expectation value $\braket{X \otimes O_k}$. Combining the real and imaginary parts, we obtain $M_{k}^{i_{k},i'_{k}}$, and $M_{k}^{i'_{k},i_{k}}=M_{k}^{i_{k},i'_{k}\ast}$. To estimate the matrix $S_{k}$, we set the diagonal terms as 1 and estimate off-diagonal elements by substituting $O_{k}$ for $I^{\otimes n_k}$ of the above procedure. 

\section{Detailed Description of Techniques relevant to Hybrid tensor Networks }\label{apdx:detail_in_relevant_techniques}

In Sec.~\ref{sec:preliminaries}, addressing Deep VQE~\cite{fujii2022deep} and entanglement forging~\cite{eddins2022doubling} methods, we briefly explain the problem setting considered in each technique and their relations with the HTN framework~\cite{yuan2021quantum}. 
In this section, we provide a more detailed explanation of each technique to supplement the discussion in the main text. After revisiting each method's problem setting (as already provided in the main text but included here for completeness), we describe their calculation steps in the following subsections. 

\subsection{Deep VQE}\label{apdx:deep_vae}

The Deep VQE algorithm provides a procedure for constructing an ansatz in scenarios where the number of available qubits is limited. Specifically, the algorithm considers a quantum state of $\bar{n}_{\rm tot}$ qubits composed of $N$ subsystems. Each subsystem, indexed by $k$, consists of $\bar{n}_k$ qubits, satisfying $\bar{n}_{\rm tot} = \sum_{k=1}^{N} \bar{n}_k$. Additionally, the algorithm is particularly designed for cases where the entire system can be grouped into $N$ clusters with strong interactions within each cluster but weak interactions between different clusters. In this setting, the Hamiltonian is expressed as:
\begin{equation}\label{apeq: deepVQE hamiltonian}
    H = \sum_{k=1}^{N} H^{[1]}_{k} + \sum_{k,l=1}^{N} H^{[2]}_{kl},\quad
    H^{[2]}_{kl} = \sum_{m} \alpha_{kl}^{m}\, W^{m}_{k}\otimes W^{m}_{l}
\end{equation}
 where $H^{[1]}_{k}$ is the $k$-th subsystem's Hamiltonian and $H^{[2]}_{kl}$ the interaction Hamiltonian acting on the $k$-th and $l$-th subsystems. The interaction term $H^{[2]}_{kl}$ is expressed by a linear combination of the tensor products of $W^{m}_{k}$ and $W^{m}_{l}$, each acting on the $k$-th and $l$-th subsystems, with the coefficients $\alpha_{kl}^{m}$. Refer to Fig.~\ref{fig:deep_vqe}(a) for the tensor representation of $H$. The Deep VQE algorithm roughly consists of four steps, outlined as follows.

 The first step of this algorithm is to find the ground state of each subsystem's Hamiltonian $H_{k}^{[1]}$ by the conventional VQE approach, minimizing the expectation value of $H_{k}^{[1]}$ by optimizing the parameters $\vec{\theta}^{(k)}$ in each quantum state $\ket{\phi_{k}}=U^{(k)}(\vec{\theta}^{(k)})\ket{0^{\bar{n}_{k}}}$. Hereafter, let 
 \begin{equation}
     \ket{\phi_{k}^{\mathrm{G}}}=U^{(k)}(\vec{\theta}^{(k) \star})\ket{0^{\bar{n}_{k}}}
 \end{equation}
 be the approximated ground state, where $\vec{\theta}^{(k)\star}$ is an optimized parameter set of the subsystem $k$. 

 In the second step, we generate the $d_{k}$-dimensional orthonormal local basis $\{\ket{\widehat{\phi}_{k}^{m}}\}_{m=0}^{\bar{d}_{k}-1}$ on the basis of the ground state $\ket{\phi_{k}^{\rm G}}$. First, we construct non-orthonormal local basis $\{\ket{{\phi}_{k}^{m}}\}_{m=0}^{\bar{d}_{k}-1}$ as
\begin{equation}\label{eq:def_of_deep_vqe_basis}
    \ket{\phi_{k}^{m}} = D_{k}^{m} \ket{\phi_{k}^{\mathrm{G}}},
\end{equation}
 where $D_{k}^{m}$ denotes the $2^{\bar{n}_{k}} \times 2^{\bar{n}_{k}}$ local excitation operator on the $k$-th subsystem, and we set $D_{k}^{0}$ as an identity operator. \textcite{fujii2022deep} pointed out that when the number of dimensions $\bar{d}_{k}$ is fixed, one of the proper choices of $D_{k}^{0}$ is assigning the local excitation $D_{k}^{m}$ to $W_{k}^{m}$ at the boundary~\footnote{The appropriate choice of a basis set is highly important because the selection determine the fundamental limits of accuracy that an ansatz can perform and the number of qubits $\bar{b}_k=\log_2{\lceil \bar{d}_k \rceil}$ required for the implementation~\cite{erhart2022constructing}. Accordingly, several methods to prepare the basis sets have recently been proposed~\cite{mizuta2021deep, erhart2022constructing}.}. In this review part, following the original research~\cite{fujii2022deep}, we choose $D_{k}^{m}$ as $W_{k}^{m}$. Then, we define the orthonormal local basis $\{\ket{\widehat{\phi}_{k}^{s}}\}_{s=0}^{\bar{d}_{k}-1}$ using the Gram-Schmidt procedure for $\{\ket{{\phi}_{k}^{m}}\}_{m=0}^{\bar{d}_{k}-1}$:
\begin{equation}\label{eq: gram-schmidt}
    \ket{\widehat{\phi}_{k}^{s}}=\frac{\ket{{\phi}_{k}^{s}} - \sum_{t<s} \braket{\widehat{\phi}_{k}^{t}|{\phi}_{k}^{s}}\ket{\widehat{\phi}_{k}^{t}}}{\|\ket{{\phi}_{k}^{s}} - \sum_{t<s} \braket{\widehat{\phi}_{k}^{t}|{\phi}_{k}^{s}}\ket{\widehat{\phi}_{k}^{t}}\|}.
\end{equation}
From Eq.~(\ref{eq: gram-schmidt}), the matrix $\beta_{k}$, which transforms the basis $\{\ket{{\phi}_{k}^{m}}\}_{m=0}^{\bar{d}_{k}-1}$ to the orthonormalized basis $\{\ket{\widehat{\phi}_{k}^{s}}\}_{s=0}^{\bar{d}_{k}-1}$, can be determined as 
\begin{equation}\label{eq: transformation matrix}
    \ket{\widehat{\phi}_{k}^{s}}=\sum_{m=0}^{\bar{d}_{k}-1} \beta^{s,m}_{k} \ket{\phi_{k}^{m}}.
\end{equation}
 Each matrix element $\beta^{s,m}_{k}$ can be calculated by exploiting a set of inner products of local basis $\{\braket{\phi_{k}^{m}|\phi_{k}^{m'}}\}_{m,m'}$. Here, each inner product $\braket{\phi_{k}^{m}|\phi_{k}^{m'}}$ can be calculated with direct measurement because it can be expressed as
\begin{equation}
    \braket{\phi_{k}^{\mathrm{G}}|{W_{k}^{m}}^{\dagger}W_{k}^{m'}|\phi_{k}^{\mathrm{G}}}
\end{equation}
 and the operator ${W_{k}^{m}}^{\dagger}W_{k}^{m'}$ can be decomposed into a linear combination of Hermitian operators. The matrix $\beta_{k}$ formulated above is needed to normalize the effective Hamiltonian in the next step.

 In the third step, we construct the normalized effective Hamiltonian on the basis of the orthonormal basis $\{\ket{\widehat{\phi}_{k}^{s}}\}_{s=0}^{d_{k}-1}$. The normalized effective Hamiltonian $H_{k}^\mathrm{eff}$ for the $k$-th subsystem is defined as
\begin{equation}\label{eq: definition of effective local hamitonian}
    (\widehat{H}_{k}^{[1],\mathrm{eff}})^{s,s'} := \braket{\widehat{\phi}^{s}_{k}|H_{k}^{[1]}|\widehat{\phi}^{s'}_{k}}.
\end{equation}
 Note that terms that can be directly measured on quantum computers are unnormalized ones $(H_{k}^\mathrm{[1],eff})^{m,m'} := \braket{\phi^{m}_{k}|H_{k}^{[1]}|\phi^{m'}_{k}}$. Thus, we first evaluate 
\begin{equation}\label{eq: calculation of Heff}
    (H_{k}^{[1],\mathrm{eff}})^{m,m'} =\braket{\phi_{k}^{\mathrm{G}}|{W_{k}^{m}}^{\dagger} H_{k}^{[1]} W_{k}^{m'}|\phi_{k}^{\mathrm{G}}},
\end{equation}
 by decomposing the operator ${W_{k}^{m}}^{\dagger} H_{k}^{[1]} W_{k}^{m'}$ into a linear combination of Hermitian operators and performing the direct measurement. Then, we transform $(H_{k}^{[1],{\rm eff}})^{m,m'}$ into the normalized effective Hamiltonian $(\widehat{H}_{k}^{[1],\mathrm{eff}})^{s,s'}$, by applying the matrix $\beta_{k}$ to $(H_{k}^{[1],\mathrm{eff}})^{m,m'}$ as
\begin{equation}\label{eq: orthonormal effective hamiltonian}
\begin{split}
    (\widehat{H}_{k}^{[1],\mathrm{eff}})^{s,s'} &= \sum_{m=0}^{\bar{d}_{k}-1} \sum_{m'=0}^{\bar{d}_{k}-1} {\beta^{s,m}_{k}}^{\ast} \braket{\phi^{m}_{k}|H_{k}^{[1]}|\phi^{m'}_{k}} \beta^{s',m'}_{k}\\
    &= \left( {\beta_{k}}^{\ast} H_{k}^{[1],\mathrm{eff}} {\beta_{k}}^{\rm T} \right)^{s,s'},
\end{split}
\end{equation}
 where $\ast$ and $\mathrm{T}$ denote the complex conjugate and transposition of matrices. Similarly, the normalized effective interaction Hamiltonian $\widehat{H}_{kl}^{[2],\mathrm{eff}}$ is defined as
\begin{equation}\label{eq: definition of orthonormal effective interaction hamitonian}
    (\widehat{H}_{kl}^{[2],\mathrm{eff}})^{s,s',t,t'} := \bra{\widehat{\phi}^{s}_{k}}\bra{\widehat{\phi}^{t}_{l}} H_{kl}^{[2]} \ket{\widehat{\phi}^{s'}_{k}}\ket{\widehat{\phi}^{t'}_{l}}.
\end{equation}
 Since $H_{kl}^{[2]}$ can be decomposed into the tensor products of the local Hamiltonian acting on the $k$-th and $l$-th subsystems as Eq.~(\ref{apeq: deepVQE hamiltonian}), un-normalized effective interaction Hamiltonian $(H^{[2],\mathrm{eff}}_{kl})^{m,u,m',u'} := \bra{\phi^{m}_{k}}\bra{\phi^{u}_{l}} H_{kl}^{[2]} \ket{\phi^{m'}_{k}}\ket{\phi^{u'}_{l}}$ can be derived from integrating the expectation value of each subsystem. Then, applying the matrices $\beta_{k}$ and $\beta_{l}$ to $(H^{[2],\mathrm{eff}}_{kl})^{m,u,m',u'}$, we obtain the normalized effective interaction Hamiltonian as
\begin{equation}\label{eq: orthonormal effective interaction hamiltonian}
    (\widehat{H}_{kl}^{[2],\mathrm{eff}})^{s,t,s',t'} = \sum_{m} \alpha_{kl}^{m} (\widehat{W}_{k}^{m,{\mathrm{eff}}})^{s,s'} (\widehat{W}_{l}^{m,{\mathrm{eff}}})^{t,t'}.
\end{equation}
 The normalized effective Hamiltonian $(\widehat{W}_{k}^{m,{\mathrm{eff}}})^{s,s'}$ and $(\widehat{W}_{l}^{m,{\mathrm{eff}}})^{t,t'}$ can be calculated in the same way as Eq.~(\ref{eq: orthonormal effective hamiltonian}). After calculating $(\widehat{H}_{k}^{[1],\mathrm{eff}})^{s,s'}$ and $(\widehat{H}_{kl}^{[2],\mathrm{eff}})^{s,t,s',t'}$, we can convert the original Hamiltonian $H$ into the normalized effective one as
\begin{equation}\label{eq: effective Hamiltonian}
    \widehat{H}^\mathrm{eff} = \sum_{k=1}^{N} \widehat{H}_{k}^{[1],\mathrm{eff}} + \sum_{k,l=1}^{N} \widehat{H}_{kl}^{[2],\mathrm{eff}}.
\end{equation}

 In the fourth step, we estimate the ground state for the effective Hamiltonian $\widehat{H}^\mathrm{eff}$ by the typical VQE with the $\bar{b}_{\rm tot}$-qubit parameterized quantum circuit $V(\vec{\eta})$ in which $\bar{b}_{\rm tot}:=\sum_{k=1}^{N} \bar{b}_{k}$ with $\bar{b}_{k}:=\lceil \log_2 \bar{d}_{k} \rceil$. The expectation value of the effective Hamiltonian $\widehat{H}^\mathrm{eff}$ for the approximated ground state $V(\vec{\eta}^\star)\ket{0^{\bar{b}}}$ approximates the ground state energy $E_{\rm min}$ for the original Hamiltonian $H$:
\begin{equation}\label{eq: enegy}
    E_{\rm min} \simeq \braket{0^{\bar{b}_{\rm tot}}|V(\vec{\eta}^\star)^\dagger \widehat{H}^\mathrm{eff} V(\vec{\eta}^\star) | 0^{\bar{b}_{\rm tot}}}.
\end{equation}

 One of the notable points of this technique is that it only takes $\max(\bar{n}_1,...,\bar{n}_{N},\bar{b}_{\rm tot})$ qubits to simulate a $\bar{n}_{\rm tot}$-qubit quantum system. Furthermore, we can simulate even larger-scale systems by repeating the series of steps.

 Lastly, we remark that by replacing the subscripts $m,m',s,s',t,t',u,u'$ used in the explanation of Deep VQE so far with 
\begin{equation}
\begin{aligned}
    m  &\rightarrow \vec{j}_k,~~~ & m' &\rightarrow \vec{j}'_k,~~~ & s  &\rightarrow \vec{i}_k,~~~ & s' &\rightarrow \vec{i}'_k, \\
    t  &\rightarrow \vec{i}_l, & t' &\rightarrow \vec{i}'_l, & u  &\rightarrow \vec{j}_l, & u' &\rightarrow \vec{j}'_l,
\end{aligned}
\end{equation}
 they correspond to the notation used in the main text.

\subsection{Entanglement forging}\label{qpdx:entanglement_forging}

 In this subsection, we review \textit{entanglement forging} proposed by \textcite{eddins2022doubling}, which allows the size of the simulatable quantum system on an available quantum processor to be doubled with the help of classical processing. The algorithm focuses on a state of a bipartite $2n$-qubit system $\ket{\phi_{\rm EF}}$. First, we start by decomposing $\ket{\phi_{\rm EF}}$ into $n$-qubit subsystems using Schmidt decomposition, i.e.,
\begin{equation}\label{apeq: entanglement forging ansatz}
    \ket{\phi_{\rm EF}} = (V_{1} \otimes V_{2}) \sum_{x=0}^{2^n-1} \lambda_{x} \ket{l_{x}} \otimes \ket{l_{x}},
\end{equation}
 where $V_{1}$ and $V_{2}$ are the $n$-qubit unitary operators acting on the 1st and 2nd subsystems respectively, $\lambda_{x}$ are the Schmidt coefficients, and $\ket{l_{x}} \in \{\ket{0},\ket{1}\}^{\otimes n}$ are the computational basis corresponding to indices $x$. As mentioned in the main text, the structure of the hybrid ansatz can be adjusted by truncating the Schmidt coefficients $\lambda_{x}$.
 
Now, we consider using the ansatz defined above to calculate the expectation value of the $2n$-qubit observable $O=O_{1} \otimes O_{2}$. The expected value $\braket{O}_{\phi_{\rm EF}}$ for the state $\ket{\phi_{\rm EF}}$ can be expressed as follows:
\begin{equation}\label{eq: O in entanglement forging ansatz}
    \braket{O}_{\phi_{\rm EF}} = \sum_{x=0}^{2^n-1} \lambda_{x}^2 \braket{l_{x}|\widehat{O}_{1}|l_{x}} \braket{l_x|\widehat{O}_{2}|l_{x}} + \sum_{x=0}^{2^n-1} \sum_{y<x} \Bigl( \lambda_{x}\lambda_{y} \sum_{p\in{\mathbb{Z}_4}} (-1)^p \braket{\phi^{p}_{l_{x}l_{y}}|\widehat{O}_{1}|\phi^{p}_{l_{x}l_{y}}} \braket{\phi^{p}_{l_{x}l_{y}}|\widehat{O}_{2}|\phi^{p}_{l_{x}l_{y}}} \Bigr),
\end{equation}
 where $\widehat{O}_{1} = V_{1}^\dagger O_{1} V_{1}$, $\widehat{O}_{2} = V_{2}^\dagger O_{2} V_{2}$, and $\ket{\phi^{p}_{l_{x}l_{y}}} = (\ket{l_{x}}+i^p\ket{l_{y}})/\sqrt{2}$ with $p \in \{0,1,2,3\} = \mathbb{Z}_4$. Note that Eq.~(\ref{eq: O in entanglement forging ansatz}) indicates that the expectation value $\braket{O}_{\phi_{\rm EF}}$ can be estimated by linearly combing the calculation results in each subsystem, all of which can be evaluated on only $n$-qubit quantum processors. 

 In the actual implementation, one may estimate $\braket{O}_{\phi_{\rm EF}}$ using the weighted sampling strategy. To adopt this approach, it is convenient to rewrite Eq.~(\ref{eq: O in entanglement forging ansatz}) as follows:
\begin{equation}
    \braket{O}_{\phi_{\rm EF}} =  \|\mu\|_{1}  
 \sum_{a} p_{\mu_{a}} \mathrm{sgn}(\mu_{a}) \cdot \mathrm{Tr}( O_{1} \rho_{a,1} ) \mathrm{Tr}( O_{2} \rho_{a,2} ),
\end{equation}
 where $\mu_{a}$ are real coefficients, $\|\mu\|_{1}=\sum_{a} |\mu_{a}|$ is a 1-norm of the coefficients $\mu_{a}$, $p_{a}=|\mu_{a}|/\|\mu\|_{1}$ are probabilities, and $\rho_{a,k}$ are the $n$-qubit states on the $k$-th subsystem, represented as $V_{k}\ket{l_{x}}$ or $V_{k}\ket{\phi_{l_{x}l_{y}}^{p}}$. The basic procedure of this sampling strategy is as follows. For each shot, we randomly prepare the states $\rho_{a,k}$ on each subsystem with probability $p_{a}$, measure each qubit related to the observable $O_{k}$($k=1,2$), and store the product of measurement outcomes of the 1st, and 2nd subsystems and $\mathrm{sgn}(\mu_{a})$. By repeating the above process for $N$ shots and taking the total sample averages with the scaling factor $\|\mu\|_{1}$, we can obtain the expectation values $\braket{O}_{\phi_{\rm EF}}$. For further details, refer to (SM.5) in \textcite{eddins2022doubling}.

 Here, note that not all states with the doubled system size can be efficiently estimated with this method because the total required number of samples is $O(\| \vec{\lambda} \|_1^4 / \epsilon^2)$ with $\epsilon$ being the desired accuracy and $\| \vec{\lambda} \|_{1}$ depends on the entanglement structure between the $n$-qubit subsystems.
 The worst sampling overhead happens when the entanglement between the 1st and 2nd subsystems is maximal (i.e., $\ket{\phi_{\rm EF}}$ is maximally entangled state), and in this case $\| \vec{\lambda} \|_1$ is $\sqrt{2^n}$. As the entanglement between the clusters becomes weaker, $\| \vec{\lambda} \|_1$ approaches 1, corresponding to no entanglement among them. Thus, the series of procedures that implements Eq.~(\ref{eq: O in entanglement forging ansatz}) is the most practical in simulating systems with weak inter-subsystem interaction.

 When employing the $\ket{\phi_{\rm EF}}$ as an ansatz for ground state calculation in the variational method, we must parameterize the entanglement forging ansatz. For each subsystem $k\in\{1,2\}$, parameters $\vec{\theta}_{k}$ can be incorporated into each unitary gate as $V_{k}(\vec{\theta}_{k})$, while for the main system representing non-local correlation, the coefficients $\lambda_x$ can be treated as the variational parameters. These coefficients $\lambda_x$ can be updated through the variational calculation along with the parameters $\vec{\theta}_{1}, \vec{\theta}_{2}$. When the problem size is relatively small, we can eliminate the coefficients $\lambda_x$ from the optimizer's search space by precisely calculating the minimum eigenvalue of the matrix that is constructed by the calculation results of the subsystems~\cite{eddins2022doubling}.

\section{Equivalence between Eq.~(\ref{eq: HTTN state_2}) and Eq.~(\ref{eq: Op-based rep})}
\label{sec: appendix A}

The aim of this section is 
to show the equivalence of Eq.~(\ref{eq: HTTN state_2}) and Eq.~(\ref{eq: Op-based rep}).
To begin with, let us review the structure of the 2-layer HTTN state $\ket{\psi_{\rm HT}}$ in Eq.~(\ref{eq: HTTN state_2}):
\begin{equation}\label{eq:A1}
    \ket{\psi_{\rm HT}}=\frac{1}{C}\sum_{\vec{i}_{1},...,\vec{i}_{N}} \psi_{\vec{i}_{1},...,\vec{i}_{N}} \ket{\psi_{1}^{\vec{i}_{1}}}\otimes \cdots \otimes \ket{\psi_{N}^{\vec{i}_{N}}},
\end{equation}
 where $\vec{i}_{k}=(i_{k,1},...,i_{k,b_{k}})\in\{0,1\}^{b_{k}} (k=1,...,N)$ is $b_k$ two-dimensional indices, $\psi_{\vec{i}_{1},...,\vec{i}_{N}}=\braket{\vec{i}_{1},...,\vec{i}_{N}|\psi}$ is the probability amplitude of an $(\sum_{k=1}^{N} b_{k})$-qubit system, $\{\ket{\psi_{k}^{\vec{i}_{k}}}\}$ is the $k$-th set of $n_{k}$-qubit states, $C$ is a normalization constant, and $\ket{\psi_{\rm HT}}$ is a $(\sum_{k} n_{k})$-qubit state. 

 Focusing on the non-local tensor $\psi_{\vec{i}_{1},...,\vec{i}_{N}}$ and local state $\ket{\psi_k^{\vec{i}_{k}}}$ in subsystem $k$, the following connecting relation exists:
\begin{equation}\label{eq:A2}
    \sum_{\vec{i}_{k}} \psi_{\vec{i}_{k},...} \ket{\psi^{\vec{i}_{k}}_{k}}  = \sum_{\vec{i}_{k}} \braket{\vec{i}_{k}|\psi} \ket{\psi_{k}^{\vec{i}_{k}}}.
\end{equation}
 In this expression, the index $\vec{i}_{k}$ of the non-local tensor $\psi_{\vec{i}_{k},...}$ is linked with the labeled state $\ket{\psi_{k}^{\vec{i}_{k}}}$ from the set of state $\{ \ket{\psi_{k}^{\vec{i}_{k}}} \}_{\vec{i}_{k}}$. Here, we can rewrite Eq.~(\ref{eq:A2}) as follows:
\begin{equation}
    \sum_{\vec{i}_{k}} \psi_{\vec{i}_{k},...} \ket{\psi^{\vec{i}_{k}}_{k}} = A_{k}^{(\xi_{k})} \ket{\psi}.
\end{equation}
 where $A_{k}^{(\xi_{k})} := \sum_{\vec{i}_{k}} \ket{\psi^{\vec{i}_{k}}_{k}} \bra{\vec{i}_{k}}$.
 By introducing the operators $A_{k}^{(\xi_{k})}(k=1,...,N)$ into Eq.~(\ref{eq:A1}), we have
\begin{equation}\label{eq:A4}
    \ket{\psi_{\rm HT}} = \frac{1}{C} (A_{1}^{(\xi_{1})} \otimes ... \otimes A_{N}^{(\xi_{N})}) \ket{\psi}.
\end{equation}
 Then, representing Eq.~(\ref{eq:A4}) in the form of a density matrix, we obtain the following expression:
\begin{equation}\label{eq:Op-based rep_2}
    \rho_{\rm HT} =\frac{(A_{1}^{(\xi_{1})} \otimes ... \otimes A_{N}^{(\xi_{N})}) \ket{\psi}\bra{\psi} (A_{1}^{(\xi_{1})} \otimes ... \otimes A_{N}^{(\xi_{N})})^\dagger}{C^2}.
\end{equation}
By taking the trace of both sides of Eq.~(\ref{eq:Op-based rep_2}) and using the fact that $\mathrm{Tr}(\rho_{\rm HT})=1$, we have $C^2=\mathrm{Tr}[(A_{1}^{(\xi_{1})} \otimes ... \otimes A_{N}^{(\xi_{N})}) \ket{\psi}\bra{\psi} (A_{1}^{(\xi_{1})} \otimes ... \otimes A_{N}^{(\xi_{N})})^\dagger]$. Thus, the above expression is equal to
\begin{equation}
    \rho_{\rm HT} =\frac{(A_{1}^{(\xi_{1})} \otimes ... \otimes A_{N}^{(\xi_{N})}) \ket{\psi}\bra{\psi} (A_{1}^{(\xi_{1})} \otimes ... \otimes A_{N}^{(\xi_{N})})^\dagger}{\mathrm{Tr}[(A_{1}^{(\xi_{1})} \otimes ... \otimes A_{N}^{(\xi_{N})}) \ket{\psi}\bra{\psi} (A_{1}^{(\xi_{1})} \otimes ... \otimes A_{N}^{(\xi_{N})})^\dagger]},
\end{equation}
which implies the equivalence between Eq.~(\ref{eq: HTTN state_2}) and Eq.~(\ref{eq: Op-based rep}).

\section{Supplementary calculations involving the SWAP operator and Bell states}

We introduce a transformation related to the trace and the SWAP operator (Lemma~\ref{lemma:swap_trace}) used in the proof of Appendix~\ref{sec:proof_of_prop1}, and a transformation related to the partial trace and the Bell-state projection (Lemma~\ref{lemma:partial_bell}) used in the proof of Appendix~\ref{sec:relation_A_and_circuit}.

\subsection{A transformation related to the trace and the SWAP operator}\label{sec:partial_swap}

\begin{lemma}\label{lemma:swap_trace}
    Let $A, B$ be arbitrary $2^n \times 2^n$ Hermitian matrices and $\rm SWAP_{12}$ be the swap operator acting on two $n$-qubit systems $1$ and $2$, then the following relation holds:
    \begin{equation}\label{apeq:swap_trace}
        \mathrm{Tr}\left[ \, {\rm SWAP_{12}}\, (A_{1} \otimes B_{2}) \, \right] = \mathrm{Tr}[AB].
    \end{equation}
\end{lemma}

\begin{proof}
By expanding $\rm SWAP$ in the computational basis $\{\ket{i}\},\{\ket{j}\}=\{\ket{0},\ket{1}\}^{\otimes n}$ as $\mathrm{SWAP}_{12}=\sum_{i,j} \ket{i}\bra{j}_{1}\otimes \ket{j}\bra{i}_{2}$, the left-hand side of Eq.~(\ref{apeq:swap_trace}) is equal to
\begin{eqnarray}
   \mathrm{Tr}\left[ \, {\rm SWAP_{12}}\, (A_{1} \otimes B_{2}) \, \right] 
   &=& \sum_{i,j} \mathrm{Tr} \left[ \, \ket{i}\bra{j}_{1} \otimes \ket{j}\bra{i}_{2} \, \left(A_{1} \otimes B_{2}\right) \right]\label{apeq:swap_trace_eq1}\\
   &=& \sum_{i,j} \mathrm{Tr}_{1}  \left[ A_{1} \, \left( \braket{i|B|j}_{2} \ket{i}\bra{j}_{1} \right) \right]\\
   &=& \mathrm{Tr}[AB],
\end{eqnarray}
where the second equality comes from taking partial trance over system 2 in the trace over systems 1 and 2 in Eq.~(\ref{apeq:swap_trace_eq1}).
\end{proof}

\subsection{A transformation related to partial trace and the Bell-state projection}\label{sec:partial_bell}

\begin{lemma}\label{lemma:partial_bell}
    Let $A, B$ be arbitrary $2^n \times 2^n$ Hermitian matrices and $\ket{\mathrm{Bell}}$ be an unnormalized  $2n$-qubit Bell state $\ket{\rm Bell}_{12}:=\sum_{i,j\in\{0,1\}^n}\ket{i}\bra{j}_{1}\otimes\ket{i}\bra{j}_{2}$,  acting on two $n$-qubit systems $1$ and $2$, the following relation holds:
    \begin{equation}\label{apeq:partial_bell}
        \mathrm{Tr}_{1}\left[ \, \ket{\rm Bell}\bra{\rm Bell}_{12}\, (A_{1} \otimes B_{2}) \, \right]  = A^{\rm T}B,
    \end{equation}
    where $\mathrm{Tr}_{1}[\bullet]$ dentoe a partial trace of system 1. 
\end{lemma}
\begin{proof}
Using the expansion of the unnormalized  $2n$-qubit Bell state $\ket{\rm Bell}_{12}=\sum_{i,j}\ket{i}\bra{j}_{1}\otimes\ket{i}\bra{j}_{2}$, we have
\begin{eqnarray}
   \mathrm{Tr}_{1}\left[ \, \ket{\rm Bell}\bra{\rm Bell}_{12}\, (A_{1} \otimes B_{2}) \, \right] 
   &=& \sum_{i,j} \mathrm{Tr}_{1} \left[ \, \ket{i}\bra{j}_{1} \otimes \ket{i}\bra{j}_{2} \, \left(A_{1} \otimes B_{2}\right) \right]\\
   &=& \left( \sum_{i,j} \braket{j|A|i}\ket{i}\bra{j} \right) B \\
   &=& A^{\rm T} B,
\end{eqnarray}
where the last equality uses $\braket{j|A|i}=\braket{i|A^{\rm T}|j}$.
\end{proof}

\section{Detailed derivation of the expansion maps in types (i)--(iv)}\label{sec:detail_derive}

In this section, we provide detailed calculations for deriving the description of the expansion maps $\mathcal{A}_k^{(\xi)}$ ($\xi=1,2,3,4$) from its definition:
\begin{equation}\label{apeq:def_of_A}
    \mathcal{A}_k^{(\xi)}(\bullet) := A_k^{(\xi)} \, \bullet \, A_k^{(\xi), \dagger}=\sum_{\vec{i}_k,\vec{j}_k}\ket{\psi_k^{\vec{i}_k}}\braket{\vec{i}_k|\bullet|\vec{j}_k}\bra{\psi_k^{\vec{j}_k}}.
\end{equation}
To proceed further from the above equation, we substitute the following equalities
\begin{equation}
    \ket{\psi_k^{\vec{i}_k}} = 
    \begin{cases}
        U_k^{(1)}\ket{\vec{i}_k}\ket{0}^{\otimes (n_k-b_k)} & \xi=1, \\
        \braket{\vec{i}_k|\psi_k^{(2)}} & \xi=2, \\
        P_k^{\vec{i}_k}\ket{\psi_k^{(3)}} & \xi=3,\\
        U_k^{\vec{i}_k}\ket{0}^{\otimes n_k} & \xi=4,
    \end{cases}
\end{equation}
into Eq.~(\ref{apeq:def_of_A}) for each $\xi$. For further details on the calculations for each type, refer to the following subsections.

\subsection{Type (i)}

Introducing $\ket{\psi_k^{\vec{i}_k}}=U_k^{(1)}\ket{\vec{i}_k}\ket{0}^{\otimes (n_k-b_k)}$ into Eq.~(\ref{apeq:def_of_A}), we have
\begin{equation}
\begin{split}
    \mathcal{A}_k^{(1)}(\bullet) 
    &= \sum_{\vec{i}_k,\vec{j}_k}U_k^{(1)}\ket{\vec{i}_k}\ket{0}^{\otimes (n_k-b_k)}\braket{\vec{i}_k|\bullet|\vec{j}_k}\bra{0}^{\otimes (n_k-b_k)} \bra{\vec{j}_k} U_k^{(1),\dagger}\\
    &= \sum_{\vec{i}_k,\vec{j}_k} \mathcal{U}_k^{(1)}\biggl( \ket{\vec{i}_k}\bra{\vec{j}_k} \otimes \ket{0}\bra{0}^{\otimes (n_k-b_k)} \biggr) \braket{\vec{i}_k|\bullet|\vec{j}_k},
\end{split}
\end{equation}
where $\mathcal{U}_k^{(1)}=U_k^{(1)}\bullet U_k^{(1),\dagger}$. Using the decomposion of $\bullet$ by computational basis as $\bullet = \sum_{\vec{i}_k,\vec{j}_k} \braket{\vec{i}_k|\bullet|\vec{j}_k} \ket{\vec{i}_k}\bra{\vec{j}_k}$, we obtain
\begin{equation}
    \mathcal{A}_k^{(1)}(\bullet) 
    = \mathcal{U}_k^{(1)}\left( \bullet \otimes \ket{0}\bra{0}^{\otimes (n_k-b_k)} \right).
\end{equation}

\subsection{Type (ii)}

Introducing $\ket{\psi_k^{\vec{i}_k}}=\braket{\vec{i}_k|\psi_k^{(2)}}$ into Eq.~(\ref{apeq:def_of_A}), we have
\begin{equation}
\begin{split}
    \mathcal{A}_k^{(2)}(\bullet) 
    &= \sum_{\vec{i}_k,\vec{j}_k} \braket{\vec{i}_k|\psi_k^{(2)}}\braket{\vec{i}_k|\bullet|\vec{j}_k}\braket{\psi_k^{(2)}|\vec{j}_k}\\
    &= \sum_{\vec{i}_k,\vec{j}_k} \mathrm{Tr}\left[ \ket{\vec{j}_k}\bra{\vec{i}_k} \, \bullet \, \right] \mathrm{Tr}_{2}\left[ \left( \ket{\vec{j}_k}\bra{\vec{i}_k}_{2} \otimes I_{k,3} \right) \ket{\psi_{k}^{(2)}}\bra{\psi_k^{(2)}}_{23} \right] \\
    &= \mathrm{Tr}_{12}\left[ \Biggl( \, \sum_{\vec{i}_k,\vec{j}_k} \ket{\vec{j}_k}\bra{\vec{i}_k}_{1} \otimes  \ket{\vec{j}_k}\bra{\vec{i}_k}_{2} \otimes I_{k,3} \Biggr) \Biggl( \, \bullet_{1} \, \otimes \ket{\psi_{k}^{(2)}}\bra{\psi_k^{(2)}}_{23} \Biggr) \right].
\end{split}
\end{equation}
Now, by introducing the definition of the unnormalized Bell state $\ket{\rm Bell}_{12}=\sum_{\vec{j}_k} \ket{\vec{j}_k}_{1}  \ket{\vec{j}_k}_{2}$, we obtain 
\begin{equation}
    \mathcal{A}_k^{(2)}(\bullet)
    = \mathrm{Tr}_{12}\left[ \ket{\rm Bell}\bra{\rm Bell}_{12}  \bigl( \, \bullet_{1} \, \otimes \ket{\psi_{k}^{(2)}}\bra{\psi_{k}^{(2)}}_{23} \bigr) \right].
\end{equation}

\subsection{Type (iii)}\label{apsec:a3}

Introducing $\ket{\psi_k^{\vec{i}_k}}=P_k^{\vec{i}_k}\ket{\psi_k^{(3)}}$ into Eq.~(\ref{apeq:def_of_A}), we have
\begin{equation}
    \mathcal{A}_k^{(3)}(\bullet) 
    = \sum_{\vec{i}_k,\vec{j}_k} P_k^{\vec{i}_k}\ket{\psi_k^{(3)}}\braket{\vec{i}_k|\bullet|\vec{j}_k}\bra{\psi_{k}^{(3)}} P_{k}^{\vec{j}_k,\dagger}.
\end{equation}
Here, by defining the controlled Pauli channel
\begin{equation}\label{apeq:type3_1}
    \mathcal{CP}_{23}(\bullet):= \biggl(\sum_{\vec{i}_k} \ket{\vec{i}_k}\bra{\vec{i}_k}_{2} \otimes P_{k,3}^{\vec{i}_k}\biggr) \bullet_{23} \biggl(\sum_{\vec{j}_k} \ket{\vec{j}_k}\bra{\vec{j}_k}_{2} \otimes P_{k,3}^{\vec{j}_k,\dagger}\biggr),
\end{equation}
the following relation holds:
\begin{equation}\label{apeq:type3_2}
    2^{b_k}\,\mathrm{Tr}_{2} \left[ \left( \ket{\vec{j}_k}\bra{\vec{i}_k}_{2} \otimes I_{k,3} \right) \, \mathcal{CP}_{23}\Bigl(\, \ket{+}\bra{+}^{\otimes b_k}_{2} \otimes \bullet_{3} \Bigr) \right] = P_{k}^{\vec{i}_k} \bullet P_{k}^{\vec{j}_k,\dagger}.
\end{equation}
This can be easily verified as
\begin{align}
    \lefteqn{2^{b_k}\,\mathrm{Tr}_{2} \left[ \left( \ket{\vec{j}_k}\bra{\vec{i}_k}_{2} \otimes I_{k,3} \right) \, \mathcal{CP}_{23}\Bigl(\, \ket{+}\bra{+}^{\otimes b_k}_{2} \otimes \bullet_{3} \Bigr) \right]} \\
    &~~~~~= \sum_{\vec{l}_k,\vec{l}'_k \in \{0,1\}^{b_k}} \mathrm{Tr}_{2} \left[ \left( \ket{\vec{j}_k}\bra{\vec{i}_k}_{2} \otimes I_{k,3} \right) \, \mathcal{CP}_{23}\Bigl(\, \ket{\vec{l}_k}\bra{\vec{l}'_k}_{2} \otimes \bullet_{3} \Bigr) \right] \\
    &~~~~~= \sum_{\vec{l}_k,\vec{l}'_k} \mathrm{Tr}_{2} \left[ \left( \ket{\vec{j}_k}\bra{\vec{i}_k}_{2} \otimes I_{k,3} \right) \, \Bigl(\sum_{\vec{i}'_k} \ket{\vec{i}'_k}\bra{\vec{i}'_k}_{2} \otimes P_{k,3}^{\vec{i}'_k}\Bigr) \Bigl(\, \ket{\vec{l}_k}\bra{\vec{l}'_k}_{2} \otimes \bullet_{3} \Bigr) \Bigl(\sum_{\vec{j}'_k} \ket{\vec{j}'_k}\bra{\vec{j}'_k}_{2} \otimes P_{k,3}^{\vec{j}'_k,\dagger}\Bigr) \right] \\
    &~~~~~= \sum_{\vec{l}_k,\vec{l}'_k} \delta_{\vec{i}_k,\vec{i}'_k} \delta_{\vec{j}_k,\vec{j}'_k} \delta_{\vec{i}'_k,\vec{l}_k} \delta_{\vec{l}'_k,\vec{j}'_k} P_{k}^{\vec{i}'_k} \bullet P_{k}^{\vec{j}'_k,\dagger} \\
    &~~~~~= P_{k}^{\vec{i}_k} \bullet P_{k}^{\vec{j}_k,\dagger},
\end{align}
where $\delta_{i,j}$ is the Kronecker delta. Substituting Eq.~(\ref{apeq:type3_2}) into Eq.~(\ref{apeq:type3_1}) yields
\begin{equation}
\begin{split}
    \mathcal{A}_k^{(3)}(\bullet) 
    &= \sum_{\vec{i}_k,\vec{j}_k} 2^{b_k}\,\mathrm{Tr}_{2} \left[ \left( \ket{\vec{j}_k}\bra{\vec{i}_k}_{2} \otimes I_{k,3} \right) \, \mathcal{CP}_{23}\Bigl(\, \ket{+}\bra{+}^{\otimes b_k}_{2} \otimes \ket{\psi_{k}^{(3)}}\bra{\psi_k^{(3)}}_{3} \Bigr) \right] \mathrm{Tr}\left[ \ket{\vec{j}_k}\bra{\vec{i}_k} \, \bullet \, \right] \\
    &= 2^{b_k}\,\mathrm{Tr}_{12} \left[ \ket{\rm Bell}\bra{\rm Bell}_{12} \, \Bigl( \,\bullet_{1}\, \otimes \mathcal{CP}_{23}\Bigl(\, \ket{+}\bra{+}^{\otimes b_k}_{2} \otimes \ket{\psi_{k}^{(3)}}\bra{\psi_k^{(3)}}_{3} \Bigr) \Bigr) \right].
\end{split}
\end{equation}

\subsection{Type (iv)}
By setting $P_k^{\vec{i}_k}$ as $U_k^{\vec{i}_k}$ and $\ket{\psi_{k}^{(3)}}\bra{\psi_k^{(3)}}_{3}$ as $\ket{0}\bra{0}^{\otimes n_k}$ in Sec.~\ref{apsec:a3}, we have
\begin{equation}
    \mathcal{A}_k^{(4)}(\bullet) 
    = 2^{b_k}\,\mathrm{Tr}_{12} \left[ \ket{\rm Bell}\bra{\rm Bell}_{12} \, \Bigl( \,\bullet_{1}\, \otimes \mathcal{CU}_{23}\Bigl(\, \ket{+}\bra{+}^{\otimes b_k}_{2} \otimes \ket{0}\bra{0}^{\otimes n_k}_{3} \Bigr) \Bigr) \right],
\end{equation}
where we have defined the controlled unitary channel $\mathcal{CU}$ as
\begin{equation}\label{apeq:type4_1}
    \mathcal{CU}_{23}(\bullet):= \biggl(\sum_{\vec{i}_k} \ket{\vec{i}_k}\bra{\vec{i}_k}_{2} \otimes U_{k,3}^{\vec{i}_k}\biggr) \bullet_{23} \biggl(\sum_{\vec{j}_k} \ket{\vec{j}_k}\bra{\vec{j}_k}_{2} \otimes U_{k,3}^{\vec{j}_k,\dagger}\biggr).
\end{equation}

\section{Proof of Proposition~\ref{prop:fact_1}}\label{sec:proof_of_prop1}

\newtheorem*{T0}{Proposition~\ref{prop:fact_1}}
\begin{T0}\label{ap_prop:fact_1}

When a single map $\mathcal{M}$ can express a given estimation procedure in the form of 
\begin{eqnarray}
    \tilde{M}_{k}^{(\xi)} &=& \sum_l \mathrm{Tr}[ O_{k} \, \mathcal{M}(\omega^{l})] \,\bar{\omega}^{l},\label{apeq:local observable estimation_2_1}\\
    \tilde{S}_{k}^{(\xi)} &=& \sum_l \mathrm{Tr}[ \mathcal{M}(\omega^{l})] \,\bar{\omega}^{l},\label{apeq:local observable estimation_2_2}
\end{eqnarray}
using a set of operators $\{\omega_{l},\bar{\omega}^{l}\}_{l}$ that satisfies $\mathrm{SWAP}=\sum_{l} \omega^{l}\otimes \bar{\omega}^{l}$, then the map $\mathcal{M}$ can be identified with the noisy expansion map $\tilde{\mathcal{A}}_k^{(\xi)}$ defined in Eq.~(\ref{eq: transformation}), i.e., 
the version of Eq.~(\ref{eq: transformation}) with $\tilde{\mathcal{A}}^{(\xi)}_k$ replaced by $\mathcal{M}$ holds. 
\end{T0}

\begin{proof}
To begin proof, we compute $\mathrm{Tr}[ \tilde{M}_k^{(\xi)} \, \bullet \, ]$, where $\tilde{M}_k^{(\xi)}$ satisfies Eq.~(\ref{apeq:local observable estimation_2_1}) and $\bullet$ is an arbitrary $2^{b_k}\times2^{b_k}$ Hermitian matrix:
\begin{eqnarray}
    \mathrm{Tr}[ \tilde{M}_k^{(\xi)} \, \bullet \, ] 
    &=& \sum_l \mathrm{Tr}[\, O_{k} \, \mathcal{M}(\omega^{l})] \,\mathrm{Tr}[ \,\bar{\omega}^{l} \, \bullet \,]\\
    &=& \sum_l \mathrm{Tr}[\, \omega^{l} \, \mathcal{M}^{\dagger}(O_k)] \,\mathrm{Tr}[ \,\bar{\omega}^{l} \, \bullet \,]\\
    &=& \mathrm{Tr}\Bigl[\, \mathrm{SWAP}_{12} \, \Bigl(\mathcal{M}_{1}^{\dagger}(O_{k,1}) \otimes \bullet_2 \Bigr) \Bigr] \\
    &=& \mathrm{Tr} \Bigl[ \mathcal{M}^{\dagger}(O_{k}) \, \bullet \, \Bigr]\\
    &=& \mathrm{Tr}[\, O_k \, \mathcal{M}(\bullet) ],
\end{eqnarray}
where the third equality comes from the definition of the matrices $\{\omega^l,\bar{\omega}^l\}_l$, i.e., $\mathrm{SWAP}=\sum_{l} \omega^{l}\otimes \bar{\omega}^{l}$, and the forth equality follows from Lamma~\ref{lemma:swap_trace}. Similarly, by setting $O_k=I_k$ in the above calculations, $\mathrm{Tr}[ \tilde{S}_k^{(\xi)} \, \bullet \, ]$ can be evaluated as follows, where $\tilde{S}_k^{(\xi)}$ satisfies Eq.~(\ref{apeq:local observable estimation_2_2}):
\begin{equation}
    \mathrm{Tr}[\, \tilde{S}_k^{(\xi)} \, \bullet \, ] = \mathrm{Tr}[ \, \mathcal{M}(\bullet) ].
\end{equation}
\end{proof}

\section{The relations between the expansion maps and the estimation procedures for \texorpdfstring{$M_k$}{Mk} and \texorpdfstring{$S_k$}{Sk}}\label{sec:relation_A_and_circuit}

In this section, we prove that the estimation procedures for $M_k^{(\xi)}$ and $S^{(\xi)}_k$ ($\xi=1,2,3,4$) can be expressed using the corresponding expansion map $\mathcal{A}_{k}^{(\xi)}$ as follows:
\begin{eqnarray}
    M_{k}^{(\xi)} &=& \sum_{l} \mathrm{Tr}[ O_k \mathcal{A}_{k}^{(\xi)} (\omega^{l})]~ \bar{\omega}^l,\label{apeq:estimation_form_m}\\
    S_{k}^{(\xi)} &=& \sum_{l} \mathrm{Tr}[ \mathcal{A}_{k}^{(\xi)} (\omega^{l})]~ \bar{\omega}^l,\label{apeq:estimation_form_s}
\end{eqnarray}
where $\{\omega^{l},\bar{\omega}^{l}\}_{l}$ satisfies $\mathrm{SWAP}=\sum_{l} \omega^{l}\otimes \bar{\omega}^{l}$. As discussed in Sec.~\ref{contraction_of_local_observables}, $S_k^{(\xi)}$ is computed using the same estimation procedure as $M_k^{(\xi)}$, except that $O_k$ is replaced with $I_k$. For this reason, if the estimation procedure for $M_k^{(\xi)}$ can be expressed in the form of Eq.~(\ref{apeq:estimation_form_m}), the procedure for $S_k^{(\xi)}$ can also be described in the form of Eq.~(\ref{apeq:estimation_form_s}). Thus, we devote the following subsections to showing that the estimation procedures for $M_k^{(\xi)}$ given in Eqs.(\ref{eq: Case1, M_k, 1}), (\ref{eq: Case2, M_k, 2}), (\ref{eq: Case3, M_k, 2}), and (\ref{eq: Case4, M_k, 2}), can be expressed in the form of Eq.~(\ref{apeq:estimation_form_m}) for each type.

\subsection{Type (i)}
Let $\{P_x\}_{x=1}^{4^{b_k}}=\{I,X,Y,Z\}^{\otimes b_k}$ be the set of $4^{b_k}$ $b_k$-qubit Pauli strings, and let $P_x=\sum_{y} \lambda_{x,y}\ket{v_{x,y}}\bra{v_{x,y}}$ ($\lambda_{x,y}\in\{\pm1\}$) be the spectral decomposition of $P_x$. According to Eq.~(\ref{eq: Case1, M_k, 1}), the estimation procedure for $M_k^{(1)}$ is given by
\begin{equation}\label{apeq: Case1, M_k, 1}
    M_{k}^{(1)} = \sum_{x,y} \mathrm{Tr}\left[ O_{k}\,\mathcal{U}_{k}^{(1)} \left( \ket{v_{x,y}}\bra{v_{x,y}} \otimes \ket{0}\bra{0}^{\otimes (n_k-b_k)} \right) \right] \frac{\lambda_{x,y}}{2^{b_k}} P_x.
\end{equation}
Here, introducing $\mathcal{A}_k^{(1)}(\bullet) = \mathcal{U}_k^{(1)}\left( \bullet \otimes \ket{0}\bra{0}^{\otimes (n_k-b_k)} \right)$ into the above equation, we have
\begin{eqnarray}
    M_{k}^{(1)} 
    = \sum_{l} \mathrm{Tr}\left[ O_{k}\,\mathcal{A}_{k}^{(1)} \left( \omega^{(1),l} \right) \right] \bar{\omega}^{(1),l},
\end{eqnarray}
where we have defined $\omega^{(1),l} := \ket{v_{x,y}}\bra{v_{x,y}}$ and $\bar{\omega}^{(1),l} := \frac{\lambda_{x,y}}{2^{b_k}} P_x$. Then, the set $\{\omega^{(1),l},\bar{\omega}^{(1),l}\}_l$ satisfies
\begin{eqnarray}
    \sum_l \omega^{(1),l} \otimes \bar{\omega}^{(1),l} 
    &=& \sum_{x,y} \ket{v_{x,y}}\bra{v_{x,y}} \otimes \frac{\lambda_{x,y}}{2^{b_k}} P_x\\
    &=& \frac{1}{2^{b_k}} \sum_x P_x \otimes P_x\\
    &=& \mathrm{SWAP},
\end{eqnarray}
which completes the proof in the case of $\xi=1$.

\subsection{Type (ii)}
According to Eq.~(\ref{eq: Case2, M_k, 2}), the estimation procedure for $M_k^{(2)}$ is given by
\begin{equation}
    M_{k}^{(2)} = \sum_{x} \mathrm{Tr}\left[ \left( P_x \otimes O_{k} \right) \, \mathcal{U}_{k}^{(2)}\bigl(\ket{0}\bra{0}^{\otimes (n_k+b_k)}\bigr) \right] \frac{1}{2^{b_k}} P_x^{\rm T}.\label{apeq: Case2, M_k, 2}
\end{equation}
Here, as we will show later, because Eq.~(\ref{eq:trans_a2}) holds, i.e.,
\begin{equation}\label{apeq:trans_a2}
    \mathrm{Tr}\left[\,O_k \, \mathcal{A}_k^{(2)}(P_x^{\rm T})\right]=\mathrm{Tr}\left[ \left( P_x \otimes O_{k} \right) \, \mathcal{U}_{k}^{(2)}\bigl(\ket{0}\bra{0}^{\otimes (n_k+b_k)}\bigr) \right],
\end{equation}
Eq.~(\ref{apeq: Case2, M_k, 2}) can be modified as
\begin{eqnarray}
    M_{k}^{(2)} 
    = \sum_{l} \mathrm{Tr}\left[ O_{k}\,\mathcal{A}_{k}^{(2)} \left( \omega^{(2),l} \right) \right] \bar{\omega}^{(2),l},
\end{eqnarray}
where we have defined $\omega^{(2),l} := P_x^{\rm T}$ and $\bar{\omega}^{(2),l} := P_x^{\rm T}/2^{b_k}$, and the set $\{\omega^{(2),l},\bar{\omega}^{(2),l}\}_l$ satisfies
\begin{equation}
    \sum_l \omega^{(2),l} \otimes \bar{\omega}^{(2),l} 
    = \frac{1}{2^{b_k}} \sum_x P_x^{\rm T} \otimes P_x^{\rm T}\\
    = \mathrm{SWAP},
\end{equation}
implying that the estimation procedure for $M_k^{(2)}$ can be represented in the form of Eq.~(\ref{apeq:estimation_form_m}). 
Therefore, it suffices to show that Eq.~(\ref{eq:trans_a2}) holds. Recalling that $\mathcal{A}_k^{(2)}$ is represented as $\mathcal{A}_k^{(2)}(\bullet)= \mathrm{Tr}_{12}\left[ \ket{\rm Bell}\bra{\rm Bell}_{12}  \bigl( \, \bullet_{1} \, \otimes \ket{\psi_{k}^{(2)}}\bra{\psi_{k}^{(2)}}_{23} \bigr) \right]$, we find that the left-hand side of Eq.~(\ref{eq:trans_a2}) (or Eq.~(\ref{apeq:trans_a2})) is equal to
\begin{equation}\label{apeq:prof_of_eq35_1}
    \mathrm{Tr}\left[\,O_k \, \mathcal{A}_k^{(2)}(P_x^{\rm T})\right] =  \mathrm{Tr}\left[ \Bigl( \ket{\rm Bell}\bra{\rm Bell}_{12} \otimes O_{k,3}\Bigr) \Bigl( \, P_x^{\rm T} \, \otimes \ket{\psi_{k}^{(2)}}\bra{\psi_{k}^{(2)}}_{23} \Bigr) \right].
\end{equation}
Here, by splitting the trace over systems 1, 2, and 3 in the right-hand side of Eq.~(\ref{apeq:prof_of_eq35_1}) into separate traces over system 1 and over systems 2 and 3, we have
\begin{equation}\label{apeq:prof_of_eq35_2}
    \mathrm{Tr}\left[\,O_k \, \mathcal{A}_k^{(2)}(P_x^{\rm T})\right] =  \mathrm{Tr}_{23}\left[ \left( \mathrm{Tr}_{1}\left[ \, \ket{\rm Bell}\bra{\rm Bell}_{12} \, (P_{x,1}^{\rm T}\otimes I_{k,2}) \right] \otimes O_{k,3} \right)  \, \mathcal{U}_{k,23}^{(2)}\bigl(\ket{0}\bra{0}^{\otimes (n_k+b_k)}\bigr) \right].
\end{equation}
Finally, since Lemma~\ref{lemma:partial_bell} leads to
\begin{equation}
    \mathrm{Tr}_{1}\left[ \, \ket{\rm Bell}\bra{\rm Bell}_{12} \, (P_{x,1}^{\rm T}\otimes I_{k,2}) \right] = P_x,
\end{equation}
we have
\begin{equation}\label{apeq:prof_of_eq35_3}
    \mathrm{Tr}\left[\,O_k \, \mathcal{A}_k^{(2)}(P_x^{\rm T})\right] =  \mathrm{Tr}\left[ \left( P_x \otimes O_{k} \right)  \, \mathcal{U}_{k}^{(2)}\bigl(\ket{0}\bra{0}^{\otimes (n_k+b_k)}\bigr) \right],
\end{equation}
which concludes the proof in the case of $\xi=2$.

\subsection{Type (iii)}\label{apsec:a3_2}
The estimation procedure for $M_k^{(3)}$ is given by Eq.~(\ref{eq: Case3, M_k, 2}) as
\begin{equation}\label{apeq: Case3, M_k, 2}
    M_{k}^{(3)} = \sum_{\vec{i}_k,\vec{i}'_k} \mathrm{Tr}\left[ \hat{O}_{k}^{\vec{i}_{k},\vec{i}'_{k}}\, \mathcal{U}_{k}^{(3)} \bigl(\ket{0}\bra{0}^{\otimes n_k}\bigr)\right] \ket{\vec{i_{k}}}\bra{\vec{i}'_{k}}.
\end{equation}
As we will show later, because Eq.~(\ref{eq:trans_a3}) holds, i.e.,
\begin{equation}\label{apeq:trans_a3}
    \mathrm{Tr}\left[\,O_k \, \mathcal{A}_k^{(3)}(\ket{\vec{i}'_k}\bra{\vec{i}_k})\right]
    =\mathrm{Tr}\left[\,\hat{O}_k^{\vec{i}_k,\vec{i}'_k} \, \mathcal{U}_k^{(3)}(\ket{0}\bra{0}^{\otimes n_k})\right],
\end{equation}
Eq.~(\ref{apeq: Case3, M_k, 2}) can be equivalently rewritten as
\begin{equation}
    M_{k}^{(3)} = \sum_{l} \mathrm{Tr}\left[\,O_k \, \mathcal{A}_k^{(3)}(\omega^{(3),l})\right] \bar{\omega}^{(3),l},
\end{equation}
where we have defined $\omega^{(3),l} := \ket{\vec{i}'_k}\bra{\vec{i}_k}$ and $\bar{\omega}^{(3),l} := \ket{\vec{i}_k}\bra{\vec{i}'_k}$. Here, since the set $\{\omega^{(3),l},\bar{\omega}^{(3),l}\}_l$ clearly satisfies $\sum_l \omega^{(3),l} \otimes \bar{\omega}^{(3),l} = \mathrm{SWAP}$, it suffices to prove Eq.~(\ref{eq:trans_a3}) for showing that the estimation procedure for $M_k^{(3)}$ can be repsented in the form of Eq.~(\ref{apeq:estimation_form_m}).
Recalling that $\mathcal{A}_k^{(3)}$ is represented as 
\begin{equation}
    \mathcal{A}_k^{(3)}(\bullet) = 2^{b_k}\,\mathrm{Tr}_{12} \left[ \ket{\rm Bell}\bra{\rm Bell}_{12} \, \Bigl( \,\bullet_{1}\, \otimes \mathcal{CP}_{23}\Bigl(\, \ket{+}\bra{+}^{\otimes b_k}_{2} \otimes \mathcal{U}_k^{(3)}(\ket{0}\bra{0}^{\otimes n_k}) \Bigr) \Bigr) \right],
\end{equation}
we notice that the left-hand side of Eq.~(\ref{eq:trans_a3}) (or Eq.~(\ref{apeq:trans_a3})) is equal to
\begin{equation}\label{apeq:prof_of_eq36_1}
    \mathrm{Tr}\left[\,O_k \, \mathcal{A}_k^{(3)}(\ket{\vec{i}'_k}\bra{\vec{i}_k})\right] = 2^{b_k}\,\mathrm{Tr} \left[ \Bigl(  \ket{\rm Bell}\bra{\rm Bell}_{12} \otimes O_{k,3} \Bigr) \, \Bigl( \,\ket{\vec{i}'_{k}}\bra{\vec{i}_k}_{1}\, \otimes \mathcal{CP}_{23}\Bigl(\, \ket{+}\bra{+}^{\otimes b_k}_{2} \otimes \mathcal{U}_k^{(3)}(\ket{0}\bra{0}^{\otimes n_k}) \Bigr) \Bigr) \right].
\end{equation}
Similar to the calculation in type (iii), by applying the split of the trace as $\mathrm{Tr}_{123}=\mathrm{Tr}_{23}\mathrm{Tr}_{1}$ to the right-hand side of Eq.~(\ref{apeq:prof_of_eq36_1}), we have
\begin{equation}\label{apeq:prof_of_eq36_4}
    \mathrm{Tr}\left[\,O_k \, \mathcal{A}_k^{(3)}(\ket{\vec{i}'_k}\bra{\vec{i}_k})\right] = 2^{b_k}\, \mathrm{Tr}_{23}\left[ \left( \mathrm{Tr}_{1}\left[ \, \ket{\rm Bell}\bra{\rm Bell}_{12} \,(\ket{\vec{i}'_k}\bra{\vec{i}_k} \otimes I_{k,2}) \right] \otimes O_{k,3} \right) \mathcal{CP}_{23}\Bigl(\, \ket{+}\bra{+}^{\otimes b_k}_{2} \otimes \mathcal{U}_k^{(3)}(\ket{0}\bra{0}^{\otimes n_k}) \Bigr) \, \right].
\end{equation}
Using the following fact derived from Lemma~\ref{lemma:partial_bell} that
\begin{equation}
    \mathrm{Tr}_{1}\left[ \, \ket{\rm Bell}\bra{\rm Bell}_{12} \,(\ket{\vec{i}'_k}\bra{\vec{i}_k} \otimes I_{k,2}) \right] = \ket{\vec{i}_k}\bra{\vec{i}'_k},
\end{equation}
Eq.~(\ref{apeq:prof_of_eq36_4}) can be written as
\begin{equation}\label{apeq:prof_of_eq36_2}
    \mathrm{Tr}\left[\,O_k \, \mathcal{A}_k^{(3)}(\ket{\vec{i}'_k}\bra{\vec{i}_k})\right] = 2^{b_k}\, \mathrm{Tr}_{23}\left[ \left( \ket{\vec{i}_k}\bra{\vec{i}'_k} \otimes O_{k,3} \right) \mathcal{CP}_{23}\Bigl(\, \ket{+}\bra{+}^{\otimes b_k}_{2} \otimes \ket{\psi_{k}^{(3)}}\bra{\psi_k^{(3)}}_{3} \Bigr) \, \right].
\end{equation}
Here, by further splitting the trace of the right-hand side of Eq.~(\ref{apeq:prof_of_eq36_2}) as $\mathrm{Tr}_{23}=\mathrm{Tr}_{3}\mathrm{Tr}_{2}$, we have
\begin{eqnarray}\label{apeq:prof_of_eq36_3}
    \mathrm{Tr}\left[\,O_k \, \mathcal{A}_k^{(3)}(\ket{\vec{i}'_k}\bra{\vec{i}_k})\right] 
    &=& \mathrm{Tr}_{3}\left[ O_{k,3}\,2^{b_k}\mathrm{Tr}_{2}\left[ \bigl( \ket{\vec{i}_k}\bra{\vec{i}'_k}_{2} \otimes I_{k,3} \bigr) \, \mathcal{CP}_{23}\Bigl(\, \ket{+}\bra{+}^{\otimes b_k}_{2} \otimes \mathcal{U}_k^{(3)}(\ket{0}\bra{0}^{\otimes n_k}) \Bigr) \, \right]\right]\\
    &=& \mathrm{Tr}\left[\,O_k \, P_k^{\vec{i}'_k}\, \mathcal{U}_k^{(3)}(\ket{0}\bra{0}^{\otimes n_k}) \, P_k^{\vec{i}_k,\dagger} \right]\\
    &=& \mathrm{Tr}\left[\,\hat{O}_k^{\vec{i}_k,\vec{i}'_k} \, \mathcal{U}_k^{(3)}(\ket{0}\bra{0}^{\otimes n_k})\right],
\end{eqnarray}
where the second equality follows from Eq.~(\ref{apeq:type3_2}). Thus, we complete the proof in the case of $\xi=3$.

\subsection{Type (iv)}\label{apsec:a4_2}
By replacing $P_k^{\vec{i}_k}$ with $U_k^{\vec{i}_k}$ and $\ket{\psi_{k}^{(3)}}\bra{\psi_k^{(3)}}$ with $\ket{0}\bra{0}^{\otimes n_k}$ in Sec.~\ref{apsec:a3_2}, we have
\begin{equation}\label{apeq:est_form_a4}
M_{k}^{(4)} = \sum_{l} \mathrm{Tr}\left[\,O_k \, \mathcal{A}_k^{(4)}(\omega^{(4),l})\right] \bar{\omega}^{(4),l},
\end{equation}
where we have defined $\omega^{(4),l} := \ket{\vec{i}'_k}\bra{\vec{i}_k}$ and $\bar{\omega}^{(4),l} := \ket{\vec{i}_k}\bra{\vec{i}'_k}$, thus completing the proof for the case $\xi=4$.

We also establish the equivalence between the quantum circuits in Fig.~\ref{Fig:exp_map_relation}(e1-1)/(e1-2) and the quantum circuit in Fig.~\ref{Fig:exp_map_relation}(e2). This equivalence can be readily verified by comparing Eq.~(\ref{apeq:est_form_a4}) and Eq.~(\ref{eq: Case4, M_k, 2}), which represents the estimation procedure for $M_k^{(4)}$. As a result, we obtain
\begin{eqnarray}
    \mathrm{Tr}[O_k \,\mathcal{U}_{k}^{(4),\vec{i}_{k}}(\ket{0}\bra{0}^{\otimes n_k})] &=&
    \mathrm{Tr}\left[\,O_k \, \mathcal{A}_k^{(4)}(\ket{\vec{i}_k}\bra{\vec{i}_k})\right],\label{eq:equiv_type4_1}\\
    \sum_{\theta=0,\frac{\pi}{2}} \mathrm{Tr}[(X\otimes O_k)\, \mathcal{C}_k^{\vec{i}_k,\vec{i}'_k} (\ket{\psi(\theta)}\bra{\psi(\theta)})] &=& \mathrm{Tr}\left[\,O_k \, \mathcal{A}_k^{(4)}(\ket{\vec{i}'_k}\bra{\vec{i}_k})\right],\label{eq:equiv_type4_2}
\end{eqnarray}
where the left-hand sides of Eqs.~(\ref{eq:equiv_type4_1}) and (\ref{eq:equiv_type4_2}) correspond to the quantum circuits in Fig.~\ref{Fig:exp_map_relation}(e1-1) and (e1-2), respectively, while the right-hand sides correspond to the quantum circuits in Fig.~\ref{Fig:exp_map_relation}(e2).

\section{Proof of the noisy expansion map in types (i) -- (iii) being a completely positive map}\label{sec:physicality_app}

 This section shows that each expansion map for types (i) -- (iii) is a completely positive (CP) map. To simplify the notation, we omit the subscript $k$, which denotes the index of the subsystem in the following subsections.

\subsection{Type (i)}

The noise-affected expansion map $\tilde{\mathcal{A}}^{(1)}:\mathrm{L}(\mathbb{C}^{d_{\rm in}})\rightarrow \mathrm{L}(\mathbb{C}^{d_{\rm out}})$ ($d_{\rm in}:=2^b$ and $d_{\rm out}:=2^n$) of type (i) is represented as 
\begin{equation}
        \tilde{\mathcal{A}}^{(1)}(\bullet) = \mathcal{W}^{(1)}(\,\bullet\, \otimes \,\ket{0}\bra{0}^{\otimes (n-b)}\,).
\end{equation}
Since the map is clearly a quantum channel, $\tilde{\mathcal{A}}^{(1)}(\bullet)$ satisfies the property of CP.

\subsection{Type (ii)}

The noise-affected expansion map $\tilde{\mathcal{A}}^{(2)}:\mathrm{L}(\mathbb{C}^{d_{\rm in}})\rightarrow \mathrm{L}(\mathbb{C}^{d_{\rm out}})$ ($d_{\rm in}:=2^b$ and $d_{\rm out}:=2^n$) of type (ii) is represented as 
\begin{equation}
        \tilde{\mathcal{A}}^{(2)}_{1\rightarrow 3}(\bullet) = \mathrm{Tr}_{12} \left[ \ket{\rm Bell}\bra{\rm Bell}_{12} \left( \bullet_1 \otimes \sigma_{23}^{(2)}\right) \right],
\end{equation}
where $\ket{\rm Bell}_{12}:=\sum_{\vec{i}\in\{0,1\}^b} \ket{\vec{i}}_1 \otimes \ket{\vec{i}}_{2}$ is an unnormalized $2b$-qubit Bell state on systems 1 and 2.
Note that we add the subscript $1\rightarrow3$ to $\tilde{\mathcal{A}}^{(2)}$ to emphasize that the input system is 1 and the output system is 3. 
To prove the map $\tilde{\mathcal{A}}^{(2)}$ is a CP map, it suffices to show that the following condition: 
\begin{equation}\label{eq:eq_condition_2}
    (\tilde{\mathcal{A}}_{1\rightarrow3}^{(2)}\otimes \mathrm{Id}_{4\rightarrow4})(\ket{\rm Bell}\bra{\rm Bell}_{14}) \geq 0
\end{equation}
where $\mathrm{Id}_{4\rightarrow4}$ is the identity channel from system 4 to 4.
By calculating the left-hand side of Eq.~(\ref{eq:eq_condition_2}), we have
\begin{eqnarray}
    (\tilde{\mathcal{A}}_{1\rightarrow3}^{(2)}\otimes \mathrm{Id}_{4\rightarrow4})(\ket{\rm Bell}\bra{\rm Bell}_{14}) 
    &=& \mathrm{Tr}_{12}\left[ \left( \ket{\rm Bell}\bra{\rm Bell}_{12} \otimes I_{34} \right) \left( \ket{\rm Bell}\bra{\rm Bell}_{14} \otimes \sigma_{23}^{(2)} \right) \right] \\
    &=& \sum_{\vec{i},\vec{j},\vec{k},\vec{l}} \mathrm{Tr}_{12}\left[ \left( \ket{\vec i}\bra{\vec j}_{1} \otimes \ket{\vec{i}}\bra{\vec{j}}_{2} \otimes I_{34} \right) \left( \ket{\vec{k}}\bra{\vec{l}}_{1} \otimes \ket{\vec{k}}\bra{\vec{l}}_{4} \otimes \sigma_{23}^{(2)} \right) \right] \\
    &=& \sum_{\vec{i},\vec{j}, \vec{k},\vec{l}} \braket{\vec{j}|\vec{k}} \braket{\vec{l}|\vec{i}} \, \mathrm{Tr}_{2}\left[ \left( \ket{\vec{i}}\bra{\vec{j}}_{2} \otimes I_{3} \right)\, \sigma_{23}^{(2)}  \right] \ket{\vec k}\bra{\vec l}_{4}
    \\
    &=& \sum_{\vec{i},\vec{j}} \bra{j}_{2}\, \sigma_{23}^{(2)} \,\ket{\vec{i}}_{2} \ket{\vec{j}}\bra{\vec{i}}_{4}\\
    &=& \sigma_{43}^{(2)},
\end{eqnarray}
where the third line follows by taking the partial trace of system 1. From the noise assumption in Sec.~\ref{sec:2-layer}, $\sigma_{43}^{(2)}$ is a density operator and thereby $\sigma_{43}^{(2)}\geq 0$. Thus, the condition for  $\tilde{\mathcal{A}}^{(2)}$ to be a CP map is satisfied.

\subsection{Type (iii)}

The noise-affected expansion map $\tilde{\mathcal{A}}^{(3)}:\mathrm{L}(\mathbb{C}^{d_{\rm in}})\rightarrow \mathrm{L}(\mathbb{C}^{d_{\rm out}})$ ($d_{\rm in}=2^b$ and $d_{\rm out}=2^n$) of type (iii) is represented as 
\begin{equation}
    \tilde{A}^{(3)}_{1\rightarrow3}(\bullet) 
    = 2^{b}\,\mathrm{Tr}_{12} \left[ \ket{\rm Bell}\bra{\rm Bell}_{12} \, \left(\bullet_{1} \, \otimes \mathcal{CP}_{23}\Bigl(\, \ket{+}\bra{+}^{\otimes b}_{2} \otimes \sigma_{3}^{(3)} \Bigr) \right)  \right],
\end{equation}
where $\sigma_{2}^{(3)}$ is an $n$-qubit quantum state acting on system 3. Similar to the proof in the previous subsection, we show that $\tilde{\mathcal{A}}^{(3)}$ is a CP map by proving 
\begin{equation}\label{eq:eq_condition_3}
    (\tilde{\mathcal{A}}_{1\rightarrow3}^{(3)}\otimes \mathrm{Id}_{4\rightarrow4})(\ket{\rm Bell}\bra{\rm Bell}_{14}) \geq 0.
\end{equation}
The left-hand side of Eq.~(\ref{eq:eq_condition_3}) is equal to 
\begin{eqnarray}
    (\tilde{\mathcal{A}}_{1\rightarrow3}^{(3)}\otimes \mathrm{Id}_{4\rightarrow4})(\ket{\rm Bell}\bra{\rm Bell}_{14}) 
    &=& 2^{b}\,\mathrm{Tr}_{12} \left[ \left( \ket{\rm Bell}\bra{\rm Bell}_{12} \otimes I_{34} \right) \, \left(\ket{\rm Bell}\bra{\rm Bell}_{14} \, \otimes \mathcal{CP}_{23}\Bigl(\, \ket{+}\bra{+}^{\otimes b}_{2} \otimes \sigma_{3}^{(3)} \Bigr) \right)  \right]\\
    &=& \sum_{\vec{i},\vec{j},\vec{k},\vec{l}} 2^{b}\,\mathrm{Tr}_{12} \left[ \left( \ket{\vec i}\bra{\vec j}_{1} \otimes \ket{\vec{i}}\bra{\vec{j}}_{2} \otimes I_{34} \right) \, \left(\ket{\vec k}\bra{\vec l}_{1} \otimes \ket{\vec{k}}\bra{\vec{l}}_{4} \, \otimes \mathcal{CP}_{23}\Bigl(\, \ket{+}\bra{+}^{\otimes b}_{2} \otimes \sigma_{3}^{(3)} \Bigr) \right)  \right]\notag\\
    &\\
    &=& \sum_{\vec{i},\vec{j}} 2^{b}\,\mathrm{Tr}_{2} \left[  \ket{\vec{i}}\bra{\vec{j}}_{2} \,  \mathcal{CP}_{23}\Bigl(\, \ket{+}\bra{+}^{\otimes b}_{2} \otimes \sigma_{3}^{(3)} \Bigr)  \right]\ket{\vec{j}}\bra{\vec{i}}_{4}\\
    &=& 2^{b}\, \mathcal{CP}_{43}\Bigl(\, \ket{+}\bra{+}^{\otimes b}_{4} \otimes \sigma_{3}^{(3)} \Bigr) \geq 0,\label{eq:eq_proof_3}
\end{eqnarray}
where the third line comes from $\vec{i}=\vec{l}$ and $\vec{j}=\vec{k}$.
Thus, $\tilde{\mathcal{A}}^{(3)}$ is a CP map.

\subsection{Note on the relation between types (iii) and (iv)}

By setting $P^{\vec{i}}$ as general unitary operators $U^{\vec{i}}$ and $\ket{\psi_k^{(3)}}$ as $\ket{0}^{\otimes n}$, i.e.,
\begin{equation}
    \ket{\psi^{\vec{i}}}=P^{\vec{i}} \ket{\psi^{(3)}} \rightarrow U^{\vec{i}}\ket{0}^{\otimes n},
\end{equation}
it is also possible to construct the same ansatz as in type (iv), and the noise-affected expansion map becomes a CP map. It seems to contradict the argument in Sec.~\ref{sec:2-layer} that the noisy expansion map in type (iv) is not a CP map in general.
However, there exists an essential assumption for using the estimation procedure in type (iii) that $\hat{O}^{\vec{i},\vec{j}}=P^{\vec{i},\dagger}O P^{\vec{j}}$ (in the case of $P^{\vec{i}}=U^{\vec{i}}$, it is $U^{\vec{i},\dagger}O U^{\vec{j}}$) can be decomposed into a linear combination of Pauli operators through classical computations. This assumption is reasonable when $P^{\vec{i}}$ and $O$ are both Pauli strings. However, such decomposition is challenging when $P^{\vec{i}}$ are general unitary operators, becoming the major obstacle to applying the estimation method of type (iii), which allows for retaining the physicality even under the influence of noise, to the estimation in type (iv). Instead, we can choose a more practical estimation method for type (iv) described in Sec.~\ref{sec: 2d}, but the expansion map may lose its CP property due to the noise.

\section{Unphysicality in types (iv)}\label{sec:unphysical_in_type_4}

\begin{figure}[tp]
\begin{center}
\centering
 \includegraphics[width=85mm]{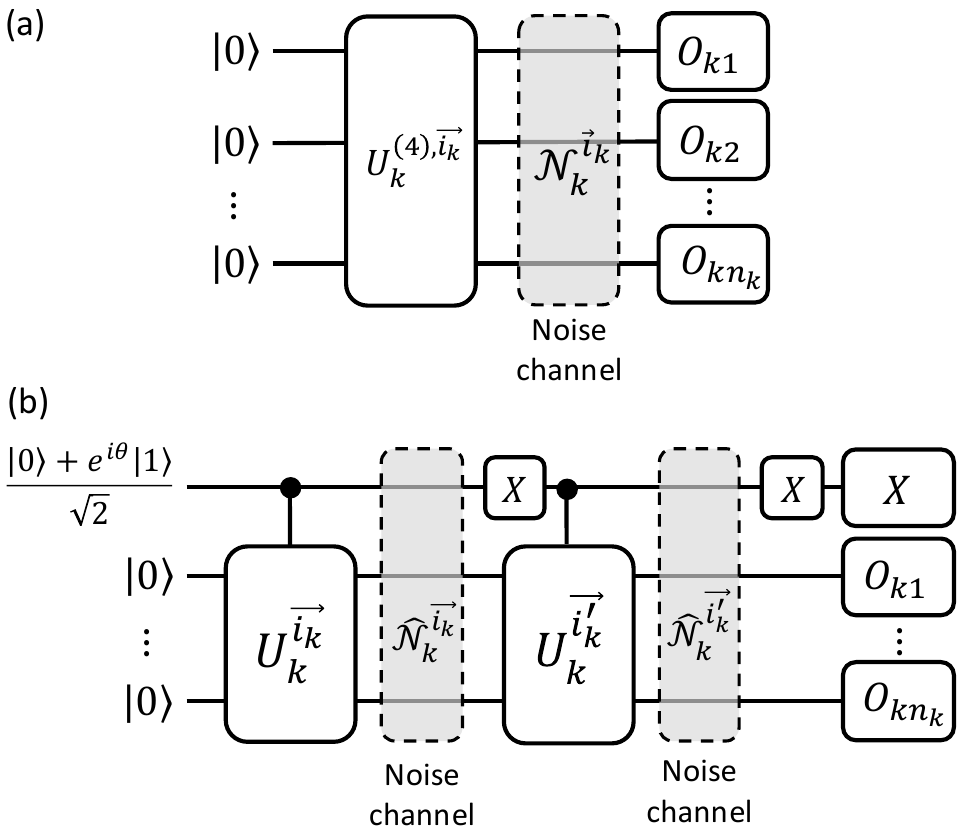}    
\end{center}
\caption{\label{fig: noise model in case iv}Noise models for the analysis in type (iv). (a) In the circuit to calculate the diagonal element $M_{k}^{\vec{i}_{k},\vec{i}_{k}}(S_{k}^{\vec{i}_{k},\vec{i}_{k}})$, we consider the noise model where the global depolarizing channel $\mathcal{N}_{k}^{\vec{i}_{k}}$ acts just after the unitary gate $U_{k}^{\vec{i}_{k}}$. (b) In the circuit to calculate the off-diagonal element $M_{k}^{\vec{i}_{k},\vec{i}'_{k}}(S_{k}^{\vec{i}_{k},\vec{i}'_{k}})$, we consider the noise model where the global depolarizing channel $\widehat{\mathcal{N}}_{k}^{\vec{i}_{k}}$ and $\widehat{\mathcal{N}}_{k}^{\vec{i}'_{k}}$ act just after the controlled-$U_{k}^{\vec{i}_{k}}$ gate and the controlled-${U}_{k}^{\vec{i}'_{k}}$ gate, respectively.}
\end{figure}

In type (iv), if we naively perform the contraction of local subsystems in the manner introduced in Sec.~\ref{sec: 2d}, the noisy HTNs state $\tilde{\rho}_{\rm HT}^{(4)}$ does not always become a positive operator in general. To observe this, we show that $\tilde{\rho}_{\rm HT}^{(4)}$ could have negative eigenvalues under the assumption of the simple noise model and observable.
As seen in Sec.~\ref{sec: 2d}, we need two types of quantum circuits to contract the local observables: the circuits to calculate the diagonal elements of $M_{k}^{(4)}$($S_{k}^{(4)}$) in Fig.~\ref{fig: Quantum circuits for M_k} (d1) and circuits to calculate the off-diagonal elements of $M_{k}^{(4)}$($S_{k}^{(4)}$) in Fig.~\ref{fig: Quantum circuits for M_k} (d2). 
 We set the noise model for two types of quantum circuits as follows. For the circuit to calculate the diagonal elements of $M_{k}^{(4)}$($S_{k}^{(4)}$), we consider the noise model where the global depolarizing noise acts just after the unitary channel $\mathcal{U}_{k}^{(4),\vec{i}_{k}}$. We consider the corresponding depolarizing noise channel $\mathcal{N}_{k}^{\vec{i}_{k}}$ as follows:
\begin{equation}
    \mathcal{N}_{k}^{\vec{i}_{k}}(\sigma_{\rm in}) = p_{\,\vec{i}_{k}}\frac{I^{\otimes n_{k}}}{2^{n_{k}}} + (1-p_{\,\vec{i}_{k}})~\sigma_{\rm in},
\end{equation}
\color{black}
 where $p_{\,\vec{i}_{k}}$ are noise rates that take $0\leq p_{\,\vec{i}_{k}} \leq 1$, and $\sigma_{\rm in}$ is an input state. For the circuits to estimate the off-diagonal elements of $M_{k}^{(4)}$($S_{k}^{(4)}$), we consider the noise model where the global depolarizing noise acts after each controlled-$U_{k}^{\vec{i}_{k}}$ operation. We assume that the corresponding noise channel $\widehat{\mathcal{N}}_{k}^{\vec{i}_{k}}$ operates as
\begin{equation}
    \widehat{\mathcal{N}}_{k}^{\,\vec{i}_{k}}(\sigma_{\rm in}) = q_{\,\vec{i}_{k}}\frac{I^{\otimes (n_{k}+1)}}{2^{(n_{k}+1)}} + (1-q_{\,\vec{i}_{k}})~\sigma_{\rm in},
\end{equation}
\color{black}
 where $q_{\,\vec{i}_{k}}$ are noise rates that take $0\leq q_{\,\vec{i}_{k}} \leq 1$, and $\sigma_{\rm in}$ is an input state. Noise models for these two types of quantum circuits to calculate $M_{k}^{(4)}$($S_{k}^{(4)}$) are graphically depicted in Fig.~\ref{fig: noise model in case iv}. 
 For the observable $O$, we consider the case where $O = \otimes_{k=1}^{N} O_{k}$ is a Pauli string composed of a tensor product of subsystem's observables $O_{k}$, and $O_{k} \in \{I,X,Y,Z\}^{\otimes n_{k}}\setminus I^{\otimes n_{k}}$ are Pauli strings excluding $I^{\otimes n_{k}}$.

 Then, by simple calculations, the noisy Hermitian operators $\tilde{M}_{k}$ and $\tilde{S}_{k}$ can be represented as $\tilde{M}_{k}^{\vec{i}_{k},\vec{j}_{k}}=r_{k}^{\vec{i}_{k},\vec{j}_{k}} M_{k}^{\vec{i}_{k},\vec{j}_{k}}$ and $\tilde{S}_{k}^{\vec{i}_{k},\vec{j}_{k}}=s_{k}^{\vec{i}_{k},\vec{j}_{k}} S_{k}^{\vec{i}_{k},\vec{j}_{k}}$ where $s_{k}^{\vec{i}_{k},\vec{j}_k}$ and $r_{k}^{\vec{i}_{k},\vec{j}_k}$ are respectively defined as 
\begin{equation}\label{apeq:r}
    r_{k}^{\vec{i}_{k},\vec{j}_{k}} = 
    \begin{cases}
        1-p_{\,\vec{i}_{k}}, & \vec{i}_{k} = \vec{j}_{k}, \\
        (1-q_{\,\vec{i}_{k}})(1-q_{\,\vec{j}_{k}}), & \vec{i}_{k} \neq \vec{j}_{k},
    \end{cases}
\end{equation}
and
\begin{equation}\label{apeq:s}
    s_{k}^{\vec{i}_{k},\vec{j}_{k}} = 
    \begin{cases}
        1, & \vec{i}_{k} = \vec{j}_{k}, \\
        (1-q_{\,\vec{i}_{k}})(1-q_{\,\vec{j}_{k}}), & \vec{i}_{k} \neq \vec{j}_{k}.
    \end{cases}
\end{equation}

Here, for an arbitrary complex matrix $E:=\sum_{\vec{i}_k,\vec{j}_{k}} e^{\vec{i}_k,\vec{j}_k} \ket{\vec{i}_k}\bra{\vec{j}_k}$ of the same size as $M_k^{(4)}$ and $S_k^{(4)}$, let us define the maps $\Gamma_{k}(E)$ and $\Lambda_{k}(E)$, which scale the matrix elements $e^{\vec{i}_k,\vec{j}_k}$ by the factors $r_k^{\vec{i}_k,\vec{j}_k}$ and $s_k^{\vec{i}_k,\vec{j}_k}$, as follows:
\begin{equation}\label{eq:shrink_map}
    \Gamma_{k}\left(E \right) := \sum_{\vec{i}_{k},\vec{j}_{k}} r_{k}^{\vec{i}_{k},\vec{j}_{k}} \,e^{\vec{i}_{k},\vec{j}_{k}} \,\ket{\vec{i}_{k}}\bra{\vec{j}_{k}}, 
\end{equation}
and
\begin{equation}
    \Lambda_{k}\left(E \right) := \sum_{\vec{i}_{k},\vec{j}_{k}} s_{k}^{\vec{i}_{k},\vec{j}_{k}} \,e^{\vec{i}_{k},\vec{j}_{k}} \,\ket{\vec{i}_{k}}\bra{\vec{j}_{k}},
\end{equation}
 where $e^{\vec{i}_{k},\vec{j}_{k}}$ are arbitrary complex numbers for all $\vec{i}_{k}$ and $\vec{j}_{k}$. It is important to remark that the functions $\Gamma_{k}$ and $\Lambda_{k}$ transform the operator $M_{k}^{(4)}$ and $S_{k}^{(4)}$ into noisy ones $\tilde{M}_{k}^{(4)}=\Gamma_{k}(M_{k}^{(4)})$ and $\tilde{S}_{k}^{(4)}=\Lambda_{k}(S_{k}^{(4)})$, respectively. Introducing $\Gamma_{k}$ and $\Lambda_{k} ~(k=1,...,N)$ into Eq.~(\ref{eq: Noisy effective hamiltonian in Op-based rep}), we have
\begin{eqnarray}
    \braket{\tilde{O}}_{\rho_{\rm HT}^{(4)}} 
    &=& \frac{\mathrm{Tr}[\,\bigotimes_{k=1}^{N}\Gamma_{k}(M_{k}^{(4)})~\rho~]}{\,\mathrm{Tr}[\,\bigotimes_{k=1}^{N}\Lambda_{k}(S_{k}^{(4)})~\rho~]}\\
    &=& \frac{\mathrm{Tr}[\,\bigotimes_{k=1}^{N} M_{k}^{(4)}~ \bigotimes_{k=1}^{N}\Gamma_k(\rho)~]}{\mathrm{Tr}[\,\bigotimes_{k=1}^{N} N_{k}^{(4)}~ \bigotimes_{k=1}^{N}\Lambda_k(\rho)~]}\\
    &=& \mathrm{Tr}\left[ O \cdot \frac{ (\bigotimes_{k=1}^{N} A_{k}^{(4)})\bigotimes_{k=1}^{N} \Gamma_{k}(\rho) (\bigotimes_{k=1}^{N} A_{k}^{(4)})^{\dagger}}{(\bigotimes_{k=1}^{N} A_{k}^{(4)})\bigotimes_{k=1}^{N} \Lambda_{k}(\rho) (\bigotimes_{k=1}^{N} A_{k}^{(4)})^{\dagger}} \right],
\end{eqnarray}
where the second equation follows by making use of the adjoint map of $\Gamma_k$ and $\Lambda_k$ and the fact that $\Gamma_k^{\dagger}=\Gamma_k$ and $\Lambda_k^{\dagger}=\Lambda_k$, as Eqs.~(\ref{apeq:r}) and (\ref{apeq:s}) establish that $r_{k}^{\vec{i}_k,\vec{j}_k}=r_{k}^{\vec{j}_k,\vec{i}_k}$ and $s_{k}^{\vec{i}_k,\vec{j}_k}=s_{k}^{\vec{j}_k,\vec{i}_k}$. The last equality uses $M_k^{(4)}=A_k^{(4),\dagger}O_kA_k^{(4)}$ and $S_k^{(4)}=A_k^{(4),\dagger}A_k^{(4)}$
 Regarding the component separated from the observable $O$ as effective density operator $\tilde{\rho}_{\rm HT}^{(4)}$, we obtain
\begin{equation}
    \tilde{\rho}_{\rm HT}^{(4)} = \frac{ (\bigotimes_{k=1}^{N} A_{k}^{(4)})\bigotimes_{k=1}^{N} \Gamma_{k}(\rho) (\bigotimes_{k=1}^{N} A_{k}^{(4)})^{\dagger}}{\mathrm{Tr}\left[(\bigotimes_{k=1}^{N} A_{k}^{(4)})\bigotimes_{k=1}^{N} \Lambda_{k}(\rho) (\bigotimes_{k=1}^{N} A_{k}^{(4)})^{\dagger}\right]}.
\end{equation}
Here, we show that the map $\Gamma_k$ defined in Eq.~(\ref{eq:shrink_map}) is not necessarily a positive map. Suppose the following noise model, where $1-p_{\vec{i}_k}=p'$, $(1-q_{\,\vec{i}_{k}})(1-q_{\,\vec{j}_{k}})=q'$ for any $\vec{i}_k, \vec{j}_k$, and $0\leq p'<q'\leq1$. Then, consider the quantum state $\bar{\rho}$ given by $\bar{\rho}:=(I^{\otimes b_{k}}+X^{\otimes b_k})/2^{b_{k}}$, the map $\Gamma_k$ transforms the input state $\bar{\rho}$ into 
\begin{equation}
    \Gamma_k(\bar{\rho}) = \frac{1}{2^{b_k}}\left( p' I^{\otimes b_k} + q' X^{\otimes b_k} \right),
\end{equation} 
and it is straightforward to observe that the operator $\Gamma_k(\bar{\rho})$ has a negative eigenvalue, by $\bra{-}^{\otimes b_k} \Gamma_k(\bar{\rho}) \ket{-}^{\otimes b_k}=(p'-q')/2^{b_k}<0$. Thus, $\Gamma_{k}$ may not be a positive map, depending on the magnitude of the noise rates.

From the above discussion, $\otimes_{k=1}^{N} \Gamma_{k}(\rho)$ is not always a positive-semidefinite operator, in general, implying that $\tilde{\rho}_{\rm HT}$ could have negative eigenvalues. 

\section{Effects of measurement noise and statistical noise on expectation calculation in HTN}\label{sec:effect_of_meas_and_stat_noise}

In this supplementary note, we discuss the effect of measurement and statistical noise on the computational results in HTNs. In Sec.~\ref{sec:measurement_noise}, we first present a measurement noise model we consider, and in Sec.~\ref{sec:noise_model}, we discuss the physicality of the expansion map in each type under the assumption. In Sec.~\ref{sec:noise_analysis}, we consider the effect of statistical noise, analyzing its impact on expectation value calculations in HTTNs from the perspective of the bias introduced to the estimators rather than the physicality of the state. For related discussions, refer to Appendix~B3 in Ref.~\cite{yuan2021quantum} and Section~2A in Ref.~\cite{fujii2022deep}.

\subsection{The effect of measurement noise}\label{sec:measurement_noise}

\subsubsection{Measurement noise model}\label{sec:noise_model}

First, we begin by modeling the measurement noise on a computational basis.
As a model for the measurement noise, we consider the following situation: when performing a projective measurement in the computational basis on each qubit, the ideal POVM $\{E_{0},E_{1}\}$ is transformed into a different POVM $\{\tilde{E}_{0}, \tilde{E}_{1}\}$ defined as follows:
\begin{equation}\label{eq:change_of_povm}
\begin{aligned}
      E_{0}&=\ket{0}\bra{0}&\rightarrow~~~~~~ \tilde{E}_{0}&:=p_{00}\ket{0}\bra{0}+p_{01}\ket{1}\bra{1},\\
      E_{1}&=\ket{1}\bra{1}&\rightarrow~~~~~~ \tilde{E}_{1}&:=p_{10}\ket{0}\bra{0}+p_{11}\ket{1}\bra{1},
\end{aligned}
\end{equation}
where $p_{ij}$($i,j\in\{0,1\}$) are the transition probabilities from the true value $i$ to a measured one $j$ and $p_{00}+p_{01}=p_{10}+p_{11}=1$. Now, introducing the noise channel $\mathcal{N}(\bullet)=\sum_{i=0}^{3}K_{i}\bullet K_{i}^{\dagger}$ where the Kraus operators are defined as $K_{0}= \sqrt{p_{00}} \ket{0}\bra{0},~K_{1}= \sqrt{1-p_{00}} \ket{1}\bra{0},~K_{2}= \sqrt{1-p_{11}} \ket{0}\bra{1},~K_{3}= \sqrt{p_{11}} \ket{1}\bra{1}$,
each altered POVM element $\tilde{E}_{i}$ can be represented as $\tilde{E}_{i}=\mathcal{N}(E_{i})$. Thus, using the expression of the dual map $\mathcal
{N}^{\dagger}$ of $\mathcal{N}$, the probability of obtaining the outcome $i$ with the noise-affected POVM $\{\tilde{E}_{i}\}$ for some input state $\rho$ can be expressed as:
\begin{equation}
    \mathrm{Tr}[ \tilde{E}_{i} \rho ] = \mathrm{Tr}[E_i \mathcal{N}^{\dagger}(\rho)].
\end{equation}
This equivalence implies that performing a measurement with the noise-affected POVM $\{\tilde{E}_{i}\}$ can be equivalently regarded as performing the ideal computational-basis measurement with $\{E_i\}$ just after the action of the channel $\mathcal{N}^{\dagger}$. We use this transformation to analyze the effect of the measurement noise in Sec.~\ref{sec:noise_analysis}.

Also, we emphasize that the channel $\mathcal{N}$ is not necessarily a quantum operation because the condition
\begin{equation}
    \sum_{i=0}^{3} K_i K_i^{\dagger}=\begin{bmatrix}
   1+(p_{00}-p_{11}) & 0 \\
   0 & 1-(p_{00}-p_{11})
\end{bmatrix} \leq I
\end{equation}
does not holds when $p_{00} \neq p_{11}$. However, if we assume that the noise channel $\mathcal{N}$ is unital (as in the measurement noise model considered in Ref.\cite{wang2021noise}), it imposes additional constraints $\tilde{E}_{0}+\tilde{E}_{1}=I$, implying $p_{00}+p_{10}=p_{01}+p_{11}=1$ and thereby $p_{00}= p_{11}$. In this case, the change in the POVM in Eq.~(\ref{eq:change_of_povm}) can be rewritten using a single noise parameter $0\leq p \leq1$ as
\begin{equation}\label{eq:change_of_povm_2}
\begin{aligned}
      E_{0}&=\ket{0}\bra{0},&\rightarrow~~~~~~ \tilde{E}_{0}&:=(1-p)\ket{0}\bra{0}+p\ket{1}\bra{1},\\
      E_{1}&=\ket{1}\bra{1},&\rightarrow~~~~~~ \tilde{E}_{1}&:=p\ket{1}\bra{1}+(1-p)\ket{1}\bra{1},
\end{aligned}
\end{equation}
and the noise channel $\mathcal{N}$ and its adjoint map $\mathcal{N}^{\dagger}$ can be described as a quantum bit-flip channel:
\begin{equation}
    \mathcal{N}(\rho) = \mathcal{N}^{\dagger}(\rho) = (1-p)\rho + p X\rho X,
\end{equation}
which implies the map $\mathcal{N}$ is a quantum operation. 

The measurement noise model discussed above focuses solely on measurements on a computational basis; however, the contraction of local tensors in HTNs requires calculating the expectation value of general Hermitian observables. Therefore, we now generalize the measurement noise model for $Z$-basis to the noise model for the measurement of an $n$-qubit Hermitian observable $O$. Before providing the noise assumption, we model the computational procedure for calculating the expectation value of $O$. We assume that we have the spectral decomposition of $O$: $O=V^{\dagger}\Lambda V$ where $V$ is a unitary operator and $\Lambda$ is a diagonal matrix. Using the spectral decomposition, we measure the expectation value of $O$ as follows: we apply the unitary $V$, perform the computational-basis measurement, and then assign the elements of $\Lambda$ to the measurement outcomes. Under this computational model, we model the measurement noise by extending the discussion at the beginning of this section: a noise channel 
\begin{equation}
    \mathcal{\bar{N}}^{\dagger}:=\mathcal{N}^{\dagger}_{1}\otimes\mathcal{N}_{2}^{\dagger}\otimes ... \otimes \mathcal{N}^{\dagger}_{n}
\end{equation}
acts immediately before the computational-basis measurement, where the subscript indicates the qubit to which the channel $\mathcal{N}^{\dagger}$ is applied. In addition to the noise assumption, we assume that the unitary rotation $V$ is performed incompletely due to some coherent noise. To model this assumption mathematically, we express the applied unitary as $\tilde{V}$ instead of $V$.

Under this noise model, the expression for the expectation value of $O$ for an arbitrary $n$-qubit quantum state $\rho$ changes as follows:
\begin{equation}
    \text{(Without noise)}~~\mathrm{Tr}[\Lambda\mathcal{V}(\rho)]~~~~\rightarrow ~~~~\text{(With noise)}~~\mathrm{Tr}[\Lambda ( \bar{\mathcal{N}} \circ \mathcal{\tilde V})(\rho)]~~~~
\end{equation}
where we have defined $\mathcal{V}(\bullet):=V\bullet V^{\dagger}$ and $\mathcal{\bar V}(\bullet):=\tilde{V}\bullet \tilde{V}^{\dagger}$.
To simplify the expression with noise assumption, we define a new channel $\hat{\mathcal{N}}_{\mathcal{V}, \mathcal{\tilde{V}}}:=\mathcal{V}^{\dagger} \circ \mathcal{\bar N}^{\dagger} \circ \bar{\mathcal{V}}$ where $\mathcal{V}^{\dagger}(\bullet):=V^{\dagger}\bullet V$. Then, it is straightforward to verify that:
\begin{equation}
    \mathcal{\bar N}^{\dagger} \circ \mathcal{\tilde{V}} = \mathcal{V} \circ \hat{\mathcal{N}}_{\mathcal{V}, \mathcal{\tilde{V}}}.
\end{equation}
Using the transformation above, the expectation value under the noise model can be expressed as:
\begin{equation}\label{eq:noise_trans}
    \mathrm{Tr}[\Lambda(\mathcal{\bar N}^{\dagger} \circ \mathcal{\tilde{V}})(\rho)] = \mathrm{Tr}[ \Lambda (\mathcal{V} \circ \hat{\mathcal{N}}_{\mathcal{V}, \mathcal{\tilde{V}}})(\rho) ] = \mathrm{Tr}[ O \hat{\mathcal{N}}_{\mathcal{V}, \mathcal{\tilde{V}}}(\rho) ].
\end{equation}
Thus, the assumptions of the measurement noise occurring at the stage of the computational-basis measurement and coherent noise, described by $\tilde{V}$, can be equivalently modeled as the action of the channel $\hat{\mathcal{N}}_{\mathcal{V}, \mathcal{\tilde{V}}}$ immediately before the ideal measurement of $O$. We also note that the channel $\hat{\mathcal{N}}_{\mathcal{V}, \mathcal{\tilde{V}}}$ is not guaranteed to be a quantum channel. Similar to the discussion of $\mathcal{N}^{\dagger}$, if $\mathcal{N}_{i}$ is unital for all $i$, then $\hat{\mathcal{N}}_{\mathcal{V}, \mathcal{\tilde{V}}}$ also becomes a valid quantum channel. 

\subsubsection{Noise analysis in type (i) - (iii)}\label{sec:noise_analysis}

We now apply the assumption of the measurement noise in Sec.~\ref{sec:noise_model} to the noise analysis in type (i) - (iii) in Sec.~\ref{sec:2-layer}. In the main text, physical noise in quantum circuits is modeled under the assumption that the ideal unitary channel $\mathcal{U}_{k}^{(i)}$ is transformed into a different quantum channel $\mathcal{W}^{(i)}_{k}$. In addition to the noise assumption, we apply the measurement noise model introduced in Sec.~\ref{sec:measurement_noise}. We consider the spectral decomposition of $O_k$ as $O_k=V_k \lambda_k V_k^{\dagger}$ and the application of the noise channel $\mathcal{\bar{N}}^{\dagger}:=\mathcal{N}^{\dagger}_{1}\otimes\mathcal{N}_{2}^{\dagger}\otimes ... \otimes \mathcal{N}^{\dagger}_{n}$ immediately before the computational basis measurement, and define the transformation of a unitary channel as $\mathcal{V}_k(\bullet)=V_k \bullet V_k^{\dagger}\rightarrow \mathcal{\tilde{V}}_k(\bullet)=\tilde{V}_k\bullet \tilde{V}_k^{\dagger}$ due to the coherence noise. 

\vspace{1\baselineskip}
\noindent\textbf{Type(i)~~} From the transformation of Eq.~(\ref{eq:noise_trans}), the effect of measurement noise can be mapped to the noise channel $\hat{\mathcal{N}}_{\mathcal{V}_k, \mathcal{\tilde{V}}_k}:=\mathcal{V}_k^{\dagger} \circ \mathcal{\bar N}^{\dagger} \circ \tilde{\mathcal{V}}_k$ on the quantum circuits in Fig.~\ref{fig: Quantum circuits for M_k}(a) where $\mathcal{V}_k^{\dagger}(\bullet):=V_k^{\dagger}\bullet V_k$. Combing this with the discussion in Sec.~\ref{sec:2-layer}, we can modify the noise-affected expansion map in Table~\ref{tab:noisy_expansion_maps} as 
\begin{equation}
    \tilde{\mathcal{A}}_k^{(1)}(\bullet) = \hat{\mathcal{N}}_{\mathcal{V}_k, \mathcal{\tilde{V}}_k} \circ \mathcal{W}_k^{(1)}(\,\bullet\, \otimes \ket{\bar 0}\bra{\bar 0}).
\end{equation}
Due the presence of the channel $\hat{\mathcal{N}}_{\mathcal{V}_k, \mathcal{\tilde{V}}_k}$, the resulting map $\tilde{\mathcal{A}}_{k}^{(1)}$ is an unphysical map in general; however, under the assumption of $\mathcal{N}_i$ to be unital, it becomes a physical map, thereby implying $\tilde{\rho}_{\rm HT}^{(1)}$ is a density operator. 

\vspace{1\baselineskip}
\noindent\textbf{Type(ii)~~} Similar to the discussion for type (i), the noisy expansion map is represented as 
\begin{equation}
    \tilde{\mathcal{A}}_{k}^{(2)}(\bullet) = \mathrm{Tr}_{12} \left[ \ket{\rm Bell}\bra{\rm Bell}_{12} \left( \bullet_1 \otimes \hat{\mathcal{N}}_{\mathcal{V}_k, \mathcal{\tilde{V}}_k}(\sigma_{k,23}^{(2)})\right) \right].
\end{equation}
Again, due to the influence of $\hat{\mathcal{N}}_{\mathcal{V}_k, \mathcal{\tilde{V}}_k}$, the resulting map $\tilde{\mathcal{A}}_{k}^{(2)}$ is generally an unphysical map. However, it becomes a physical map under the unital noise assumption, resulting in $\tilde{\rho}_{\rm HT}^{(2)}$ being a density operator. 

\vspace{1\baselineskip}
\noindent\textbf{Type(iii)~~} 
In the contraction of local tensors in type (iii), we estimate the expectation value of $\hat{O}^{\vec{i}_k,\vec{i}'_k}_k$ by expanding $\hat{O}^{\vec{i}_k,\vec{i}'_k}_k=P_{k}^{\vec{i}_k} O_k P_k^{\vec{i}'_k}$ into a linear combination of Pauli observables $\hat{O}_k=\sum_{l_k} f_{l_k} P_{l_k}$, performing direct measurements for each $P_{l_k}$, and obtaining the expectation value of $\hat{O}^{\vec{i}_k,\vec{i}'_k}_k$ by combing the outcomes with $f_{l_k}$. Thus, the estimation of $M_k^{(3)}$ in type (iii) requires the implementation of different unitary operators $V_{l_k}$ for measuring  $P_{l_k}:=V_{l_k}\Lambda_{l_k}V_{l_k}^{\dagger}$.
As a result, different coherent errors occurring in the implementation of $V_{l_k}$ lead to an unphysical-noise effect on the expansion map $\tilde{\mathcal{A}}_k^{(3)}$, resulting in $\tilde{\rho}_{\rm HT}^{(3)}$ becoming an unphysical state. This emergence of unphysical noise arising from the assumption of measurement noise in type (iii) is analogous to the emergence of unphysical noise observed in type (iv) in Sec.~\ref{sec: noise analysis, case4}, where different quantum circuits were used in the estimation of $M_k$, leading to different types of noise effects being mapped onto the expansion map.

This conclusion holds even if the measurement errors in the computational basis are modeled as unital noise. 

\subsection{The effect of statistical noise}
We now discuss the differences in the impact of gate noise and statistical noise in expectation value estimation tasks using hybrid tensor networks. For simplicity, consider the scenario where noise occurs in the 2-layer HTTN state described by Eq.~(\ref{eq: Op-based rep}). In the contraction of local systems $k=1,...,N$ and a main system, we model two types of noise-induced modifications:
\begin{itemize}
    \item[1] Gate noise, which changes $M_k^{(\xi)}$ into $\tilde{M}_k^{(\xi)}$, $S_k^{(\xi)}$ into $\tilde{S}_k^{(\xi)}$, and $\ket{\psi}\bra{\psi}$ into $\rho$.
    \item[2] Statistical noise arising from the estimation of $M_k^{(\xi)}$ and $S_k^{(\xi)}$, futher altering $\tilde{M}_k^{(\xi)}$ into $\tilde{M}_k^{(\xi)}+\delta \tilde{M}_k^{(\xi)}$ and $\tilde{S}_k^{(\xi)}$ into $\tilde{S}_k^{(\xi)}+\delta \tilde{S}_k^{(\xi)}$.
\end{itemize}
To analyze these effects, we evaluate the absolute difference between 
\begin{equation}
    E_{\rm min} := \min_{\sigma\,\text{is a quantum state}} \mathrm{Tr}[\,O\,\sigma\,]~~~\text{and}~~~~\frac{\mathrm{Tr}\left[\otimes_{k=1}^{N}(\tilde{M}_k^{(\xi)}+\delta \tilde{M}_k^{(\xi)}) \, \rho \right]}{\mathrm{Tr}\left[\otimes_{k=1}^{N}(\tilde{S}_k^{(\xi)}+\delta \tilde{S}_k^{(\xi)}) \, \rho \right]} 
\end{equation}
yielding the following inequality.
\begin{equation}
    \left\lvert E_{\rm min}  - \frac{\mathrm{Tr}\left[\otimes_{k=1}^{N}(\tilde{M}_k^{(\xi)}+\delta \tilde{M}_k^{(\xi)}) \, \rho \right]}{\mathrm{Tr}\left[\otimes_{k=1}^{N}(\tilde{S}_k^{(\xi)}+\delta \tilde{S}_k^{(\xi)}) \, \rho \right]}  \right\rvert 
    \leq \left\lvert E_{\rm min} - \mathrm{Tr}\left(O \, \tilde{\rho}_{\rm HT} \right) \right\rvert ~~~+~~~\left\lvert \mathrm{Tr}\left(O \, \tilde{\rho}_{\rm HT} \right) - \frac{\mathrm{Tr}\left[\otimes_{k=1}^{N}(\tilde{M}_k^{(\xi)}+\delta \tilde{M}_k^{(\xi)}) \, \rho \right]}{\mathrm{Tr}\left[\otimes_{k=1}^{N}(\tilde{S}_k^{(\xi)}+\delta \tilde{S}_k^{(\xi)}) \, \rho \right]}  \right\rvert.
\end{equation}
The first term on the right-hand side reflects gate noise. As shown in the main text,
$\tilde{\rho}_{\rm HT}$ is guaranteed to be a valid quantum state, so its error contribution can be mitigated through variational updates. On the other hand,
the second term reflects the effect of statistical noise, which can be viewed as a fluctuation around the first term. To keep this term within $\epsilon\in[0,1]$ with high probability, the required number of shots $x_{\rm tot}=\sum_{k=0}^{N} x_{k}$ where $x_k$ ($k=1,...,N$) and $x_0$ is the number of shots used in the contraction of the local systems $k$ and the main system) depends on the variety of conditions, such as the bond dimensions $2^{b_k}$ ($k=1,...,N$), the types $\xi$, the specific estimation procedures for $M_k^{(\xi)}$, and $S_k^{(\xi)}$, the locality of $O_k$. Minimizing $x_{\rm tot}$ (i.e., establishing the sample complexity for the estimation value estimation in HTNs) is a practically important problem, and we leave it as future research.

\end{document}